\documentstyle[10pt,epsf,epsfig,hangcaption,xspace,amssymb,amsfonts,amsmath,amsthm,cite,
dp_delphititle]{dp_delphi}
\makeindex
\def\DpPaperGroup{EP-PH}
\def\DpPaperRef{2005-015}
\def\DpDate{22 March 2005}
\def\DpAuthors{DELPHI Collaboration}
\def\DpSubmit{(Accepted by Euro. Phys. Journ.)}
\def\DpTitle{{Determination 
of heavy quark non-perturbative parameters from spectral moments in 
semileptonic B decays
}}
\def\DpComment{ }
\def\DpEMail{
}

\newcommand{\ps}{{\rm ps}}

\newcommand{\Dsm}{{\rm D}_s^-}
\newcommand{\Dp}{{\rm D}^+}

\newcommand{\Do}{{\rm D}^0}

\newcommand{\Dstar}{{\rm D}^{\ast}}

\newcommand{\Dstarp}{{\rm D}^{\ast +}}
\newcommand{\Dstarstar}{{\rm D}^{\ast \ast}}
\newcommand{\Dstarstarp}{{\rm D}^{\ast \ast +}}
\newcommand{\Dstarstaro}{{\rm D}^{\ast \ast 0}}

\newcommand{\pisstar}{\pi^{\ast \ast}}

\newcommand{\Vcb}{\left | {\rm V}_{cb} \right |}

\newcommand{\Bsb}{\overline{\rm{B^0_s}}}

\newcommand{\Bm}{\rm{B^{-}}}

\newcommand{\Bdb}{\overline{\rm{B^{0}_{d}}}}
\newcommand{\Lb}{\Lambda^0_b}

\newcommand{\Bstar}{{\rm B}^{\ast}}

\newcommand{\Dsstar}{\rm{D}^{\ast \ast}}
\newcommand{\Km}{\rm{K}^-}
\newcommand{\Kp}{\rm{K}^+}

\newcommand{\GeV}{\rm{GeV}}

\newcommand{\GeVc}{\rm{GeV/}{\it c}}
\newcommand{\GeVcd}{\rm{GeV/}{\it c}^2}

\newcommand{\MeVc}{\rm{MeV/}{\it c}}
\newcommand{\MeVcd}{\rm{MeV/}{\it c}^2}

\newcommand{\mumu}{\ifmmode {\mu^+\mu^-} \else ${\mu^+\mu^-} $ \fi}

\newcommand{\beq}{\begin{equation}}
\newcommand{\eeq}{\end{equation}}
\newcommand{\ba}{\begin{array}}
\newcommand{\ea}{\end{array}}
\newcommand{\bc}{\begin{center}}
\newcommand{\ec}{\end{center}}
\newcommand{\be}{\begin{eqnarray}}
\newcommand{\bes}{\begin{eqnarray*}}
\newcommand{\ees}{\end{eqnarray*}}
\newcommand{\Kz}{\ifmmode {\rm K^0_s} \else ${\rm K^0_s} $ \fi}
\newcommand{\Zz}{\ifmmode {\rm Z} \else ${\rm Z } $ \fi}
\newcommand{\qqbar}{\ifmmode {\rm q\bar{q}} \else ${\rm q\bar{q}} $ \fi}
\newcommand{\ccbar}{\ifmmode {\rm c\bar{c}} \else ${\rm c\bar{c}} $ \fi}
\newcommand{\bbbar}{\ifmmode {\rm b\bar{b}} \else ${\rm b\bar{b}} $ \fi}
\newcommand{\xxbar}{\ifmmode {\rm x\bar{x}} \else ${\rm x\bar{x}} $ \fi}
\newcommand{\rphi}{\ifmmode {\rm R\phi} \else ${\rm R\phi} $ \fi}
\begin{document}
%%%%%%%%%%%%%%%%%%%%%%%%%% They are a problem with Coll.Sty ?
\makeatletter
%\input{dp_system:coll.sty}
% Collapse citation numbers to ranges.  Non-numeric and undefined labels
% are handled.  No sorting is done.  E.g., 1,3,2,3,4,5,foo,1,2,3,?,4,5
% gives 1,3,2-5,foo,1-3,?,4,5
\newcount\@tempcntc
\def\@citex[#1]#2{\if@filesw\immediate\write\@auxout{\string\citation{#2}}\fi
  \@tempcnta\z@\@tempcntb\m@ne\def\@citea{}\@cite{\@for\@citeb:=#2\do
    {\@ifundefined
       {b@\@citeb}{\@citeo\@tempcntb\m@ne\@citea\def\@citea{,}{\bf ?}\@warning
       {Citation `\@citeb' on page \thepage \space undefined}}%
    {\setbox\z@\hbox{\global\@tempcntc0\csname b@\@citeb\endcsname\relax}%
     \ifnum\@tempcntc=\z@ \@citeo\@tempcntb\m@ne
       \@citea\def\@citea{,}\hbox{\csname b@\@citeb\endcsname}%
     \else
      \advance\@tempcntb\@ne
      \ifnum\@tempcntb=\@tempcntc
      \else\advance\@tempcntb\m@ne\@citeo
      \@tempcnta\@tempcntc\@tempcntb\@tempcntc\fi\fi}}\@citeo}{#1}}
\def\@citeo{\ifnum\@tempcnta>\@tempcntb\else\@citea\def\@citea{,}%
  \ifnum\@tempcnta=\@tempcntb\the\@tempcnta\else
   {\advance\@tempcnta\@ne\ifnum\@tempcnta=\@tempcntb \else \def\@citea{--}\fi
    \advance\@tempcnta\m@ne\the\@tempcnta\@citea\the\@tempcntb}\fi\fi}
 
\makeatother
%%%%%%%%%%%%%%%%%%%%%%%%%% ??????????????????????????????????
% Generate the title page
\begin{titlepage}
\pagenumbering{roman}
\CERNpreprint{\DpPaperGroup}{\DpPaperRef} % Reference of the paper
\date{{\small\DpDate}} % Date of the paper
\title{\DpTitle} % Title of the paper
\address{\DpAuthors} % General name of the author(s)
\begin{shortabs} % Start the abstract
\noindent
%   abstract.tex
%
\noindent

%=========================================================================%
%===================> DELPHI note abstract =====> To be filled <=====%
Moments of the hadronic invariant mass
%, above the $\Dstar$ meson mass,
and of the lepton energy spectra in semileptonic 
B decays have been determined with the data recorded by the DELPHI
detector at LEP. 
%Using six of these moments:
%\begin{eqnarray}
%<m_{D^{**}}> &=& (2.483 \pm 0.033 \pm 0.036 )~(\GeVcd) \nonumber \\
%<m_{D^{**}}^2>  & =&(6.22 \pm 0.16\pm 0.17 )~(\GeVcd)^2 \nonumber \\
%<m_{D^{**}}^4> & =&(40.1 \pm 2.0\pm 1.9 )~(\GeVcd)^4 \nonumber \\
%<m_{D^{**}}^6> & =&(270.6 \pm 20.9\pm 12.0 )~(\GeVcd)^6 \nonumber \\
%<m_{D^{**}}^8> & =&( 1932.\pm 206. \pm  )~(\GeVcd)^8 \nonumber \\
%<m_{D^{**}}^{10}> & =&( 14732.\pm 2039.\pm  )~(\GeVcd)^{10} \nonumber \\
%<m_H^2-m_{spin}^2> &= & (0.647 \pm 0.046 \pm  0.090 )~(\GeVcd)^2  \nonumber\\
%<(m_H^2-<m_H^2>)^2> &= & (1.56 \pm 0.18 \pm 0.16 )~(\GeVcd)^4  \nonumber \\
%<(m_H^2-<m_H^2>)^3> &= & (4.05 \pm 0.74 \pm 0.32 )~(\GeVcd)^6 \nonumber \\
%<E_{\ell}^*> & = &(1.380\pm 0.005 \pm 0.011)~({\rm GeV}) \nonumber  \\
%<(E_{\ell}^*-<E_{\ell}^*>)^2> & = &(0.179 \pm 0.003 \pm 0.006)~({\rm GeV}^2) \nonumber \\ 
%<(E_{\ell}^*-<E_{\ell}^*>)^3> & = &(-0.028 \pm 0.001 \pm 0.004)~({\rm GeV}^3) \nonumber
%\end{eqnarray}
From measurements of the inclusive $b$-hadron semileptonic decays, and
imposing constraints from other measurements on $b$- and $c$-quark masses,
the first three moments of the lepton energy distribution and of the hadronic 
mass distribution, have been used to determine parameters which enter 
into the extraction of $\Vcb$ from the measurement of the inclusive 
$b$-hadron semileptonic decay width. The values obtained in the kinetic scheme
are:

%Using the first three moments of the lepton energy and of the hadronic 
%mass distribution, and imposing constraints from other measurements
%on $b$- and $c$-quark masses,
% parameters entering in the extraction of $\Vcb$ from the measurement
%of the inclusive $b$-hadron partial semileptonic decay width have been 
%obtained including corrections at order $1/m_b^3$:

%Measured values of some of these moments have been used as inputs in a
%multiparameter  fit to extract the heavy quark masses, the b quark
%kinetic energy inside the heavy hadron and
%the leading coefficients of the $1/m_b^3$ term in the heavy quark expansion.

\begin{eqnarray}
m_b(1~{\rm{GeV}}) & =& 4.591 \pm 0.062  \pm 0.039 \pm 0.005~\GeVcd \nonumber \\
m_c(1~{\rm{GeV}})& =& 1.170  \pm 0.093 \pm 0.055 \pm 0.005~\GeVcd \nonumber \\
\mu_{\pi}^2(1~{\rm{GeV}})& =& 0.399  \pm 0.048 \pm 0.034 \pm 0.087~{\rm GeV}^2 \nonumber \\
\tilde{\rho}_D^3 & = &0.053 \pm 0.017 \pm 0.011 \pm 0.026~{\rm GeV}^3, \nonumber 
\end{eqnarray}
and include corrections at order $1/m_b^3$.

Using these results, and present measurements of the inclusive semileptonic
decay partial width of $b$-hadrons at LEP, an accurate determination of $\Vcb$
is obtained:
\begin{equation}
\Vcb = 0.0421 \times \left ( 1 \pm 0.014_{\,meas.}
\pm  0.014_{\,fit} 
\pm  0.015_{\,th.}
\right ). \nonumber
\end{equation}
%=========================================================================%

\end{shortabs}
\vfill
\begin{center}
\DpSubmit \ \\ % Horrible hack to allow to have DpSubmit empty
\DpComment \ \\
\DpEMail \ \\
\end{center}
\vfill
\clearpage
\headsep 10.0pt
\addtolength{\textheight}{10mm}
\addtolength{\footskip}{-5mm}
\begingroup
% Commands to process the author names
\newcommand{\DpName}[2]{\hbox{#1$^{\ref{#2}}$},\hfill}
\newcommand{\DpNameTwo}[3]{\hbox{#1$^{\ref{#2},\ref{#3}}$},\hfill}
\newcommand{\DpNameThree}[4]{\hbox{#1$^{\ref{#2},\ref{#3},\ref{#4}}$},\hfill}
\newskip\Bigfill \Bigfill = 0pt plus 1000fill
\newcommand{\DpNameLast}[2]{\hbox{#1$^{\ref{#2}}$}\hspace{\Bigfill}}
%\small
\footnotesize
\noindent
\DpName{J.Abdallah}{LPNHE}
\DpName{P.Abreu}{LIP}
\DpName{W.Adam}{VIENNA}
\DpName{P.Adzic}{DEMOKRITOS}
\DpName{T.Albrecht}{KARLSRUHE}
\DpName{T.Alderweireld}{AIM}
\DpName{R.Alemany-Fernandez}{CERN}
\DpName{T.Allmendinger}{KARLSRUHE}
\DpName{P.P.Allport}{LIVERPOOL}
\DpName{U.Amaldi}{MILANO2}
% ...
\normalsize
\endgroup
\titlefoot{Department of Physics and Astronomy, Iowa State
     University, Ames IA 50011-3160, USA
    \label{AMES}}
\titlefoot{Physics Department, Universiteit Antwerpen,
     Universiteitsplein 1, B-2610 Antwerpen, Belgium \\
     \indent~~and IIHE, ULB-VUB,
     Pleinlaan 2, B-1050 Brussels, Belgium \\
     \indent~~and Facult\'e des Sciences,
     Univ. de l'Etat Mons, Av. Maistriau 19, B-7000 Mons, Belgium
    \label{AIM}}
\titlefoot{Physics Laboratory, University of Athens, Solonos Str.
     104, GR-10680 Athens, Greece
    \label{ATHENS}}
\titlefoot{Department of Physics, University of Bergen,
     All\'egaten 55, NO-5007 Bergen, Norway
    \label{BERGEN}}
\titlefoot{Dipartimento di Fisica, Universit\`a di Bologna and INFN,
     Via Irnerio 46, IT-40126 Bologna, Italy
    \label{BOLOGNA}}
\titlefoot{Centro Brasileiro de Pesquisas F\'{\i}sicas, rua Xavier Sigaud 150,
     BR-22290 Rio de Janeiro, Brazil \\
     \indent~~and Depto. de F\'{\i}sica, Pont. Univ. Cat\'olica,
     C.P. 38071 BR-22453 Rio de Janeiro, Brazil \\
     \indent~~and Inst. de F\'{\i}sica, Univ. Estadual do Rio de Janeiro,
     rua S\~{a}o Francisco Xavier 524, Rio de Janeiro, Brazil
    \label{BRASIL}}
\titlefoot{Coll\`ege de France, Lab. de Physique Corpusculaire, IN2P3-CNRS,
     FR-75231 Paris Cedex 05, France
    \label{CDF}}
\titlefoot{CERN, CH-1211 Geneva 23, Switzerland
    \label{CERN}}
\titlefoot{Institut de Recherches Subatomiques, IN2P3 - CNRS/ULP - BP20,
     FR-67037 Strasbourg Cedex, France
    \label{CRN}}
\titlefoot{Now at DESY-Zeuthen, Platanenallee 6, D-15735 Zeuthen, Germany
    \label{DESY}}
\titlefoot{Institute of Nuclear Physics, N.C.S.R. Demokritos,
     P.O. Box 60228, GR-15310 Athens, Greece
    \label{DEMOKRITOS}}
\titlefoot{FZU, Inst. of Phys. of the C.A.S. High Energy Physics Division,
     Na Slovance 2, CZ-180 40, Praha 8, Czech Republic
    \label{FZU}}
\titlefoot{Dipartimento di Fisica, Universit\`a di Genova and INFN,
     Via Dodecaneso 33, IT-16146 Genova, Italy
    \label{GENOVA}}
\titlefoot{Institut des Sciences Nucl\'eaires, IN2P3-CNRS, Universit\'e
     de Grenoble 1, FR-38026 Grenoble Cedex, France
    \label{GRENOBLE}}
\titlefoot{Helsinki Institute of Physics and Department of Physical Sciences,
     P.O. Box 64, FIN-00014 University of Helsinki, 
     \indent~~Finland
    \label{HELSINKI}}
\titlefoot{Joint Institute for Nuclear Research, Dubna, Head Post
     Office, P.O. Box 79, RU-101 000 Moscow, Russian Federation
    \label{JINR}}
\titlefoot{Institut f\"ur Experimentelle Kernphysik,
     Universit\"at Karlsruhe, Postfach 6980, DE-76128 Karlsruhe,
     Germany
    \label{KARLSRUHE}}
\titlefoot{Institute of Nuclear Physics PAN,Ul. Radzikowskiego 152,
     PL-31142 Krakow, Poland
    \label{KRAKOW1}}
\titlefoot{Faculty of Physics and Nuclear Techniques, University of Mining
     and Metallurgy, PL-30055 Krakow, Poland
    \label{KRAKOW2}}
\titlefoot{Universit\'e de Paris-Sud, Lab. de l'Acc\'el\'erateur
     Lin\'eaire, IN2P3-CNRS, B\^{a}t. 200, FR-91405 Orsay Cedex, France
    \label{LAL}}
\titlefoot{School of Physics and Chemistry, University of Lancaster,
     Lancaster LA1 4YB, UK
    \label{LANCASTER}}
\titlefoot{LIP, IST, FCUL - Av. Elias Garcia, 14-$1^{o}$,
     PT-1000 Lisboa Codex, Portugal
    \label{LIP}}
\titlefoot{Department of Physics, University of Liverpool, P.O.
     Box 147, Liverpool L69 3BX, UK
    \label{LIVERPOOL}}
\titlefoot{Dept. of Physics and Astronomy, Kelvin Building,
     University of Glasgow, Glasgow G12 8QQ
    \label{GLASGOW}}
\titlefoot{LPNHE, IN2P3-CNRS, Univ.~Paris VI et VII, Tour 33 (RdC),
     4 place Jussieu, FR-75252 Paris Cedex 05, France
    \label{LPNHE}}
\titlefoot{Department of Physics, University of Lund,
     S\"olvegatan 14, SE-223 63 Lund, Sweden
    \label{LUND}}
\titlefoot{Universit\'e Claude Bernard de Lyon, IPNL, IN2P3-CNRS,
     FR-69622 Villeurbanne Cedex, France
    \label{LYON}}
\titlefoot{Dipartimento di Fisica, Universit\`a di Milano and INFN-MILANO,
     Via Celoria 16, IT-20133 Milan, Italy
    \label{MILANO}}
\titlefoot{Dipartimento di Fisica, Univ. di Milano-Bicocca and
     INFN-MILANO, Piazza della Scienza 2, IT-20126 Milan, Italy
    \label{MILANO2}}
\titlefoot{IPNP of MFF, Charles Univ., Areal MFF,
     V Holesovickach 2, CZ-180 00, Praha 8, Czech Republic
    \label{NC}}
\titlefoot{NIKHEF, Postbus 41882, NL-1009 DB
     Amsterdam, The Netherlands
    \label{NIKHEF}}
\titlefoot{National Technical University, Physics Department,
     Zografou Campus, GR-15773 Athens, Greece
    \label{NTU-ATHENS}}
\titlefoot{Physics Department, University of Oslo, Blindern,
     NO-0316 Oslo, Norway
    \label{OSLO}}
\titlefoot{Dpto. Fisica, Univ. Oviedo, Avda. Calvo Sotelo
     s/n, ES-33007 Oviedo, Spain
    \label{OVIEDO}}
\titlefoot{Department of Physics, University of Oxford,
     Keble Road, Oxford OX1 3RH, UK
    \label{OXFORD}}
\titlefoot{Dipartimento di Fisica, Universit\`a di Padova and
     INFN, Via Marzolo 8, IT-35131 Padua, Italy
    \label{PADOVA}}
\titlefoot{Rutherford Appleton Laboratory, Chilton, Didcot
     OX11 OQX, UK
    \label{RAL}}
\titlefoot{Dipartimento di Fisica, Universit\`a di Roma II and
     INFN, Tor Vergata, IT-00173 Rome, Italy
    \label{ROMA2}}
\titlefoot{Dipartimento di Fisica, Universit\`a di Roma III and
     INFN, Via della Vasca Navale 84, IT-00146 Rome, Italy
    \label{ROMA3}}
\titlefoot{DAPNIA/Service de Physique des Particules,
     CEA-Saclay, FR-91191 Gif-sur-Yvette Cedex, France
    \label{SACLAY}}
\titlefoot{Instituto de Fisica de Cantabria (CSIC-UC), Avda.
     los Castros s/n, ES-39006 Santander, Spain
    \label{SANTANDER}}
\titlefoot{Inst. for High Energy Physics, Serpukov
     P.O. Box 35, Protvino, (Moscow Region), Russian Federation
    \label{SERPUKHOV}}
\titlefoot{J. Stefan Institute, Jamova 39, SI-1000 Ljubljana, Slovenia
     and Laboratory for Astroparticle Physics,\\
     \indent~~Nova Gorica Polytechnic, Kostanjeviska 16a, SI-5000 Nova Gorica, Slovenia, \\
     \indent~~and Department of Physics, University of Ljubljana,
     SI-1000 Ljubljana, Slovenia
    \label{SLOVENIJA}}
\titlefoot{Fysikum, Stockholm University,
     Box 6730, SE-113 85 Stockholm, Sweden
    \label{STOCKHOLM}}
\titlefoot{Dipartimento di Fisica Sperimentale, Universit\`a di
     Torino and INFN, Via P. Giuria 1, IT-10125 Turin, Italy
    \label{TORINO}}
%\titlefoot{INFN,Sezione di Torino, and Dipartimento di Fisica Teorica,
%     Universit\`a di Torino, Via P. Giuria 1,\\
%     \indent~~IT-10125 Turin, Italy
\titlefoot{INFN,Sezione di Torino and Dipartimento di Fisica Teorica,
     Universit\`a di Torino, Via Giuria 1,
     IT-10125 Turin, Italy
    \label{TORINOTH}}
\titlefoot{Dipartimento di Fisica, Universit\`a di Trieste and
     INFN, Via A. Valerio 2, IT-34127 Trieste, Italy \\
     \indent~~and Istituto di Fisica, Universit\`a di Udine,
     IT-33100 Udine, Italy
    \label{TU}}
\titlefoot{Univ. Federal do Rio de Janeiro, C.P. 68528
     Cidade Univ., Ilha do Fund\~ao
     BR-21945-970 Rio de Janeiro, Brazil
    \label{UFRJ}}
\titlefoot{Department of Radiation Sciences, University of
     Uppsala, P.O. Box 535, SE-751 21 Uppsala, Sweden
    \label{UPPSALA}}
\titlefoot{IFIC, Valencia-CSIC, and D.F.A.M.N., U. de Valencia,
     Avda. Dr. Moliner 50, ES-46100 Burjassot (Valencia), Spain
    \label{VALENCIA}}
\titlefoot{Institut f\"ur Hochenergiephysik, \"Osterr. Akad.
     d. Wissensch., Nikolsdorfergasse 18, AT-1050 Vienna, Austria
    \label{VIENNA}}
\titlefoot{Inst. Nuclear Studies and University of Warsaw, Ul.
     Hoza 69, PL-00681 Warsaw, Poland
    \label{WARSZAWA}}
\titlefoot{Fachbereich Physik, University of Wuppertal, Postfach
     100 127, DE-42097 Wuppertal, Germany \\
\noindent
{$^\dagger$~deceased}
    \label{WUPPERTAL}}
\addtolength{\textheight}{-10mm}
\addtolength{\footskip}{5mm}
\clearpage
\headsep 30.0pt
\end{titlepage}
%%%%%%%%%%%%%%%%%%%%%%%%%
%
% Change for the document body
%%\pagestyle{heading} % for page numbering
\pagenumbering{arabic} % page numbering in number
\setcounter{footnote}{0} %
\large
%%\linenumbers %%%CD
 \section {Introduction}
\label{sec:intro}

%The extraction of the $|V_{cb}|$ and $|V_{ub}|$ elements of the CKM mixing 
%matrix from inclusive semi-leptonic (s.l.) decays relies on the accurate 
%knowledge of heavy quark masses, 
%%the $b$-quark kinematics inside the heavy hadron, 
%the control of perturbative QCD corrections 
%and the determination of non-perturbative parameters arising from higher order
%$\left ( \emptyset (1/m_b^n),~n>1\right )$ 
%correction. Determining these CKM elements with small and well understood 
%errors, requires 
%%not only to precisely know these values but 
%also to test 
%the consistency of Heavy Quark Expansion predictions for inclusive s.l. 
%B decays. 
%Extracting values for the different 
% parameters, directly from the data, has become a key issue. 
Several years ago it was proposed to obtain an 
accurate value of 
$|V_{cb}|$ by comparing the measurement of the inclusive semileptonic decay 
partial width
in the process $b \rightarrow c \ell^- \overline{\nu}_{\ell}$, with the corresponding
theoretical expression, obtained using the Operator Product Expansion
(OPE) formalism, 
applied in the 
heavy quark mass limit \cite{ref:bigiold}. 
%% PR d2
A recent appraisal of the limitations of this approach can be found in 
\cite{ref:bigi2} from which the following expression, for the semileptonic 
decay width, has been taken:
\begin{eqnarray}
%{\scriptsize
\Gamma_{sl}(b \rightarrow c \ell^- \overline{\nu}_{\ell}) & =&
\frac{G_F^2 m_b^5(\mu)}{192 \pi^3} \left | V_{cb} \right |^2 
\left ( 1+ A_{ew}\right )
A^{pert}(r,\mu) \nonumber \\
 & & \left [ z_0(r) \left ( 
1- \frac{\mu_{\pi}^2(\mu)-\mu_{G}^2(\mu)+\frac{\rho_D^3(\mu)+\rho_{LS}^3(\mu)}
{m_b(\mu)}}{2 m_b^2(\mu)} \right ) \right .  \nonumber \\
 & - & \left . 2(1-r)^4 \frac{\mu_{G}^2(\mu)-\frac{\rho_D^3(\mu)+\rho_{LS}^3(\mu)}
{m_b(\mu)}}{ m_b^2(\mu)} +d(r) \frac{\rho_D^3(\mu)}{m_b^3(\mu)} + ...
\right ]. \label{eq:gammasl}
%}
\end{eqnarray}
In this expression, $z_0(r)$ is the tree-level phase space factor and 
$r=m_c^2(\mu)/m_b^2(\mu)$. Definitions for the other quantities are 
given in \cite{ref:bigi2}. Equation~(\ref{eq:gammasl}) contains an 
expansion in $\alpha_s$, corresponding to perturbative QCD corrections
expressed in $A^{pert}(r,\mu)$, and an expansion in $1/m_b$, corresponding
to non-perturbative QCD contributions. An auxiliary scale $\mu (= 1\,\GeV)$
is introduced to demark the border between long- and short-distance dynamics 
in OPE. Numerically, $A_{ew}$, corresponding to the ultraviolet 
renormalization 
of the Fermi interaction, is well-known  
and amounts to +1.4$\%$ \cite{ref:sirlin};
$A^{pert}(r,\mu)$ corresponds to $\sim -9 \%$ corrections and the 
non-perturbative terms have typically few percent 
contributions \cite{ref:bigi2}.
The smallness of these last corrections comes, partly, from the fact
that the term proportional to $1/m_b$ is absent \cite{ref:chay,Bigi:1992su} in 
Equation~(\ref{eq:gammasl}). 
The quantities $\mu_{\pi}^2$, $\mu_{G}^2$, $\rho_D^3$
and $\rho_{LS}^3$ denote the expectation values of the kinetic,
chromomagnetic, Darwin and spin-orbit operators respectively.
These parameters have to be determined by experiment. From the mass splitting 
between $\Bstar$ and B meson the following value 
$\mu_{G}^2(1\,\GeV)= (0.35 ^{+0.03}_{-0.02})\GeV^2$ has been obtained
\cite{Uraltsev:2001ih}. The value for $\mu_{\pi}^2$ is less certain;
in this regularization scheme the inequality 
$\mu_{\pi}^2(\mu) \geq \mu_{G}^2(\mu)$ holds for any normalization scale.
Constraints have been also established for terms contributing at order
$1/m_b^3$. The Darwin parameter, $\rho_D^3(\mu)$, must be positive and 
the spin-orbit operator value, $\rho_{LS}^3(\mu)$, is expected to
be negative and to satisfy $-\rho_{LS}^3 \leq\rho_D^3 $ \cite{ref:bigi2}. 
In addition,
it has been demonstrated that the value of 
$\Gamma_{sl}(b \rightarrow c \ell^- \overline{\nu}_{\ell})$ 
is rather insensitive to the exact value of $\rho_{LS}^3$ \cite{ref:amsterl}.
 
A few lessons can be drawn from the previous considerations:
\begin{itemize}
\item the largest correction, to the naive free quark decay model, 
is expected to originate from perturbative QCD.
Its evaluation  is closely connected to the definition
adopted for the running quark masses;
\item in addition to the values of heavy quark masses, $m_b$ and $m_c$,
two parameters need to be determined by experiment, 
$\mu_{\pi}^2$ and $\rho_D^3$, to have a control of non-perturbative
QCD corrections up to the $1/m_b^3$ order;
\item the definition for the quark masses has to be consistent with both 
perturbative and non-perturbative dynamics.
\end{itemize}

%As will be discussed in Section~\ref{sec:discussion}, it is now 
It is then expected  that the value of $|V_{cb}|$, 
extracted in this way from inclusive
$b$-hadron semileptonic decays, can be  determined with a relative
uncertainty from theory at the 1.5$\%$ level
(see \cite{ref:bigi2} for a detailed breakdown of contributing
sources in this number). 
%%end PR d2
To match this accuracy, an experimental control of the parameters governing
non-perturbative QCD corrections, even at a modest level,
is required and is the main purpose of the present analysis.
% at the same level of precision. 

The determination of moments of the lepton energy and of the hadronic
mass spectra in ${\rm B} \rightarrow {\rm X}_c \ell
\overline{\nu}_{\ell}$ decays provides important information on these
parameters since they can be analysed using the same formalism, and
since corresponding expressions depend on the same non-perturbative
parameters entering in Equation~(\ref{eq:gammasl}).  
The following notations for moments of the lepton energy spectrum 
have been used:
\begin{equation}
M_1^{\ell}=<E_{\ell}^*> ~{\rm and}~
M_n^{\prime \ell}=<(E_{\ell}^*-<E_{\ell}^*>)^n>,~n>1,
\end{equation}
where $E_{\ell}^*$ denotes the value of the lepton energy
in the $b$-hadron rest frame; and for moments of the hadronic mass 
system:
\begin{equation}\label{eq:defmomh}
M_n^H=<(m_H^2-m_{spin}^2)^n> ~{\rm and}~
M_n^{\prime H}=<(m_H^2-<m_H^2>)^n>.
\end{equation} 
$m_{spin}~=~1.97375~\GeVcd$ denotes the spin averaged D meson mass, which 
is equal
to the weighted average of the D and $\Dstar$ masses.
%%PR d2
The OPE expresses lepton moments through quark masses as a double
expansion in $\alpha_s$ and $1/m_b$:
\begin{equation}
M_n^{\prime \ell}=\left ( \frac{m_b}{2}\right )^n
\left [ \phi_n(r)+\overline{a}_n(r) \frac{\alpha_s}{\pi}
+\overline{b}_n(r) \frac{\mu_{\pi}^2}{m_b^2}
+\overline{c}_n(r) \frac{\mu_{G}^2}{m_b^2}
+\overline{d}_n(r) \frac{\rho_{D}^3}{m_b^3}
+\overline{s}_n(r) \frac{\rho_{LS}^3}{m_b^3} +...
\right ].
\nonumber
\end{equation}
The higher coefficient functions $\overline{b}_n(r)$,
$\overline{c}_n(r)$, ... are also perturbative series in $\alpha_s$.
Due to the kinematic definition of the hadronic invariant mass, $M_X$,
the general expression for the hadronic moments includes explicitly
the value of the {\it b}-hadron mass, $M_B$:
\begin{eqnarray}
M_n^{\prime H} = m_b^{2n} \sum_{l=0}
\left [ \frac{M_B-m_b}{m_b}\right ]^l
\left ( E_{nl}(r) \right . &+& a_{nl}(r)  \frac{\alpha_s}{\pi}
+b_{nl}(r)\frac{\mu_{\pi}^2}{m_b^2}
+c_{nl}(r)\frac{\mu_{G}^2}{m_b^2}  \nonumber \\
&+& \left .
d_{nl}(r)\frac{\rho_{D}^3}{m_b^3}
+s_{nl}(r)\frac{\rho_{LS}^3}{m_b^3} + ...  
\right ) .
\end{eqnarray}
%% end PRd2
Numerical values for all $r$-dependent functions entering into 
these expressions can be found in \cite{ref:amsterl}.

In the following, the $\mu$-scale independent third order correction term, 
$\tilde{\rho}_{D}^3$ has been fitted in place of 
$\rho_{D}^3$. The two quantities are related through the expression:
$\tilde{\rho}_{D}^3 \approx \rho_{D}^3 (1~{\rm GeV})-0.1 ~{\rm GeV}^3$
\cite{urabig}.

The consistency of the bounds set by moments of different
distributions, and with other data, tests the underlying theory
assumptions. 
%For the first time an experimental determination of
%non-perturbative corrections, contributing at the $1/m_b^3$ order, is
%provided in this article.

%% PR d2
It can be noted that the approach used here does not rely
on the validity of an expansion in $1/m_c$, as already advocated
in \cite{ref:amsterl}. The value of the charm
quark mass, entering into the expression of moments, is taken as
a parameter whose value has also been fitted using data.
%%end PR d2

Measurements of moments have been reported by the 
CLEO~\cite{cleo_mom1,cleo_mom2,cleo_mom3}, 
BaBar~\cite{babar_mom0,babar_mom1,babar_mom2} and 
BELLE~\cite{belle_mom1,belle_mom2}
Collaborations operating at the $\Upsilon(4S)$ resonance, and
by DELPHI using preliminary data taken at the $\Zz$ pole energy \cite{ref:amsterl}.
Results have also been recently published by CDF~\cite{cdf_mom}.
 
While there is an obvious advantage in measuring the energy 
spectra in events where the decaying B rest frame almost coincides with the 
laboratory frame, low energy particles cannot be identified there. It is thus 
necessary to rely on models for extrapolating the lepton energy spectrum to 
zero energy or to resort to computations for a truncated spectrum.  
Performing this analysis at energies around the $\Zz$ peak ensures sensitivity
to almost the full lepton spectrum, thus reducing modelling assumptions. 
The main challenge at the higher energy is,
for the lepton energy moments analysis, the accurate determination of 
the B rest frame. 
%The techniques adopted here are presented in Section~~\ref{sec:dstar}

This paper presents the results obtained from analyses of the data recorded 
with the DELPHI detector at LEP on moments of the 
hadronic mass and charged lepton energy distributions. 
The analysis procedures are discussed in 
Sections~\ref{sec:mx} and \ref{sec:el}. In Section~\ref{sec:fit}
these results are then used as inputs of a multi-parameter fit to determine 
the relevant corrections contributing at ${\cal O}(1/m_b^2)$ and 
${\cal O}(1/m_b^3)$, together with the heavy quark masses. 
The use of higher moments guarantees 
a sensitivity to these parameters and the simultaneous use of the hadronic and leptonic 
spectra ensures that a larger number of parameters can be kept free in the fit. 
We discuss the results both in terms of the extraction of the parameters and 
the implications for $|V_{cb}|$, and as a consistency check of the underlying 
theoretical assumptions. 

%% ** A.O.** 4/5/2004 ** I have changed a bit the sentence below **
%{\it .. add something on D** properties measurement} 
In addition, production and decay properties of broad $\Dsstar$ states
have been studied. They are reported in Section~\ref{sec:dstar}.
$\Dstarstar$ refers, in the present analysis, 
to all hadronic systems of mass higher than the $\Dstar$. 
$b$-meson semileptonic decays with charmed hadrons emitted in the final state
correspond to $\overline{{\rm B}} \rightarrow 
{\rm D},~\Dstar,~\Dstarstar \ell^- \overline \nu_{\ell}$ transitions. 
$\Dstarstar$ states are about 30$\%$ of $b$-hadron semileptonic 
decays \cite{ref:dsstarargus,ref:dsstaraleph}. 
They can be resonant or non-resonant hadronic ${\rm D}^{(*)}n\pi$ systems. 
Resonant states 
are supposed to be dominated by L=1, orbitally excited $c \overline{q}$ states.
There are four such states: ${\rm D}_0^*$, ${\rm D}_1^*$, ${\rm D}_2^*$ 
and ${\rm D}_1$ with respectively 
${\rm J^P}=0^+,~1^+,~2^+$ and $1^+$. 
The two 1$^+$ states, having common final states, mix and 
physical states are expected to decay into almost pure D or S wave
$\Dstar \pi$ or ${\rm D} \rho$ final states \cite{ref:dsstarrosner}.
The two states (${\rm D}_2^*$ and ${\rm D}_1$) 
which, because of angular
momentum and parity conservation, have to decay into a D wave
are expected to be narrow and their measured widths are of the order
of 20~$\MeVcd$ \cite{ref:PDG02}.
The contribution from these narrow
resonances has been measured \cite{ref:dsstaraleph,ref:narrowcleo}
 and it amounts to about
one third of all $\Dstarstar$ states. Non-resonant or broad resonant 
states are thus expected to have a dominant contribution in $b$-hadron 
semileptonic
decays. In addition to L=1 mesons, there could be
states corresponding to higher angular momentum values or to radial
excitations. All such states are expected to be broad \cite{ref:quarkmod}.
Contributions from non-resonant D$\pi$ or $\Dstar \pi$ final 
states have been evaluated in \cite{ref:goity} and this model has been
considered in the present analysis.

In a previous DELPHI publication \cite{ref:dsstardelphi}, production
fractions of $\Dsstar$ states in $b$-hadron semileptonic decays have been 
measured. The main focus of the present analysis is on the mass distribution 
of these states. This was made possible by improving the separation
between signal and background events with respect to the previous analysis.
%These measurements have been used to determine the moments of the 
%hadronic mass distribution in $b$-hadron semileptonic decays and to measure
%parameters of non-perturbative QCD origin which contribute in the
%evaluation of $\Vcb$ from inclusive measurements.

%This study has complementary aspects when compared with the corresponding CLEO
%measurement \cite{ref:cleo_mom2}. 
%$\Dsstar$ states have been exclusively reconstructed
%whereas an inclusive approach has been followed by CLEO.
%As a result, the accuracy on the $\Dsstar$ mass measurement is higher 
%in the present study for states which are fully reconstructed
%but some model dependence is introduced to account for final
%states with missing neutrals or involving several pions. 
%Another difference comes from the fact that
%$b$-hadrons are fastly moving at LEP whereas they are produced almost at 
%rest in CLEO.
%Seen from the $b$-hadron rest frame, the minimal energy requirement on the 
%charged lepton 
%is lower in the present study than in the CLEO measurement. 

\section{Data analysis} 
This study is based on $b$-hadron semileptonic decays
%, selected from a sample
%of about 3.4~M $e^+e^- \rightarrow \Zz \rightarrow q \bar q$ events
recorded with the DELPHI detector at LEP from 1992 to 1995. 
Since the determination of the moments of the hadronic mass and
charged lepton energy distributions have different requirements, two
analyses have been performed. The first focuses on the exclusive
reconstruction of $\Dsstar$ states.
%In order to enhance the statistics, it requires only a loose lepton 
%selection. 
For the second analysis a lepton sample with low background is
required.
%Therefore more stringent cuts have been enforced in the lepton selection. 
This section presents 
those parts of the data selection and event reconstruction procedure
which are common to both 
analyses. 

\subsection{Hadronic event selection and simulation}
\label{sec:A}

In order to select hadronic $\Zz$ decays, standard hadronic selection
cuts have been applied.  Each event has been divided into two opposite
hemispheres by a plane orthogonal to the thrust axis. The polar 
angle \footnote{In the DELPHI coordinate system, $z$ is along the electron 
beam direction, 
$\phi$ and R are the azimuthal angle and radius in the $xy$ plane, 
and $\theta$ is the 
polar angle with respect to the $z$ axis.}
of the thrust axis of the event had to satisfy the requirement $\left
|\cos{\theta} \right |<0.95$.  Charged and neutral particles have been
clustered into jets using the LUCLUS \cite{ref:luc} algorithm with the
resolution parameter $d_{join}=5~ \GeVc$.  About 3.4 million events
have been selected from the LEP1 data sets.

The JETSET 7.3 Parton Shower \cite{ref:luc} program has been used to generate 
hadronic $\Zz$ decays, which were passed through the detailed detector 
simulation DELSIM \cite{ref:delsim} and processed by the same 
analysis chain as the data. A sample of about nine million
$\Zz \rightarrow q \overline{q}$ events has been used. To increase the 
simulation statistics, an additional sample of about 3.6 million 
$\Zz \rightarrow b \overline{b}$ events, equivalent to about 17 million hadronic 
$\Zz$ decays, has also been used. Statistics of the analysed hadronic samples are given in 
Table~\ref{tab:stat}.
\begin{table}[htb]
\begin{center}
  \begin{tabular}{|c|c|c|c|}
    \hline
Year &  Real data & Simulated & Simulated\\
  &      & $\Zz \rightarrow q \overline{q}$ & $\Zz \rightarrow b \overline{b}$\\
    \hline
 1992+1993 & 1355805  & 3916050 & 1096199    \\
 1994+1995 & 2012921 &  5012881 & 2495335   \\
\hline
 Total & 3368726  & 8928931  & 3591534    \\
    \hline
  \end{tabular}
  \caption[]{\it {Analysed number of events. In 1992 and 1993
only two-dimensional vertex reconstruction was available.}
  \label{tab:stat}}
\end{center}
\end{table}

$\Zz \rightarrow b \bar b$ events have been selected using an event $b$-tagging 
technique \cite{ref:btag} based on the reconstructed impact parameters of particle 
tracks.  

%$\Do$ mesons
% have been reconstructed using their decays into $\Km \pi^+$,
%%and $\Km \pi^+ \pi^+ \pi^-$. 
%$\Km \pi^+ \pi^+ \pi^-$ and $\Km \pi^+ (\pi^0)$. This last channel
%has been used only when considering charged $\Dstarp$ cascade decays.
%The $\Km \pi^+ \pi^+$ decay channel has been used for $\Dp$ mesons.
%$\Dstarp$ are measured using the 
%$\Do \pi^+$ transition and the charged pion is denoted as $\pistar$
%in the following. 

Events for the exclusive analysis have been  selected by requiring the presence of one 
tagged lepton candidate with momentum $p >$~2~GeV$/c$  and of a $\Do$, $\Dp$ or $\Dstarp$ 
candidate~\footnote{Throughout this paper charge-conjugate states are implicitly
included.}
in the same event hemisphere. 
%In addition, the mass of the 
%${\rm D}^{(*)}-\ell$ system has been restricted to be in the range between 2.5 and 
%5.5~$\GeVcd$.

In order to measure moments of the lepton energy distribution in
inclusive $b \rightarrow X_c \ell \overline{ \nu}_{\ell}$ decays,
events have been required to contain one tagged lepton candidate with
momentum $p >$~2.5~GeV$/c$ (for muons) or $p >$~3~GeV$/c$ (for electrons).
%These criteria have selected ??  events in the 1992-93 data and  250~k events 
%in the 1994-95 data.

\subsection{Muon identification}
\label{sec:muid}

Muons have been identified based on the response of the Muon Chambers. 
Details can be found in~\cite{ref:delsim}. 

For the inclusive lepton analysis muon candidates have been accepted if they
fulfilled the ``standard'' selection criteria, their momenta in the lab frame 
exceeded 2.5~GeV$/c$ and were contained within the polar angle intervals:
$|\cos\theta_{\mu}| < 0.62$ or $0.68 < |\cos\theta_{\mu}| < 0.94$,
defining the barrel and the forward regions.
The muon identification efficiency has been measured with $\Zz \rightarrow \mu^+\mu^-$ 
events, in the decays $\tau \rightarrow \mu \nu_{\tau} \overline{\nu}_{\mu}$ 
and in two-photon 
$\gamma \gamma \rightarrow \mu^+\mu^-$ events.
A  mean efficiency within the acceptance region of $0.82 \pm 0.01$ has been found, with little dependence on 
the muon momentum and on the track polar angle. This agrees with simulation, both in 
absolute value and in the momentum dependence, within a precision of 2\%.

%{(\it Patrick: ``For muons, the RD/MC ratios are equal to 
%$96\%$ for the two periods with a $\pm 2\%$ uncertainty.'' ??)}

The probability for a hadron to fake a muon has been estimated on  
anti-$b$ tagged events. 
After subtracting the expected remaining muon content in this sample,
the misidentification probability for hadrons 
has been found to be $(0.52\pm0.03)$\% in the 
barrel and $(0.36\pm0.06)$\% in the forward regions, respectively.
Applying the same procedure to simulation events gave however values lower by
factors of $2.03 \pm 0.12$  in the barrel and of $1.22 \pm 0.20$ in the 
forward regions respectively. The simulation predictions have therefore been 
corrected for these factors.

In the exclusive $\Dsstar$ analysis, muon candidates have been accepted if they
fulfilled the ``loose'' selection criteria and their momenta exceeded 
2~GeV$/c$.  The corresponding efficiency is $\sim 80\%$  
%{(\it ?? in ~\cite{muonid} it is  90\% for loose muons in 1994)}
and the hadron misidentification probability is $\sim 1\%$.
With this selection, the correction factors to be applied to simulated events 
in which the candidate lepton is a misidentified hadron have been found to be  
$1.44 \pm 0.05$ ($1.61 \pm 0.05$) in 1992-93 (1994-95) data samples 
%{(\it ? errors much smaller than previous ones)}.

\subsection{Electron identification}
\label{sec:elecid}

Electron candidates have been tagged within the range $0.03 < |\cos\theta_{e}| < 0.72$ 
in polar angle, based on the combination of the response of the HPC, the specific 
ionization $dE/dx$ in the TPC and the RICH Cherenkov detector~\cite{ref:delsim}.
%A momentum dependent cut, 
%%%%%%%% A cut on what???? GM 2/2/04
%%%%%%%%
A momentum dependent cut applied on the neural net output variable, 
which provides a constant efficiency over the  full 
momentum range, has been applied. 

For the inclusive lepton analysis, electron candidates have been selected with a 
selection cut corresponding to 65\%~efficiency and requiring momenta greater than 
%{\bf  rearranged so that hadrons (fake l) are mentioned only in one place
% + an hadron -> a hadron  }
3~GeV$/c$. The probability for a hadron to fake an electron is about 0.4$\%$. The 
electron identification efficiency has been measured  from data by means of a sample of 
isolated electrons extracted from selected Compton events and one of electrons 
originating from photon conversions in the detector. The ratio between the efficiencies 
measured in data and simulated events has been parametrized as a function of the 
transverse momentum and polar angle of the particle track. Results are summarised in 
Table~\ref{table:electrons}. 
A corresponding correction factor has been applied to simulated $q\bar{q}$
events.
  
The probability  for mis-tagging a hadron as an electron has also been measured using data,
by selecting an anti $b$-tagged background sample, as for the muons. 

Electrons from photon conversions, mainly produced in the outer ID wall and in the inner
TPC frame, have been rejected by removing electron candidates originating at a secondary
vertex and carrying little transverse momentum relative to the direction defined from the
primary to the secondary vertex. The ratio of the measured 
misidentification probability in data 
to that in simulated events is given in Table~\ref{table:electrons}.

In the exclusive $\Dsstar$ analysis, electron candidates have been selected
with 75\%~efficiency and requiring momenta greater than  2~GeV$/c$. 
The probability for a hadron to fake an electron is about 1$\%$.
Correction factors applied to simulated events are also given in Table~\ref{table:electrons}. 

\begin{table}
\begin{center}
\begin{tabular}{|l|c|c|c|c|}
\hline
                     & 1992  & 1993 & 1994 & 1995 \\
\hline \hline
 65\% efficiency cut &   &  &  &  \\
Efficiency (data/MC) & 0.83$\pm$0.02 & 0.83$\pm$0.02 & 0.92$\pm$0.02 &
0.93$\pm$0.02 \\ 
\hline
Misid. Prob. (data/MC) & 0.57$\pm$0.04 & 0.77$\pm$0.05 & 
0.76$\pm$0.05 & 0.70$\pm$0.06 \\
\hline
%\hline
% 75\% efficiency cut &   &  &  &  \\
%Misid. Prob. (data/MC) (\%) & \mch{0.69$\pm$0.03} &\mch{0.77$\pm$0.03}\\
%\hline
\hline
%\mchh{\it 75\% efficiency cut and $p>3$ GeV/c}    \\
 75\% efficiency cut &   &  &  &  \\
Efficiency (data/MC) & 0.89$\pm$0.02 & 0.88$\pm$0.02 & 0.94$\pm$0.02 &
0.94$\pm$0.02 \\ 
\hline
Misid. Prob. (data/MC) & 0.61$\pm$0.04 & 0.77$\pm$0.03 & 
0.80$\pm$0.02 & 0.74$\pm$0.03 \\
\hline
\end{tabular}
\caption[]{\it 
The ratio between values measured in real and simulated events for electron 
identification efficiency and probability of tagging a hadron as an electron}
\label{table:electrons}
\end{center}
\end{table}

\subsection{Hadronic decay reconstruction}
The reconstruction of
%explained in \cite{ref:companion} to reconstruct the
$\Do$ and $\Dstarp$ candidates, in which the $\Do$ decays
%into $\Km \pi^+$ or $\Km \pi^+ \pi^+ \pi^-$.
into $\Km \pi^+$, $\Km \pi^+ \pi^+ \pi^-$ or $\Km \pi^+ (\pi^0)$,
is explained in detail in \cite{ref:companion}.
The reconstruction of the 
$\Dp \rightarrow \Km \pi^+ \pi^+$ decay is based on a similar
approach.
For all decay channels, the main steps of the analysis consist of:
\begin{itemize}
\item reconstructing a D decay vertex from its charged decay products;
\item selecting a mass window centred on the nominal D mass;
\item reconstructing a B decay vertex using the D trajectory and a charged lepton;
\item requiring a minimum distance between the B decay vertex and the 
main vertex and also between the B and the D decay vertices;
\item imposing a minimum momentum on the D (6 $\GeVc$) 
and on the D-$\ell$ (10 $\GeVc$) candidates;
\item requiring a D-$\ell$ mass between 2.5 and 5.5 $\GeVcd$.
\end{itemize}

\section{$\Dsstar$ production in $b$-hadron semileptonic decays}
\label{sec:dstar}

In the following sections the exclusive analysis leading to the measurement 
of hadronic moments is presented. $\Do$, $\Dp$, $\Dstarp$ and $\Dsstar$ reconstruction
is described in Section~\ref{sec:Dreco}. In Section~\ref{sec:Dback} a discriminant 
variable that has different sensitivity to $\Dsstar$ signal and background
events is defined. 
This variable is used in Section~\ref{sec:fitbr} to measure the amount of
$\Dsstar \ell^-\overline{\nu}_{\ell}$ states in the data.
%Since this fraction results compatible with zero, only D$\pi$ states have been 
%taken into account in the rest of the analysis.
From the study of the $\Dsstar$ mass distribution, branching ratios and properties 
of different $\Dsstar$  states  are measured 
in Sections~\ref{sec:Dprod} and \ref{sec:Dsres}. 
%Finally, in section~\ref{sec:mx}, moments of the  $\Dsstar$ mass distribution
%and of the hadronic mass distribution in $b$-hadron
%semileptonic decays are determined.

\subsection{$\Do$, $\Dp$, $\Dstarp$ and $\Dsstar$ reconstruction}
\label{sec:Dreco}

%{\it anything here?}
%$\Do$ and $\Dstarp$ candidates have been reconstructed through the $\Do$ decays
%%into $\Km \pi^+$ or $\Km \pi^+ \pi^+ \pi^-$.
%into $\Km \pi^+$, $\Km \pi^+ \pi^+ \pi^-$ or $\Km \pi^+ (\pi^0)$,
%following the procedure described in~\cite{ref:companion}. Similarly also the 
%$\Dp \rightarrow \Km \pi^+ \pi^+$ decay has been reconstructed. 
 
\begin{figure}[hbtp!]
  \begin{center}
    \mbox{\epsfig{file=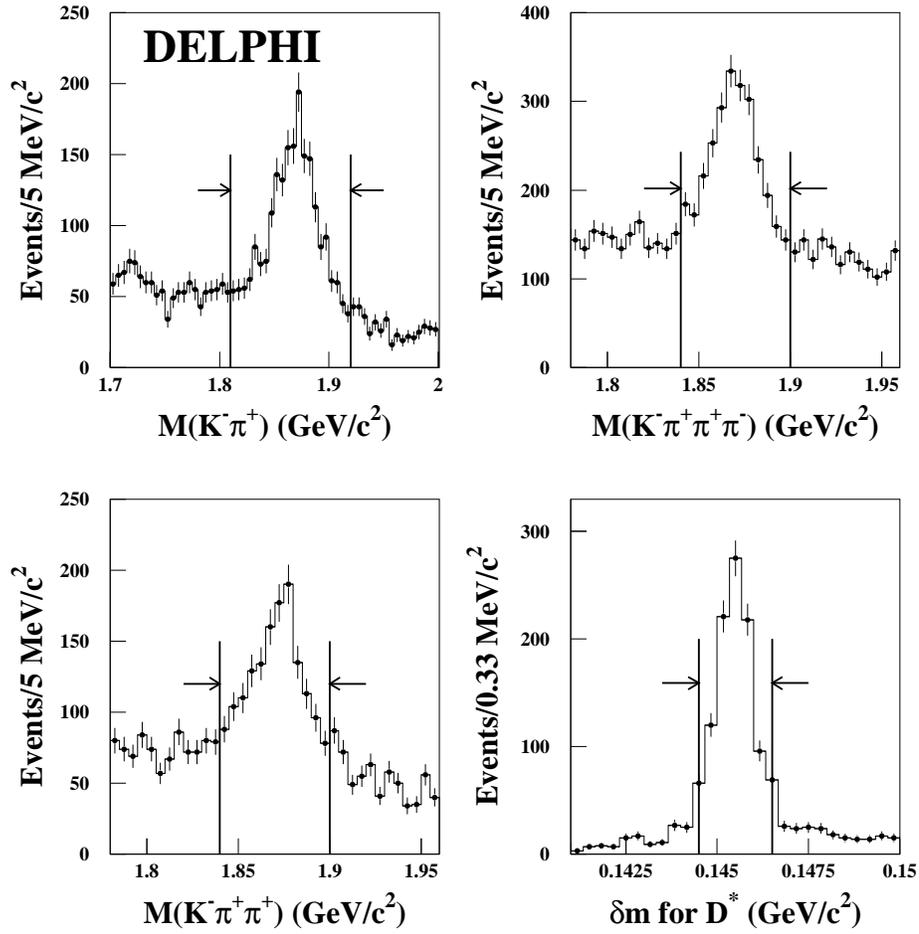,width=14cm}}
  \end{center}
  \caption[]{\it {$\Do$, $\Dp$ and $\Dstarp$ signals used in the present 
analysis, in which the $\Do$ meson decays into charged particles only
and which correspond to events registered in 1992-95. 
%There could be several entries at the same mass value depending
%on the multiplicity of $\pisstar$ candidates. 
Intervals used
to define the signal and sideband regions are indicated.}
   \label{fig:masses}}
\end{figure}

Mass distributions of $\Do$ and $\Dp$ candidates and distributions of the
mass difference between  
 the $\Do \pi^+$ and the $\Do$ candidate, in the case of channels involving a $\Dstarp$,
have been used to define signal and sidebands samples (Figure \ref{fig:masses}). 
Events from the sidebands have 
been used to evaluate the level of the combinatorial background under the D or 
$\Dstarp$ signal.
%Cuts used to define these regions are summarised in Table~\ref{tab:mcuts}.

%\begin{table}[htb]
%\begin{center}
%  \begin{tabular}{|c|c|c|}
%    \hline
% decay channel & signal ($\GeVcd$) & sidebands ($\GeVcd$)\\
%    \hline
%    \hline
%$\Dstarstar \rightarrow \Do \pi^+ X$ (no $\Dstarp$)& & \\
%\hline
% $\Do \rightarrow \Km \pi^+$& $[1.81,~1.92]$ &  
%$[1.70,~1.81]~{\rm and}~[1.92,~2.03]$    \\
%$\Do \rightarrow \Km \pi^+ \pi^+ \pi^-$  &$[1.84,~1.90] $   &   
%$[1.78,~1.84]~{\rm and}~[1.90,~1.96]$    \\
%\hline
%    \hline
%$\Dstarstar \rightarrow \Dp \pi^- X$ & & \\
%\hline
% $\Dp \rightarrow \Km \pi^+ \pi^+$&$[1.84,~1.90] $   &   
%$[1.78,~1.84]~{\rm and}~[1.90,~1.96]$    \\ 
%\hline
%    \hline
%$\Dstarstar \rightarrow \Dstarp \pi^- X$ &$\delta_m$ signal & $\delta_m$ sideband\\
%\hline
% $\Do \rightarrow \Km \pi^+$ and $\Km \pi^+ \pi^+ \pi^-$& 
%$[0.1445,~0.1465]$ &  $>0.1465$    \\
%$\Do \rightarrow \Km \pi^+ (\pi^0)$  &
%$<0.165$ &  $>0.165$    \\
%\hline
%  \end{tabular}
%  \caption[]{\it {Values of the mass or mass difference intervals 
%used to select events in the signal and sideband(s) regions.}
%  \label{tab:mcuts}}
%\end{center}
%\end{table}

For $\Dsstar$ states, decay channels into a ${\rm D}^{(*)}$ 
and, at most, two pions have been considered. 
When searching
for $\Dstarstar \rightarrow \Do \pi^+$ decays, a veto has been applied
against the $\Dstarp$  by removing candidates with a mass difference
$m(\Do \pi^+)-m(\Do)<0.1465 ~\GeVcd$, 
for $\Do$ mesons decaying
into charged particles only
\footnote{For $\Do \pi$ final states, only $\Do$ decaying
into $\Km \pi^+$ or $\Km \pi^+\pi^+\pi^-$ have been used.}.
%this cut is extended to $0.165~\GeVcd$
%for the $\Km \pi^+ (\pi^0)$ channel.
As the combinatorial background under D signals is higher than
for channels involving a $\Dstarp$, values of the cuts given
in \cite{ref:companion} have been made tighter for
$\Dsstar \rightarrow {\rm D} \pi$ events, as compared with those applied to
$\Dsstar \rightarrow {\Dstarp} \pi$ decays.

Events have been selected by 
reconstructing a lepton, a ${\rm D}^{(*)}$ and a charged pion 
whose trajectories are compatible with the hypothesis that they
originate from a common secondary vertex.  
$\Dstarstar$ decays considered in the present analysis involve always
at least one charged particle track \footnote{This 
particle is called  $\pisstar$ in the following analysis.}
emitted at the $b$-decay vertex, in addition
to the exclusively reconstructed ${\rm D}^{(*)}$ meson. 
The $\pisstar$ momentum has been required to be higher than $0.5~\GeVc$
and the track must be associated to at least one measurement in the Vertex Detector.
%Signals of ${\rm D}^{(*)}$ hadrons reconstructed and associated
%to a candidate $\pisstar$ are given in Figure \ref{fig:masses}.

The overall efficiencies for selecting signal events (see Table \ref{tab:effic}), 
not including the
 decay branching fractions of $\Dstarp$, $\Do$ or $\Dp$ mesons into their considered 
decay channels, have been estimated from simulated events. 
%% PR d2
Efficiencies are rather similar for the 92-93 and 94-95 samples when 
considering channels with a $\Dstarp$ or with $\Do \rightarrow \Km \pi^+$
in spite of the reduced performances of the VD for the 92-93 period. This is
because, as the background level is rather low for these channels, loose cuts 
on the vertex separation have been applied. For the other channels,
which require tighter cuts, efficiencies are markedly lower in 92-93.
%%end PR d2
These values have already been 
corrected for differences between the actual measured lifetimes of 
$b$-hadrons and that used in the simulation. 

\begin{table}[htb]
\begin{center}
  \begin{tabular}{|c|c|c|}
    \hline
 decay channel &  92-93 MC & 94-95 MC\\
    \hline
\hline
$\Dstarstar \rightarrow \Do \pi^+$ & & \\
\hline
 $\Do \rightarrow \Km \pi^+$& $(10.4\pm 0.5)\%$ &  $(13.0 \pm 0.5)\%$    \\
$\Do \rightarrow \Km \pi^+ \pi^+ \pi^-$  &  $(3.0 \pm 0.3)\%$  &   $(4.8 \pm 0.3)\%$  \\
\hline
\hline
$\Dstarstar \rightarrow \Dp \pi^-$ & & \\
\hline
 $\Dp \rightarrow \Km \pi^+ \pi^+$& $(5.9 \pm 0.3)\%$ &  $(9.1 \pm 0.3)\%$    \\
\hline
\hline
$\Dstarstar \rightarrow \Dstarp \pi^-$ & & \\
\hline
 $\Do \rightarrow \Km \pi^+$& $(12.0 \pm 0.3)\%$ &  $(13.9 \pm 0.4)\%$    \\
$\Do \rightarrow \Km \pi^+ \pi^+ \pi^-$  &  $(5.0 \pm 0.2)\%$  &   $(5.8 \pm 0.2)\%$  \\
$\Do \rightarrow \Km \pi^+ (\pi^0)$  &  $(6.6 \pm 0.6 )\%$  &   $(7.3 \pm0.6 )\%$  \\
\hline
  \end{tabular}
  \caption[]{\it { Global efficiencies of the analysis chain to reconstruct 
and select  simulated signal events. 
Quoted uncertainties are only
statistical. In addition to the simulated events mentioned
in Table \ref{tab:stat}, dedicated event samples corresponding to the
different channels have been used to increase the statistics.}
  \label{tab:effic}}
\end{center}
\end{table}

\subsection{Signal separation from background sources}
\label{sec:Dback}

When considering ${\rm D}^{(*)} \pi^{\pm}$ combinations,
the main sources of background can be divided into two categories
depending on whether they correspond or not to a real reconstructed ${\rm D}^{(*)}$ meson.
The latter is the  combinatorial background situated under the
charm mass signal. Background events with a real ${\rm D}^{(*)}$ can
originate from the following sources:

\begin{itemize}
\item the $\pisstar$ candidate is not produced at the $b$-decay vertex
but comes from the beam interaction point (primary pion background);

\item the lepton originates from the weak decay of another charm particle
emitted in the $b$-decay (cascade background);

\item the lepton originates from a $\tau$ decay (tau background);

\item the candidate lepton is a misidentified hadron
or a converted photon (fake lepton background);

\item the reconstructed charm meson originates from a $c\overline{c}$
event (charm background).

\end{itemize}

A variable, used to isolate the signal from these backgrounds,
has been defined from 
the probability distributions of several discriminant observables, 
whose shapes have been
obtained from the simulation and also directly from data, as in the case of the 
combinatorial background. The following observables have been used:

\begin{itemize}
\item the lifetime-signed impact parameters of the $\pisstar$ relative to the 
main vertex of the event, in ${\rm R}\phi$ and $z$ projections, normalised to their uncertainty;

\item the normalised and lifetime-signed impact parameters of the $\pisstar$ relative 
to the secondary  vertex, in ${\rm R}\phi$ and $z$ projections;

\item the normalised and lifetime-signed decay distance between the primary and the 
secondary vertices;

\item the cosine of the decay angle defined as the angle of the $\pisstar$ direction, 
boosted to the $b$-hadron rest frame,
%\cite{ref:companion}, 
relative to the B direction. 
This variable is uniformly distributed for the signal whereas it is peaked at 
%low 
negative values for backgrounds;

\item the $\chi^2$ probability for the secondary vertex, which should be uniformly 
distributed for the signal and peaked at small values for 
the main sources of background;

\item the two variables, $d_{\pm}$, which depend on the presence of additional 
charged particles at the secondary vertex. 
%The $+$ sign corresponds to
%a particle having the same sign as the charged lepton. 
They are defined in the following way:
\begin{itemize}
\item all charged particles, other than the $\Dstarstar$ decay products and the lepton,
emitted in the same event hemisphere as the $b$-candidate, with momentum larger than 500
$\MeVc$, which form a mass with the ${\rm D}^{(*)}\pisstar\ell^-$ system lower than 6 $\GeVcd$
and which have values for their impact parameters relative to the $b$-decay vertex smaller than 2
and 1.5 $\sigma$ in $\rphi$ and $z$ respectively, are considered;
\item selected particles, having the same ($+$) or the opposite ($-$) charge as the lepton
are considered separately. If there are several candidates in a class, the one with the 
largest impact parameter to the main vertex is retained and the quantity:
\begin{equation}
x_{\pm} = \epsilon(\rphi) \times nsig(\rphi)^2 + \epsilon(z) \times nsig(z)^2
\end{equation}
is evaluated, where  $\epsilon$ and $nsig$ are, respectively, the sign and the number of 
standard deviations of the track impact parameter relative to the main vertex~\footnote{The second term of this equation is not considered for 
1992-1993 data since the $z$ coordinate is not measured. }. The sign
of the impact parameter is taken to be positive (negative) if the corresponding track 
trajectory intercepts the line of the jet axis from the main vertex downstream (upstream) from that vertex.
\end{itemize}
%by considering all charged particles, emitted in the 
%same event hemisphere as the $b$-candidate, excluding the ${\rm D}^{(*)}$
%decay products, the lepton and the $\pisstar$ candidate, with momentum larger than 
%500~$\MeVc$, forming an invariant mass lower than 6 $\GeVcd$ when adding the 
%${\rm D}^{(*)}\pisstar\ell^-$ system and having impact parameter values, relative to the 
%$b$ decay vertex, smaller than 2 and 1.5 $\sigma$, in $\rphi$ and $z$ respectively.
%Candidates with the same ($d_+$) or the opposite ($d_-$) charge as the lepton have been
%selected. In the case of several candidates, that with the largest impact
%parameter relative to the main vertex, has been kept and the quantity:
%\begin{equation}
%x_{\pm} = \epsilon(\rphi) \times nsig(\rphi)^2 + \epsilon(z) \times nsig(z)^2
%\end{equation}
%has been evaluated. Here $\epsilon$ and $nsig$ indicate the lifetime sign and the number of 
%standard deviations for the track impact parameter relative to the main vertex, 
%respectively. As track offsets may extend to very large values, the variable $d_{\pm}$
%has been taken to be equal to the logarithm of $(1+x_{\pm}^2)$ signed according to the 
%sign of $x$. For events with no spectator track candidate, a fixed value of -4. has been
%attributed to $d_{\pm}$.
As the track impact parameters can extend to very large values because of the relatively
long decay time of $b$-hadrons, the variables $d_{\pm}$
are taken to be equal to the logarithm of $(1+x_{\pm}^2)$ and their sign is taken to 
be the same as $x_{\pm}$. 
For events with no spectator track candidate, that is, with no additional tracks
compatible with the $b$-decay vertex, a fixed value of -4.0 is used for $d_{\pm}$.
For ${\rm D}^{(*)} \pi$ signal events, it is expected that no additional track is 
present at the $b$-decay vertex whereas for ${\rm D}^{(*)} \pi^+ \pi^-$ candidates, 
another track with a precisely defined charge correlation with the lepton, is 
expected.
These properties have been used in the analysis in the following way.
%%PR d2 (paragraph below displaced from the start of section 3.3)
Selected events have been distributed into right-sign and wrong-sign
candidates. Right-sign events correspond to $\Do \pi^+$,
$\Dp \pi^-$ and $\Dstarp \pi^-$ pairs whereas wrong-sign events have
an opposite sign pion. Since only 
$\Dstarstarp$ or $\Dstarstaro$ states can be produced in the semileptonic decay
of a $b$-hadron, where they are accompanied by a negatively charged lepton,
wrong-sign combinations can receive contributions only from 
${\rm D}^{(*)}\pi^+\pi^-$ final states. 
%% end PR
For right-sign combinations it has been required that
signal events behave as if there is no additional charged particle track at the 
$b$-decay vertex. This implies that, in the case of ${\rm D}^{(*)} \pi \pi$ decays, 
only $\Do \pi^+ \pi^0$ decays or
those involving two charged pions and where the $\pi^-$ escapes detection,
can contribute. For wrong-sign combinations, it has been required
that signal events behave as if there was another charged particle track,
of sign opposite to the $\pisstar$ and of trajectory compatible
with the $b$-decay vertex position.
\end{itemize}

\begin{figure}[hbtp!]
  \begin{center}
    \mbox{\epsfig{file=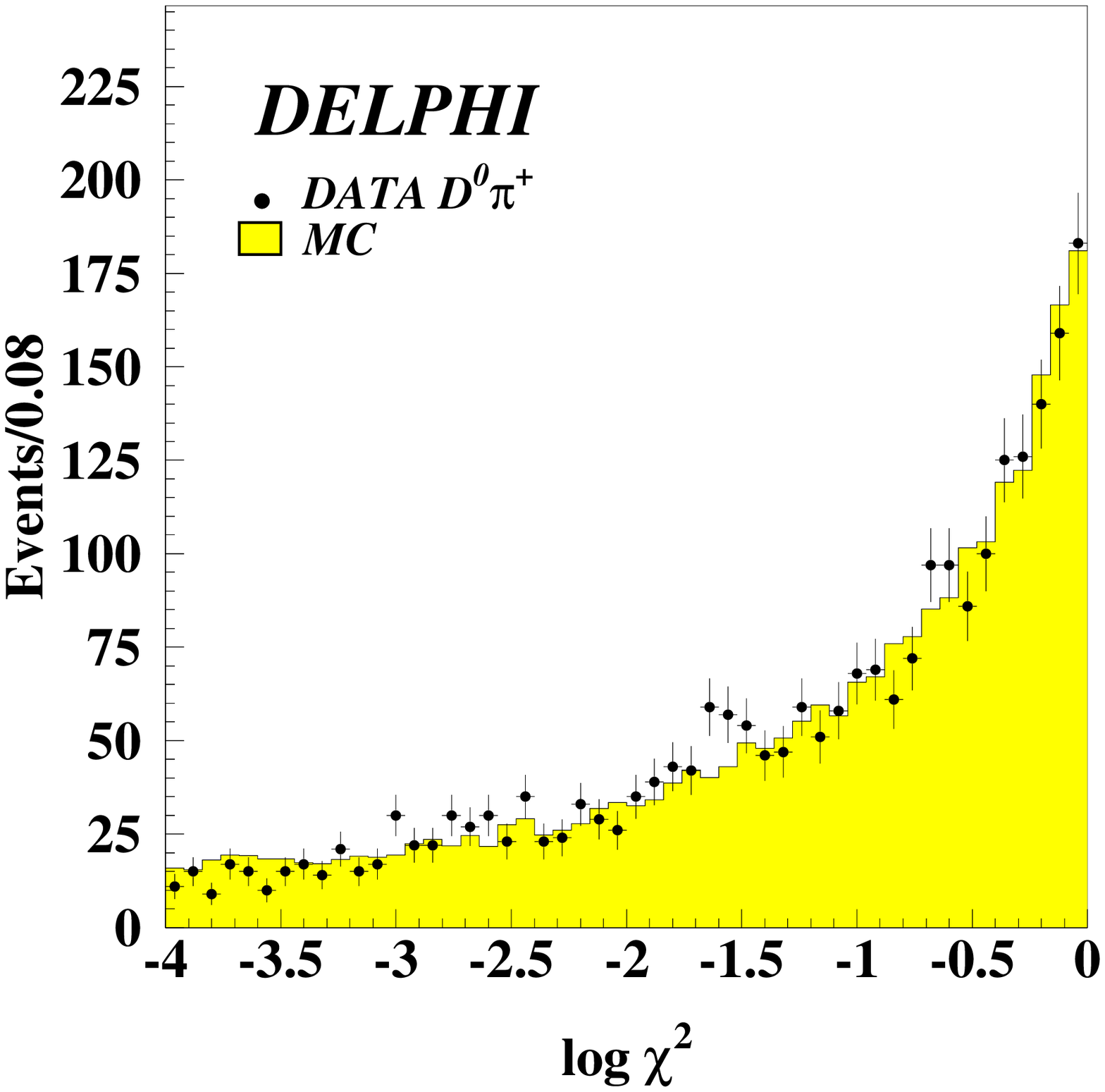,width=8.5cm}
          \epsfig{file=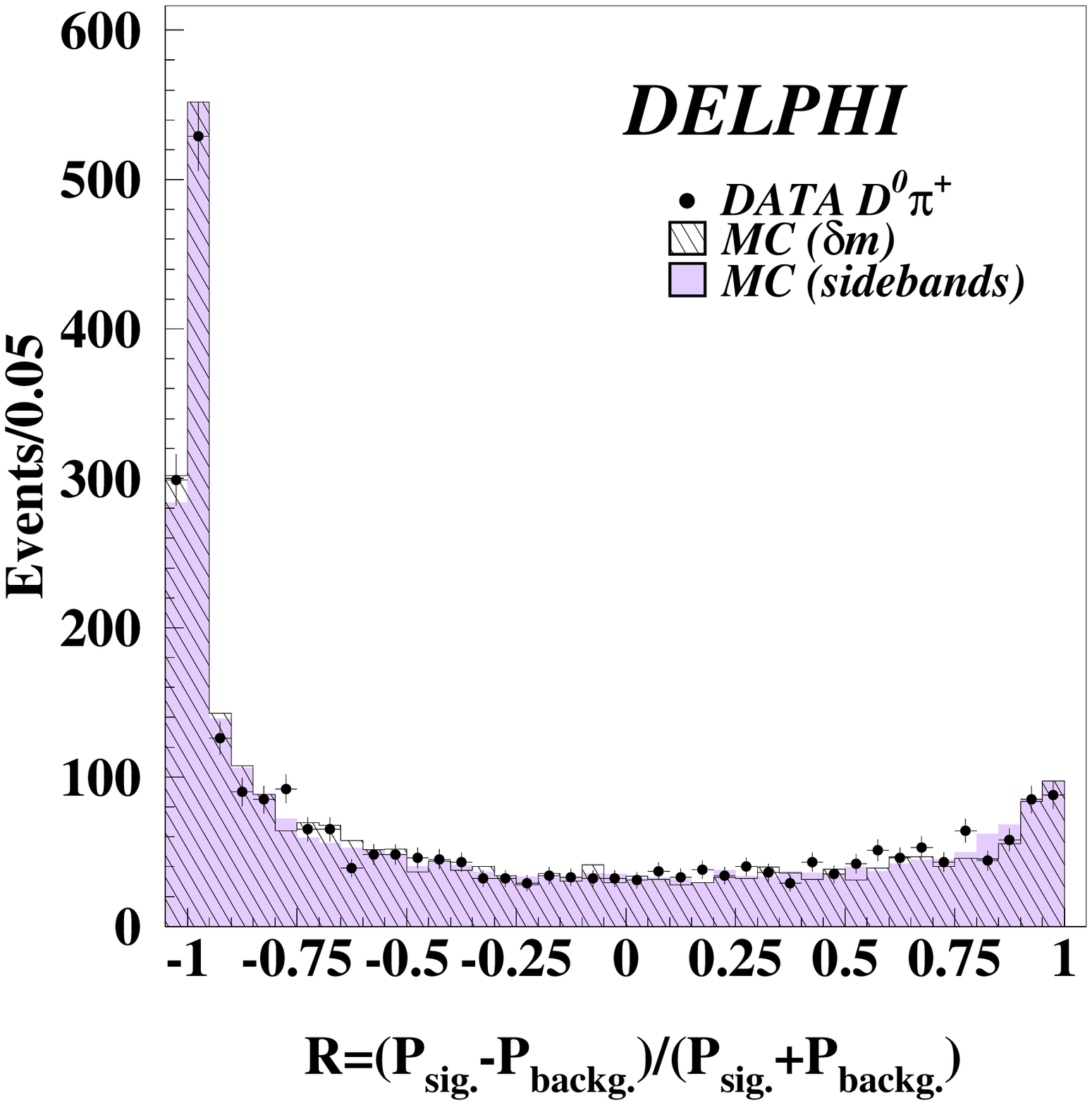,width=8.5cm}}
  \end{center}
  \caption[]{\it {
Left: Distributions of the decimal logarithm of the $\chi^2$ probability for secondary 
$\Do \pi \ell^-$ vertices in events selected in the sidebands
of the signal obtained using real (points with error bars)
and simulated (histogram) events.
Right: Distributions of the values of the discriminant variable
corresponding to events selected in the sidebands of the signal for real
data (points with error bars) and simulated events (shadowed histogram).
The other histogram (hatched) corresponds to simulated events from 
combinatorial background situated in the signal mass region.
The histograms have been normalized to the same number of entries.}
   \label{fig:rvarbkg}}
\end{figure}

Probability distributions for each of these nine (seven for 1992-1993 
data as the vertex detector measured only ${\rm R}\phi$ coordinates) variables,
have been obtained using the simulation for signal and background events.
The agreement in the shape of the distributions of 
these variables for real and simulated events is 
illustrated in Figure~\ref{fig:rvarbkg}-left
in which the $\chi^2$ probability distributions for secondary
vertices for events selected in the sidebands of the signal
 have been compared.
This distribution was selected as it is sensitive to possible 
differences between real and simulated events.

\begin{figure}[hbtp!]
% ds+pi-
\begin{center}
    \mbox{\epsfig{file=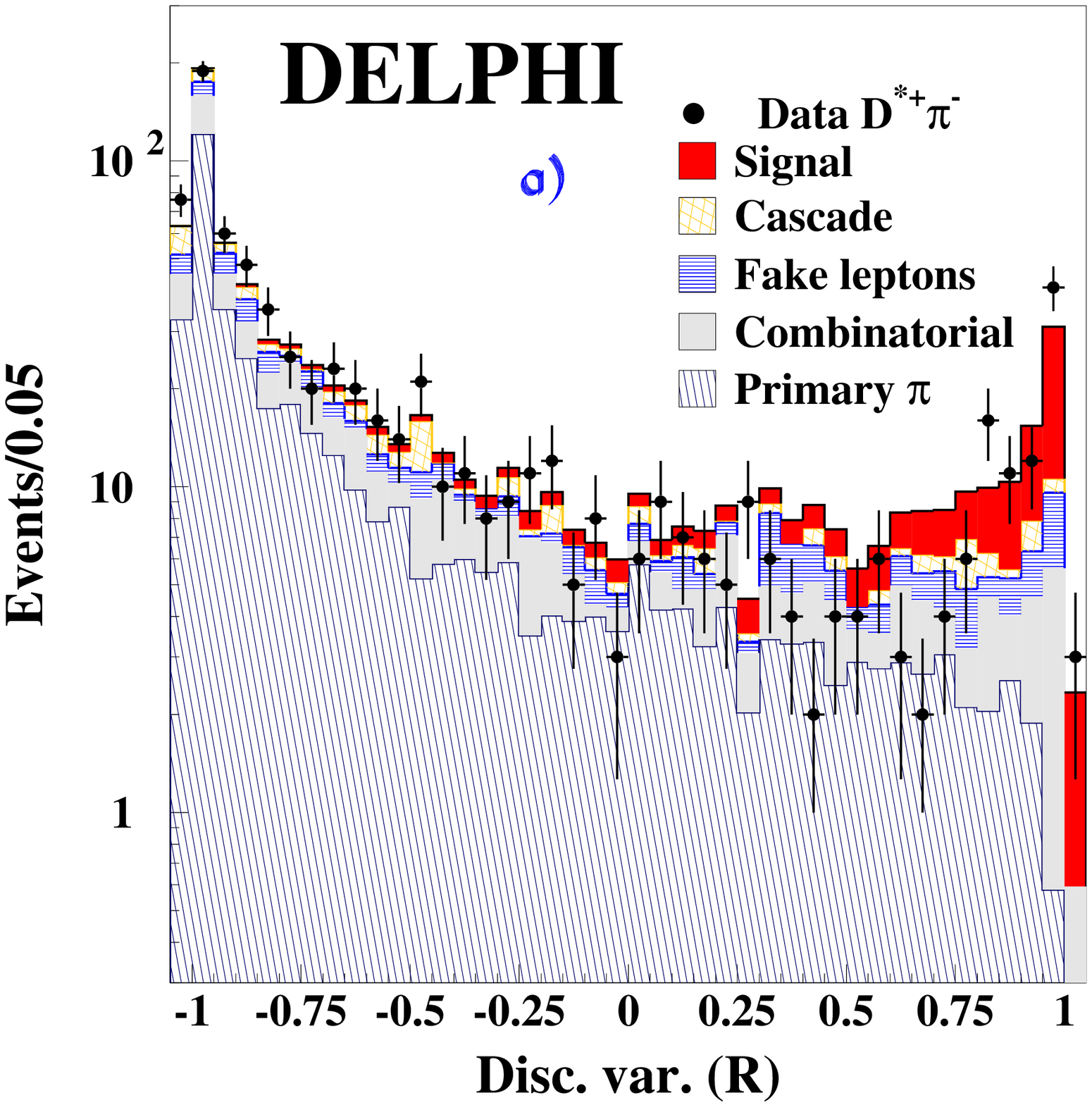,width=8.5cm}\hspace{-0.5cm}
    \epsfig{file=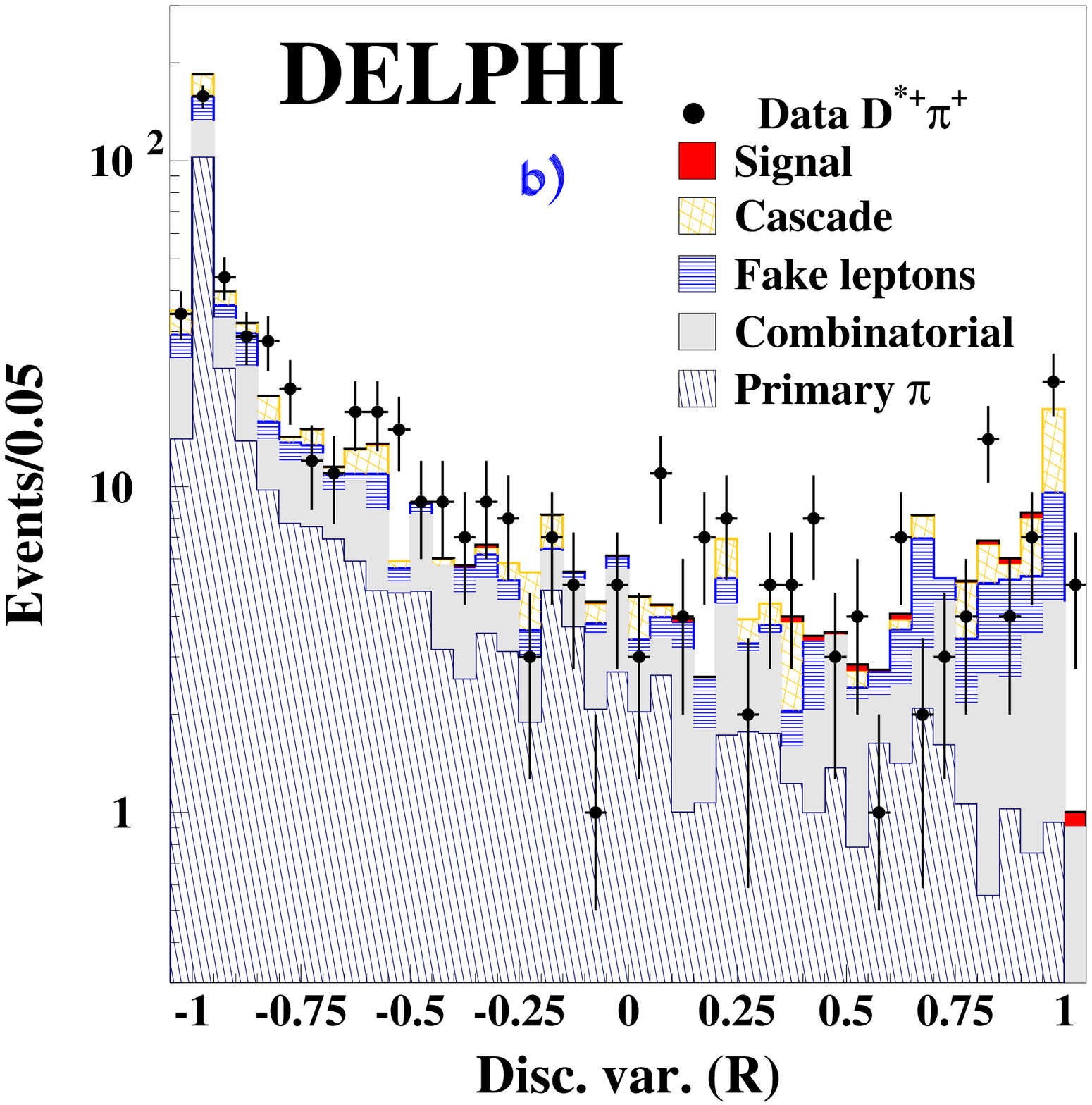,width=8.5cm}}
\end{center}
  \caption[]{\it {Discriminant variable distributions for
a) $\Dstarp \pi^-$ and b) $\Dstarp \pi^+$ candidates. 
Points with error bars correspond to real events whereas the histograms 
show the fitted components. The fake lepton background in these plot 
includes $\tau$ events.}
   \label{fig:dstarpi}}
\end{figure}

The probabilities $P_{signal}$ and $P_{backg.}$ have been obtained by 
multiplying the probability of each discriminating variable 
and a global discriminant 
$R$  has been defined as:
\begin{equation}
R=\frac{P_{signal}-P_{backg.}}{P_{signal}+P_{backg.}},
\end{equation}
%% PR d2
in which events corresponding to the two largest background
sources, namely the combinatorial and the primary pion 
background components,
%and those corresponding to a real ${\rm D}^{(*)}$ with a $\pisstar$
%candidate originating from the main vertex of the event 
have been used to determine the probabilities for background events. 
%% end PR
The distribution of the variable $R$ is peaked at +1(-1) for signal 
(background) events 
(see Figure \ref{fig:dstarpi}).
%-\ref{fig:dstarpi}). 
%Data events from the sidebands of the signal 
%have been used to determine the shape of the $R$ distribution for the 
%combinatorial background component. 

In Figure \ref{fig:rvarbkg}-right the distributions of this quantity for 
combinatorial background events selected in the sidebands of the $\Do$
signal for real and simulated events are compared. These distributions
are in agreement but, to account for possible differences between real
and simulated events and as the combinatorial background component
is the largest of the background components for $\Do \pi$ and $\Dp \pi$ final states, 
real events situated in the sidebands of the signal have
been used, in the following, to determine also the shape of the $R$ distribution 
for the combinatorial background component. 

\subsection{$\Dsstar$ production rates}
\label{sec:fitbr}

Events corresponding to
the $\Do \rightarrow \Km \pi^+ (\pi^0)$ decay channel have not been included
in this study, as they are affected by a larger combinatorial background.

%The fractions of the components contributing to 
The $R$ variable
distributions, for the different ${\rm D}^{(*)}$ decay channels, have been measured
separately for right- and wrong-sign samples and for the two data-taking periods.
A binned likelihood fit \cite{ref:hmcmll} 
has been performed \footnote{Statistical uncertainties due to the 
finite number of analyzed simulated events are included in this fit.}
in order to measure the branching 
fraction in each channel. 
The fractions of signal and primary pion background events have been left free 
to vary in the
fit whereas the fractions of the other components with a real ${\rm D}^{(*)}$ have been 
fixed to values taken from external measurements. 
The discriminant variable distribution for 
combinatorial background events has been obtained from data using the sideband events.
%(see Table \ref{tab:mcuts}). 
Distributions for the other background components 
have been taken from the simulation.

%The uncertainties adopted include the data statistical uncertainty and the finite 
%simulation statistics but also the effects of the weighting procedure applied on each 
%event to correct for the differences in the response of the lepton tagging in real and 
%simulated events as discussed above.

Fractions which have been kept fixed in the fit have then been varied,
in turn, according to their expected overall accuracy and the fit repeated 
to estimate the corresponding systematic errors. The fitting procedure has been verified 
on simulated events 
(see the last column of Tables \ref{tab:measbrrs}-\ref{tab:measbrws}).

%\begin{figure}[htb!]
%% d0pi+
%\begin{center}
%    \mbox{\epsfig{file=rd_dopiplus_rs.eps ,width=8.5cm}
%    \epsfig{file=rd_dopiplus_ws.eps ,width=8.5cm}}
%\end{center}
%  \caption[]{\it {Discriminant variable distributions for
%$\Do \pi^+$ (left) and $\Do \pi^-$ (right) candidates. Points with
%error bars correspond to real events whereas the histograms give
%the fitted components.}
%   \label{fig:dopi}}
%\end{figure}

%\begin{figure}[htb!]
%% d+pi-
%\begin{center}
%    \mbox{\epsfig{file=rd_dpluspimin_rs.eps ,width=8.5cm}
%    \epsfig{file=rd_dpluspimin_ws.eps ,width=8.5cm}}
%\end{center}
%  \caption[]{\it {Discriminant variable distributions for
%$\Dp \pi^-$ (left) and $\Dp \pi^+$ (right) candidates. Points with
%error bars correspond to real events whereas the histograms give
% the fitted components.}
%   \label{fig:dppi}}
%\end{figure}

The simulation of cascade events comprises $b \rightarrow \Dsm {\rm D} X$ events only. 
Corrections have therefore been applied to the rate and the topology of these events 
to account for the missing decay modes. Present measurements from ALEPH~\cite{twocharma}
and BaBar~\cite{twocharmb} on double charm decays of $b$-hadrons have been used,  
together with results on $c$-hadron inclusive and exclusive semileptonic decays 
from~\cite{ref:PDG02}, to determine these corrections. 
They depend on the topology of the studied channel. For each channel,
the correction factor on the branching fraction and  
the probabilities to have no track of same (${\cal P}(0)_{ss}$)
or opposite-sign (${\cal P}(0)_{os}$) as the lepton
have been evaluated (see Table \ref{tab:corcasc}).

\begin{table}[htb]
\begin{center}
{\small
  \begin{tabular}{|c|c|c|c|c|}
    \hline
 Channel & Expected rate $(\%)$ & corr. factor & ${\cal P}(0)_{ss}(\%)$ & ${\cal P}(0)_{os}(\%)$ \\
    \hline
Right sign candidates & & & & \\
\hline
 $b\rightarrow \Do \pi^+ \ell^- X$, veto on $\Dstarp$  & $1.61 \pm 0.29$& $3.5$    & $33\pm 4$ &$33\pm4$    \\
 $b\rightarrow \Dp \pi^- \ell^- X$ & $0.54 \pm 0.13$& $2.3$    & $63\pm 4$ &$15 \pm 3$    \\
 $b\rightarrow \Dstarp \pi^- \ell^- X$ & $0.65 \pm 0.26$& $2.7$    & $69\pm 5$ &$13 \pm 3$    \\
\hline
Wrong sign candidates & & & & \\
\hline
 $b\rightarrow \Do \pi^- \ell^- X$, veto on $\Dstarp$ & $0.58 \pm 0.12$& $2.6$    & $\sim 1$ &$\sim 0$    \\
 $b\rightarrow \Dp \pi^+ \ell^- X$ & $0.43 \pm 0.09$& $3.4$    & $13\pm 2$ &$90 \pm 3$    \\
 $b\rightarrow \Dstarp \pi^+ \ell^- X$ & $0.58 \pm 0.17$& $4.2$    & $22\pm 3$ &$89 \pm 2$    \\

\hline
  \end{tabular}
  \caption[]{\it {
%Correction factors, applied on simulated events,
%to account for double-charm $b$-hadron decays which were not included in the
%simulation.
Double-charm expected rates and probabilities for having no additional charged 
track in the considered decay channels.}
  \label{tab:corcasc}}
}
\end{center}
\end{table}

Measured branching fractions are summarized in Tables \ref{tab:measbrrs}
and \ref{tab:measbrws} respectively for right- and wrong-sign candidates.

\begin{table}[htb!]
\begin{center}
{\small
  \begin{tabular}{|c|c|c|c|}
    \hline
 Right sign candidates & number of &  branching  &  number of events \\
 Channel  &    fitted events &   fraction   ($\%$)         & fitted (expected) MC \\ 
    \hline
\hline
 $b\rightarrow \Do \pi^+ \ell^- X$, veto on $\Dstarp$  & & & \\
\hline
$\Do \rightarrow \Km \pi^+$ (92-93) & $109.0\pm 27.5$& $2.34 \pm 0.59$    & $515.2 \pm 67.7~(435)$  \\
$\Do \rightarrow \Km \pi^+$ (94-95) & $60.3\pm 20.3$& $0.69\pm 0.23$    & $939.6 \pm 67.5~(1001)$  \\
$\Do \rightarrow \Km \pi^+ \pi^+ \pi^-$ (92-93) & $13.2\pm 20.0$& $0.51 \pm 0.77$    & $262.8 \pm 50.9~(243)$  \\
$\Do \rightarrow \Km \pi^+ \pi^+ \pi^-$ (94-95) & $59.0\pm 20.7$& $0.93 \pm 0.33$    & $774.5 \pm 64.5~(756)$  \\
\hline
Average & &$0.89 \pm 0.18$ & \\
\hline
\hline
 $b\rightarrow \Dp \pi^- \ell^- X$  & & & \\
\hline
$\Dp \rightarrow \Km \pi^+ \pi^+$ (92-93)& $47.4\pm18.8$& $0.76 \pm 0.30$    & $350.1 \pm 47.6~(394)$ \\
$\Dp \rightarrow \Km \pi^+ \pi^+$ (94-95)& $36.2\pm16.5$& $0.25 \pm 0.11$    & $1101.1 \pm 63.0~(1132)$ \\
\hline
Average & &$0.31 \pm 0.10$ & \\
\hline
\hline
 $b\rightarrow \Dstarp \pi^- \ell^- X$   & & & \\
\hline
$\Do \rightarrow \Km \pi^+$ (92-93)& $13.2 \pm 9.3$& $0.36 \pm 0.26$    & $125.3 \pm 23.6~(128)$     \\
$\Do \rightarrow \Km \pi^+$ (94-95)& $32.8 \pm 8.8$& $0.52 \pm 0.14$    & $282.6 \pm 28.7~(303)$     \\
$\Do \rightarrow \Km \pi^+ \pi^+ \pi^-$ (92-93) & $16.4\pm 7.1$& $0.55 \pm 0.24$    & $111.2 \pm 20.2~(96)$  \\
$\Do \rightarrow \Km \pi^+ \pi^+ \pi^-$ (94-95) & $12.9\pm 7.2$& $0.25 \pm 0.14$    & $239.5 \pm 28.0~(227)$  \\
\hline
Average & &$0.40 \pm 0.09$ & \\
\hline
  \end{tabular}
  \caption[]{\it {Measured branching fractions for 
$b\rightarrow {\rm D} \pi \ell^- X$ in right-sign 
combinations using the {\rm D} decay branching fractions
taken from Table \ref{tab:external}. The last column gives the results obtained on
simulated events. Numbers within parentheses correspond to the real number of 
simulated signal.}
  \label{tab:measbrrs}}
}
\end{center}
\end{table}

\begin{table}[htb!]
\begin{center}
{\small
  \begin{tabular}{|c|c|c|c|}
    \hline
 Wrong sign candidates & number of &  branching  &  number of events \\
  Channel  &    fitted events &  fraction ($\%$)  & fitted (expected) MC \\ 
    \hline
\hline
 $b\rightarrow \Do \pi^- \ell^- X$, veto on $\Dstarp$  & & & \\
\hline
$\Do \rightarrow \Km \pi^+$ (92-93) & $34.3\pm 17.8$& $0.74 \pm 0.38$    & $57.0 \pm 41.9~(70)$  \\
$\Do \rightarrow \Km \pi^+$ (94-95) & $26.3\pm 15.3$& $0.30 \pm 0.17$    & $136.8\pm 36.6~(141)$  \\
$\Do \rightarrow \Km \pi^+ \pi^+ \pi^-$ (92-93) & $3.5\pm 14.1$& $0.14 \pm 0.56$    & $25.1 \pm 33.1~(40)$  \\
$\Do \rightarrow \Km \pi^+ \pi^+ \pi^-$ (94-95) & $1.4\pm 15.5$& $0.02 \pm 0.23$    & $143.1 \pm 41.1~(166)$  \\
\hline
Average & &$0.26 \pm 0.13$ & \\
\hline
\hline
 $b\rightarrow \Dp \pi^+ \ell^- X$  & & & \\
\hline
$\Dp \rightarrow \Km \pi^+ \pi^+$ (92-93)& $10.1\pm17.2$& $+0.16 \pm 0.27$    & $119.8 \pm 20.0~(87)$ \\
$\Dp \rightarrow \Km \pi^+ \pi^+$ (94-95)& $-9.9\pm15.8$& $-0.07 \pm 0.11$    & $190.6 \pm 44.2~(226)$ \\
\hline
Average & &$-0.04 \pm 0.10$ & \\
\hline
\hline
 $b\rightarrow \Dstarp \pi^+ \ell^- X$   & & & \\
\hline
$\Do \rightarrow \Km \pi^+$ (92-93)& $-0.3 \pm 8.3$& $-0.01 \pm 0.22$    & 
not simul.     \\
$\Do \rightarrow \Km \pi^+$ (94-95)& $0.62 \pm 6.3$& $+0.01 \pm 0.10$    &  -    \\
$\Do \rightarrow \Km \pi^+ \pi^+ \pi^-$ (92-93) & $9.0\pm 6.5$& $+0.30 \pm 0.22$    &  -  \\
$\Do \rightarrow \Km \pi^+ \pi^+ \pi^-$ (94-95) & $-8.2\pm 7.8$& $-0.16 \pm 0.15$    &  -   \\
\hline
Average & &$0.00 \pm 0.07$ & \\
\hline
  \end{tabular}
  \caption[]{\it {Measured branching fractions for the different wrong-sign
combinations. The last column gives the results obtained on
simulated events and numbers within parentheses correspond to the 
real simulated signal.}
  \label{tab:measbrws}}
}
\end{center}
\end{table}

The following contributions to the systematic uncertainties have been 
considered:
\begin{itemize}
\item uncertainties related to the values of external parameters, such as
$R_b$, and the different branching fractions of charmed hadrons into the 
reconstructed final states. The values used in the present analysis, taken from 
\cite{ref:PDG02}, are given in 
Table \ref{tab:external};
%\item uncertainty on the efficiency to signal events due to finite
%statistics in simulated events

\item detector-dependent uncertainties such as those on the tracking efficiency,
on the lepton identification efficiency and on the correction of 
differences between real and simulated events relative to the fake lepton rate;
%given in Table \ref{tab:fakel}.

\item differences between real and simulated events on track reconstruction
accuracy.
Distributions of the discriminant variable for signal-like events,
obtained in real and simulated events, have also been compared
using $\Dstarp \rightarrow \Do \pi^+$ decays, with
$\Do \rightarrow \Km \pi^+ ~{\rm or} ~\Km \pi^+ \pi^+ \pi^-$. These events
have been analysed using the same criteria as those applied to
$\Dsstar \rightarrow \Do \pi^+$ candidates. A discriminant variable ($R^*$)
is constructed using the same input quantities as for the 
$R$ variable, apart from $d_{\pm}$ whose effects have been evaluated separately.
Possible differences between real and simulated events affecting track offset
measurements, decay length reconstruction and $\chi^2$ probability of secondary
vertices, in the case of a signal, can then be studied.
Distributions of the $R^*$ variable obtained using 461 reconstructed $\Dstarp$
in real data and 4100 in simulated events are compared in Figure \ref{fig:rvardstar}.
The two distributions agree within the statistical uncertainties. 
A possible difference in the shape
of these distributions has been parametrized assuming a linear variation 
with the 
value of $R^*$. The fitted slope is equal to $-0.01 \pm 0.08$. In the following,
the effect of a variation on the shape of the discriminant variable distribution,
induced by a linear correction of slope equal to $\pm 0.1$, has been evaluated;

\begin{figure}[hbtp!]
% ds+pi-
\begin{center}
    \mbox{\epsfig{file=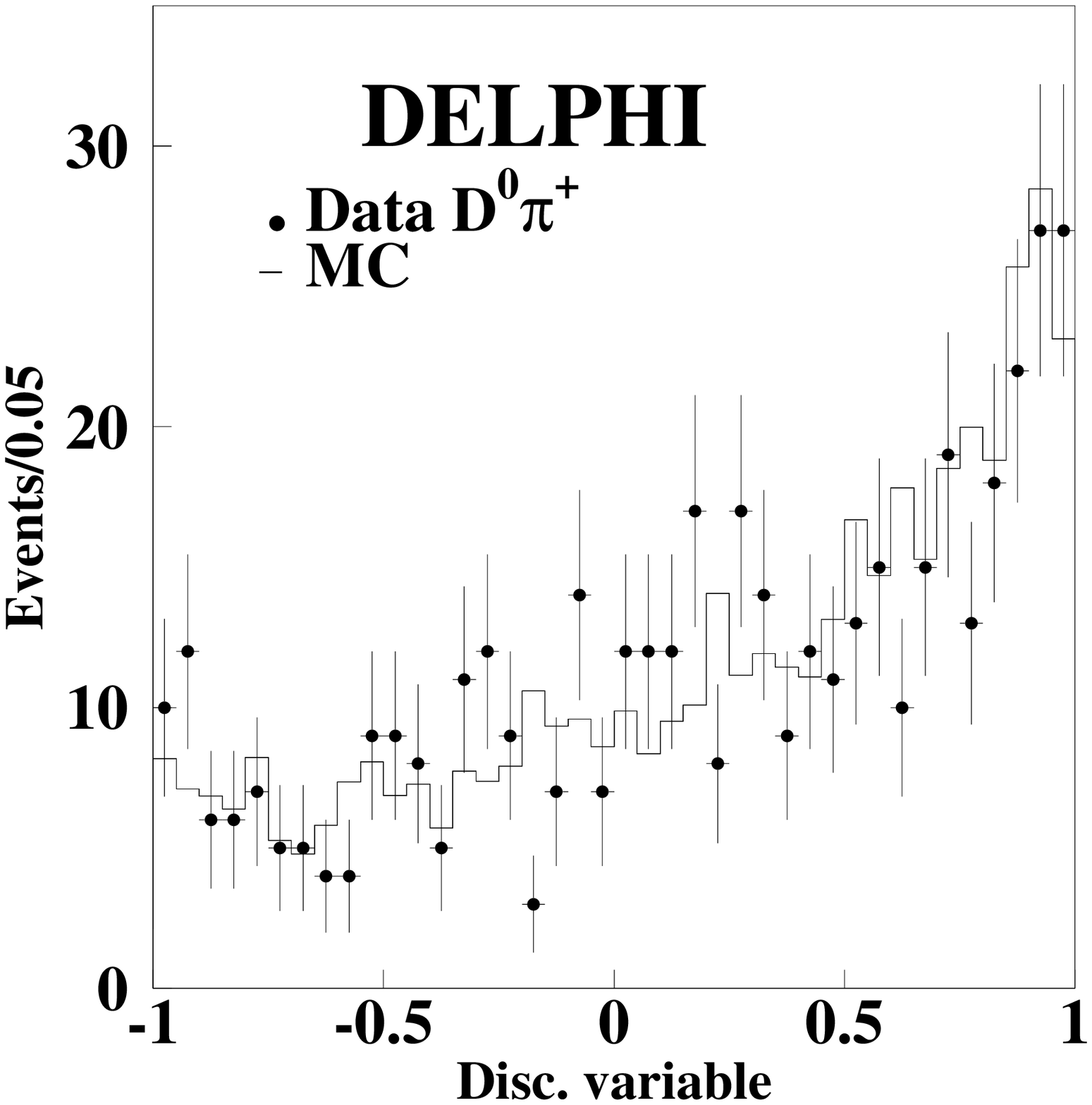 ,width=8.5cm}}
\end{center}
  \caption[]{\it {Discriminant variable distributions for
$\Dstarp \rightarrow \Do \pi^+$ candidates selected in data 
(points with error bars)
and in simulated hadronic $\Zz$ decays (histogram).}
   \label{fig:rvardstar}}
\end{figure}

\item corrections of the cascade decay rate applied to simulated events
have been evaluated with a $30\%$ relative uncertainty
(see Table \ref{tab:corcasc}). This variation is larger than 
most of the 
uncertainties attached to the correction applied to each channel and 
covers the dispersion of the values of these corrections. Variations
in the cascade rates have been applied in a correlated manner to all
channels when this was relevant.
Effects from possible differences between real and simulated 
events for the 
$d_{\pm}$ distributions have been evaluated in \cite{ref:companion}
by studying the decay channel 
$b \rightarrow \Dstarp {\rm X} \ell^- \overline{\nu}_{\ell}$.
The probability to have no additional track, for signal events
in which there is in reality no other track coming from the $b$-vertex,
is of the order
of 80$\%$ and we have taken a 5$\%$ relative error. For events, in which
there is at least one such track produced at the secondary vertex, the 
probability to miss it is estimated to be ($20\pm10\%$).
For events selected in the $\delta m$ region, situated above the $\Dstarp$
signal, in real and simulated events, the probabilities for having no
spectator track are of the order of 34$\%$ and differ by $(2.5\pm1.5)\%$.

\end{itemize}
The uncertainty related to the finite statistics of simulated events has been 
included in the statistical uncertainty of fitted signal event numbers.

\begin{table}[htb]
\begin{center}
  \begin{tabular}{|c|c|}
    \hline
parameter  & central value \\
or hypothesis & and uncertainty\\
\hline
 R$_b$ & $0.21664 \pm 0.00068$ \\
 BR$(\Dstarp \rightarrow \Do \pi^+)$ & $0.677 \pm 0.005$\\
 BR$(\Do \rightarrow \Km \pi^+)$ & $0.0380 \pm 0.0009$ \\
 BR$(\Do \rightarrow \Km \pi^+ \pi^+ \pi^-)$ & $0.0746 \pm 0.0031$\\
 BR$(\Do \rightarrow \Km \pi^+ \pi^0)$ & $0.131 \pm 0.009$\\
 BR$(\Do \rightarrow \Km \ell^+ \nu_\ell)$ & $0.070\pm 0.003$\\
 BR$(\Do \rightarrow \Km K^+)$ & $0.00412 \pm 0.00014$\\
 BR$(\Dp \rightarrow \Km \pi^+ \pi^+)$ & $0.091 \pm 0.006$\\
 P($b \rightarrow \Bdb$)  & $0.388 \pm 0.013$\\
 P($b \rightarrow \Bsb$)  & $0.106 \pm 0.013$\\
 $\tau(\Bdb)$  & $ 1.542 \pm 0.016 ~\ps$  \\
\hline
  \end{tabular}
  \caption[]{\it { Values for the external parameters used in the analysis
\cite{ref:PDG02}.}
  \label{tab:external}}
\end{center}
\end{table}

A summary of these contributions is given in Table \ref{tab:ratesyst}.

\begin{table}[htb]
\begin{center}
{\small
  \begin{tabular}{|c|c|c|c|c|c|c|c|}
    \hline
 Right sign  & External  &  Tracking  & Lepton ID & Fake lept. & Discr. & Cascade & Total\\
 candidates  & BR, $R_b$ & efficiency & efficiency  & rate & var. & modelling & \\ 
    \hline
 $b\rightarrow \Do \pi^+ \ell^- X$ &0.028 &  0.012 &0.016 &0.006 & 0.047& 0.033& 0.067 \\ 
 $b\rightarrow \Dp \pi^- \ell^- X$  & 0.021&  0.005 &0.006 & 0.003 &0.007 & 0.011 & 0.026 \\
 $b\rightarrow \Dstarp \pi^- \ell^- X$   & 0.014 & 0.007&0.007 & 0.003 &0.015 & 0.013 & 0.027 \\
\hline
 Wrong sign  & &   &  &  &  & &  \\
 candidates  &  &  &  &  &  & &  \\ 
    \hline
 $b\rightarrow \Do \pi^- \ell^- X$ &0.028 & 0.012 &0.016 & 0.006 &0.047 & 0.053 &0.079 \\ 
 $b\rightarrow \Dp \pi^+ \ell^- X$  & 0.021 & 0.005 &0.006 &0.002 &0.007 & 0.017 &0.029 \\
 $b\rightarrow \Dstarp \pi^+ \ell^- X$   & 0.014 & 0.007&0.007 & 0.004 &0.015 & 0.067 &0.071 \\

\hline
  \end{tabular}
  \caption[]{\it {Contributions of systematic uncertainties 
(in $10^{-2}$ units) to the
measured production rates of $b \rightarrow {\rm D}^{(*)}\pi X \ell^- \overline{\nu}_{\ell}$
events.}
  \label{tab:ratesyst}}
}
\end{center}
\end{table}

Measured branching fractions, for right-sign events,
are summarized in Table \ref{tab:ratesrs}, where they are compared
with similar results obtained in other experiments. The present analysis
%brings improvements, as compared with 
supersedes the previous DELPHI analysis \cite{ref:dsstardelphi}, 
in terms of statistical accuracy, as more information
has been used to separate signal and background events, and systematics.
All experimental results are compatible.

\begin{table}[htb]
\begin{center}
{\small
  \begin{tabular}{|c|c|c|c|}
    \hline
 Right sign  & DELPHI & DELPHI & ALEPH \\
 candidates  & this analysis & \cite{ref:dsstardelphi} & \cite{ref:dsstaraleph}\\ 
    \hline
 $b\rightarrow \Do \pi^+ \ell^- X$ &$8.9 \pm 1.8 \pm 0.7$ & $11.6 \pm 2.4 \pm 1.1$& 
$7.3 \pm1.8 \pm 1.0$\\ 
 $b\rightarrow \Dp \pi^- \ell^- X$ &$3.1 \pm 1.0 \pm 0.3$ & $4.9 \pm 1.8 \pm 0.7$& 
$3.0 \pm 0.7 \pm 0.5$\\ 
 $b\rightarrow \Dstarp \pi^- \ell^- X$  &$4.0 \pm 0.9 \pm 0.3$ & $4.8 \pm 0.9 \pm 0.5$& 
$4.7 \pm 0.8 \pm 0.6$\\ 
\hline
 Wrong sign  & &   &   \\
 candidates  &  &  &   \\ 
    \hline
 $b\rightarrow \Do \pi^- \ell^- X$ &$2.6\pm1.3\pm0.8$ &$2.3\pm1.5\pm0.4$ &  \\ 
 $b\rightarrow \Dp \pi^+ \ell^- X$  & $-0.4 \pm1.0\pm0.3$& $2.6\pm1.5\pm0.4$ &  \\
 $b\rightarrow \Dstarp \pi^+ \ell^- X$   & $0.0\pm0.7\pm0.7$& $0.6 \pm0.7\pm0.2$&  \\

\hline
  \end{tabular}
  \caption[]{\it {Comparison between
measured production rates (in $10^{-3}$ units) of 
$b \rightarrow {\rm D}^{(*)}\pi X\ell^- \overline{\nu}_{\ell}$
events.}
  \label{tab:ratesrs}}
}
\end{center}
\end{table}

There is no significant excess of events in wrong-sign combinations.
Measurements of the $\Do \pi^-$ and $\Dp \pi^+$ channels can be averaged,
independently of the isospin of the $\pi \pi$ system, as
the same number of events is expected in the two channels giving:
\begin{equation}
{\rm BR}(b \rightarrow \Do \pi^+ \pi^- \ell^- \overline{\nu}_{\ell})~=~
{\rm BR}(b \rightarrow \Dp \pi^+ \pi^-\ell^- \overline{\nu}_{\ell})~=~
(0.06 \pm 0.08 \pm 0.04)~\% 
\end{equation}
\begin{equation}
{\rm BR}(b \rightarrow \Dstarp \pi^+ \pi^-\ell^- \overline{\nu}_{\ell})
~=~(0.00 \pm 0.07 \pm 0.07)~\%
\end{equation}
The values corresponding to   90$\%$ C.L. upper limits are
equal to, respectively,
$0.18~\%$ and $0.13~\%$.

%As a result, in the following analysis, only ${\rm D}^{(*)} \pi$, final 
%states have been considered.

\subsection{Study of the $\Dsstar$ hadronic  mass distribution}
\label{sec:Dprod}

In order to study the mass distribution of right sign events, corresponding to the 
$\Dsstar$ signal, the cut 
$R>0.25$ has been applied on the discriminant variable to reduce the contribution from
background events. This cut corresponds to a selection efficiency which 
varies between 67$\%$ and 85$\%$ for signal events, depending on the channel
and on the data sample.
 
%\begin{table}[htb]
%\begin{center}
%  \begin{tabular}{|c|c|c|c|}
%    \hline
% Channel & optimal region & efficiency &   ``optimal'' cut\\
%         & for a cut on $R$ &   $R>0.7$ & and (efficiency) \\ 
%    \hline
% $b\rightarrow \Do \pi \ell^- X$, veto on $\Dstarp$  & & & \\
%\hline
%$\Do \rightarrow \Km \pi^+$ (92-93) & $[0,~0.5]$& $0.39$    & $0.5~(0.52)$  \\
%$\Do \rightarrow \Km \pi^+$ (94-95) & $[0.3,~0.8]$& $0.59$    & $0.7~(0.59)$  \\
%$\Do \rightarrow \Km \pi^+ \pi^+ \pi^-$ (92-93) & $[-0.25,~0.50]$& $0.36$    & $0.5~(0.53)$  \\
%$\Do \rightarrow \Km \pi^+ \pi^+ \pi^-$ (94-95) & $[0.3,~0.85]$& $0.52$    & $0.7~(0.52)$  \\
%\hline
% $b\rightarrow \Dp \pi \ell^- X$  & & & \\
%\hline
%$\Dp \rightarrow \Km \pi^+ \pi^+$ (92-93)& $[0.1,~0.7]$& $0.47$    & $0.70~(0.47)$ \\
%$\Dp \rightarrow \Km \pi^+ \pi^+$ (94-95)& $[0.6,~0.8]$& $0.60$    & $0.70~(0.60)$ \\
%\hline
% $b\rightarrow \Dstarp \pi \ell^- X$   & & & \\
%\hline
%$\Do \rightarrow \Km \pi^+$ (92-93)& $[0.5,~0.8]$& $0.49$ &$0.8~(0.41)$      \\
%$\Do \rightarrow \Km \pi^+$ (94-95)& $[0.75,~0.9]$& $0.63$ & $0.8~(0.58)$     \\
%$\Do \rightarrow \Km \pi^+ \pi^+ \pi^-$ (92-93) & $[0.4,~0.9]$& $0.54$& $0.8~(0.47)$  \\
%$\Do \rightarrow \Km \pi^+ \pi^+ \pi^-$ (94-95) & $[0.5,~0.9]$& $0.65$ & $0.8~(0.57)$  \\
%\hline
%Average & & & \\
%\hline
%  \end{tabular}
%  \caption[]{\it {
%.}
%  \label{tab:cutr}}
%\end{center}
%\end{table}

%The fits of the measured mass distributions prefers a ${\rm D} \pi$ component maximal 
%at threshold. 

A maximum likelihood fit has been performed using the 
$\Delta_m = m({\rm D}^{(*)}\pisstar) - m({\rm D}^{(*)})$ and the $R$ variables,
introducing  the following components and parameters:

\begin{itemize}
\item $b_{D_0^*}={\rm BR}(\Bdb \rightarrow {\rm D}_0^{*+} \ell^- \overline{\nu}_{\ell})$.
In practice this component is considered to be a broad resonant mass distribution
which can account for various possible states;
\item $m_{D_0^*}$: the mass of the ${\rm D}_0^*$: it is kept fixed 
at 2.4 $\GeVcd$
and a scan
of its possible values between 2.3 and 2.5 $\GeVcd$ has been made;

\item $\Gamma_{D_0^*}$: the total width of the ${\rm D}_0^*$;
\item $b_{D_1^*}={\rm BR}(\Bdb \rightarrow {\rm D}_1^{*+} \ell^- \overline{\nu}_{\ell})$;
\item $m_{D_1^*}$: the mass of the ${\rm D}_1^*$;
\item $\Gamma_{D_1^*}$: the total width of the ${\rm D}_1^*$;
\item $b_{D_1}={\rm BR}(\Bdb \rightarrow {\rm D}_1^+ \ell^- \overline{\nu}_{\ell})$;
\item $b_{D_2^*}={\rm BR}(\Bdb \rightarrow {\rm D}_2^{*+} \ell^- \overline{\nu}_{\ell})$;
\item $b_{NR}={\rm BR}(\Bdb \rightarrow {\rm D} \pi \ell^- \overline{\nu}_{\ell})$ corresponding to a possible non-resonant contribution;
\item $s_{NR}$: the slope of an assumed mass distribution for the non-resonant
component which is taken to be exponentially decreasing from threshold;
\item $b_{\pi\pi}={\rm BR}(\Dsstar \rightarrow {\rm D} \pi \pi)$ which is assumed
to be independent of any particular $\Dsstar$ state.
\end{itemize}

Constraints from external measurements of the production rate and mass 
distribution for narrow states  have been applied~\cite{ref:lephf}:
\begin{itemize}
\item $b_{D_1}= (0.64 \pm 0.11)~\%$; 
\item $b_{D_2^*}=(0.28 \pm 0.09)~\%$. 
\end{itemize}
Removing these two constraints from the fit and imposing the values for the
mass and width of the broad ${\rm D}_1^*$ state, we find  
$b_{D_1}= (0.33 \pm 0.17)~\%$ and $b_{D_2^*}=(0.37 \pm 0.17)~\%$, 
compatible with the values given above.

The following decay channels have been considered:
\begin{itemize}
\item ${\rm D}_0^* \rightarrow {\rm D} \pi$;
\item ${\rm D}_1^*~{\rm and}~{\rm D}_1 \rightarrow \Dstar \pi 
~{\rm and}~{\rm D}\rho$;
\item the ${\rm D}_2^*$ can decay both into ${\rm D}\pi$ and ${\rm D}^*\pi$. 
The value 
$0.29\pm 0.07$ \cite{ref:argcle} has been used for the decay probability into 
${\rm D}^{*}\pi$
channels;
\item the possible non-resonant component is expected to decay into
${\rm D} \pi$ only as, for the $\Dstar \pi$ channel, there is no 
contribution from the $\Dstar$ or $\Bstar$ poles which are expected to play
the main role in non-resonant production \cite{ref:goity}.

\end{itemize}

As this analysis is based on the reconstruction of charged particles, decay
channels with neutrals appear at a lower mass than the genuine 
decaying $\Dsstar$. Narrow states decaying into $\Dstar \pi$, where
the $\Dstar$ decays into a neutral pion or photon and a reconstructed 
D, still appear as  peaks in the ${\rm D} \pi$ mass distribution, 
but slightly broader and displaced from their nominal values. 
The expected accuracy of the mass reconstruction for completely exclusive
decays and the effects induced by missing neutrals have been studied
using simulated events. 
For completely reconstructed $\Dsstar$ decays, the experimental resolution on
$\Delta_m$ is equal to 4 $\MeVcd$. 
For decays with a $\Dstar$
cascading into D$\pi^0$ or D$\gamma$, and in which the neutral particle 
is lost,
an additional smearing of 4 or 15 $\MeVcd$ is expected, respectively.
For the $\Dstarp \pi^-$ exclusive decay channel, in which the $\Do$,
cascading from the $\Dstarp$, decays into $\Km \pi^+ (\pi^0)$, the
mass resolution is poorer and values of 50 and 70 $\MeVcd$
have been obtained from the simulation for 94-95 and 92-93 data samples
respectively. For $\Dsstar \rightarrow {\rm D}\pi\pi$ decays, where
only one pion is reconstructed, the ${\rm D}\pi$ mass distribution
has been modelled using  Gaussian distributions whose central value
and standard deviation have been parametrized as a function of the central
$\Dsstar$ mass. Two parametrizations have been used for
narrow and broad states, respectively. 
For broad states the variation of the standard
deviation, as a function of the width of the $\Dsstar$ state,
has also been included. 
For narrow states with a 2.4 $\GeVcd$ mass, the expected standard deviation
of the ${\rm D}\pi$ mass distribution is 50 $\MeVcd$ whereas it is 90 $\MeVcd$
for a broad state having the same mass and a 200 $\MeVcd$ width.

Events have been distributed in two-dimensional histograms of the discriminant 
variable $R$
versus $\Delta_m$ with a 10 $\MeVcd$ binning, to be sensitive to the presence of narrow 
components. For each channel
($\Do \pi^+(2)\footnote{The numbers in parentheses indicate the number of analysed samples corresponding
to different D$^{(0,+)}$ decay channels.},
~\Dp \pi^-(1)~{\rm and }~\Dstarp \pi^-(3)$) and each
data set (2), corresponding to twelve histograms in total, 
the observed number of events in each bin has been compared with the expectations assuming
a Poisson distribution with average corresponding to the sum of the expected number
of signal and background events. 

Two-dimensional distributions for the expected number of signal and background events
have been constructed from the product of one-dimensional $\Delta_m$ and $R$ distributions.

\begin{figure}[hbtp!]
  \begin{center}
    \mbox{\epsfig{file=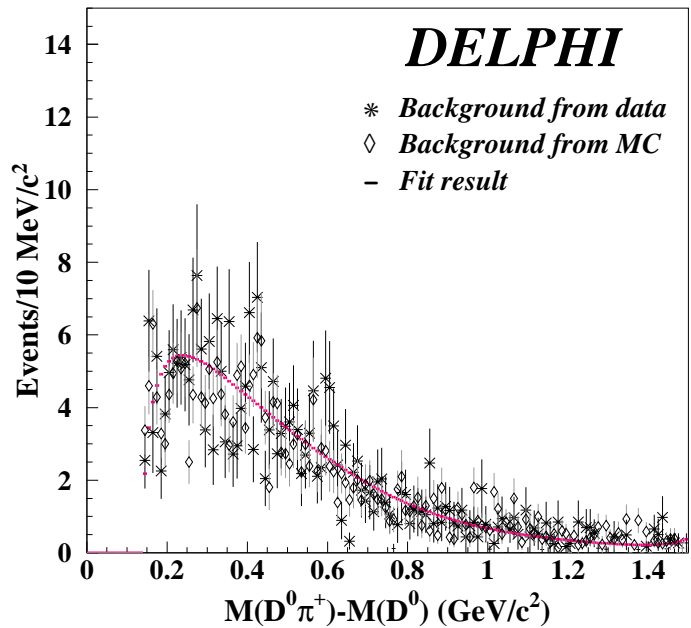,width=10cm,height=10cm}}
  \end{center}
  \caption[]{\it {Comparison between $\Delta_m$ distributions obtained 
for combinatorial background events selected in the sidebands for
real data (stars) and simulation (diamonds) corresponding to
the channel $\Do \pi^+$ in the 1994-1995 data sample.
The two distributions 
have been normalized to the same number of entries. The curve fitted
to simulated events has been superimposed.}
   \label{fig:backgdatamc}}
\end{figure}

For background events, the $\Delta_m$ distributions have been taken from 
simulation after verifying that their shape agreed with those measured with
real data events selected in the sidebands
(see Figure \ref{fig:backgdatamc}). A fit of these distributions,
using the parametrization given by the following expression:
\begin{equation}
Bckg(\Delta_m)= (\Delta_m - m_{\pi})^{\alpha} \exp{(p_0 +p_1 \Delta_m
+p_2 \Delta_m^2 +p_3 \Delta_m^3 + p_4 \Delta_m^4 )};~\alpha=0.5,
\label{eq:bckg}
\end{equation}
has been performed and the fitted distribution has been used to determine the expected
average background in each bin. The $R$ distribution corresponds to the
one resulting from the fit of the different background components done
in Section \ref{sec:fitbr}. The normalisation of background events
has also been obtained from these fits.

The $\Delta_m$ distributions for signal events have been obtained by summing all 
components whose contribution is expected in the various channels. Breit-Wigner
distributions have been used for each resonant state ($i$):

\begin{equation}\label{eq:bw}
BW(\Delta_m)= N_i \frac{\frac{\Gamma_{ij}}{2}}
{(m-m_i)^2+\left (\frac{\Gamma_{ij}}{2} \right )^2}.
\end{equation}
In this expression, $m$ is equal to 
$ m({\rm D}^{(*)}) + \Delta_m + { shift_j}$ with $shift_j$ parametrizing the 
possible displacement due to loss of neutrals in the $\Dstar$ decay, depending on 
the channel $j$ of interest. The width, $\Gamma_{ij}$, receives 
contributions from both the natural width of the physical state
$\Gamma_{i}$ and from the experimental resolution.
Expected variations of $\Gamma_{i}$, as a function of the mass ($m$) and of the angular 
momentum in the decay have been taken into account.
Finally, $N_i$ is the normalisation factor for the integral of $BW(\Delta_m)$ over the 
accessible mass range to be unity. 

For the non-resonant D$\pi$ component, a normalized 
exponential distribution has been used:
\begin{equation}
NR(\Delta_m)=\frac{s_{NR} \exp{(-s_{NR}(\Delta_m-m_{\pi}))}}
{1-\exp{(-s_{NR}(\Delta_m^{max}-m_{\pi}))}}
\end{equation}
This distribution is maximum for low values of $\Delta_m$. Such a behaviour
is expected when considering that the non-resonant D$\pi$
component is induced by the $\Dstar$ and $\Bstar$ poles, using
chiral dynamics \cite{ref:goity}. 
%This component can interfere with resonant
%states, decaying into the same final state. Using \cite{ref:goity} it
%has been verified that the interference term has a similar mass dependence.
Using the same model, in the $\Dstar \pi$ decay channel, the non-resonant contribution 
is expected to be small, as the $\Dstar$ and $\Bstar$ poles are not 
effective in this channel, and it has been neglected.

The $R$ distribution for signal events is taken from the simulation. 
%It has been verified that the cuts used to select the events do not 
%introduce a measurable bias in the $\Dsstar$ mass distribution.

\subsection{Results on production characteristics of $\Dsstar$ states}
\label{sec:Dsres}

\begin{figure}[hbtp!]
% ds+pi-
\begin{center}
%    \mbox{\epsfig{file=fit_rd_rs_dopiplus_50mev_13_3_1.eps ,width=8.5cm}
%    \epsfig{file=fit_rd_rs_dpluspimin_50mev_13_3_1.eps ,width=8.5cm}}\\
%    \mbox{\epsfig{file=fit_rd_rs_dspluspimin_50mev_13_3_1.eps ,width=8.5cm}
    \mbox{\epsfig{file=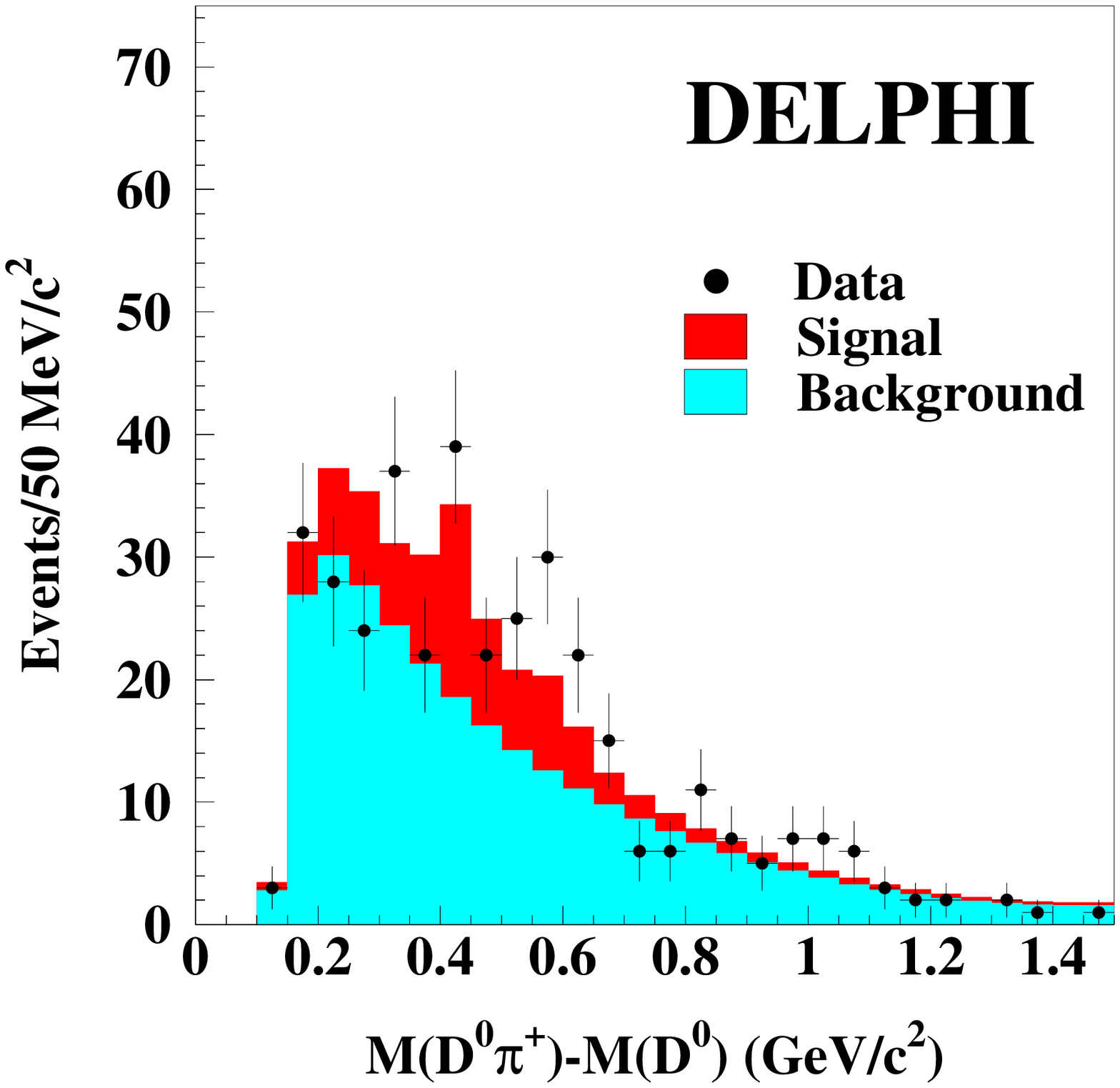 ,width=8.5cm}
    \epsfig{file=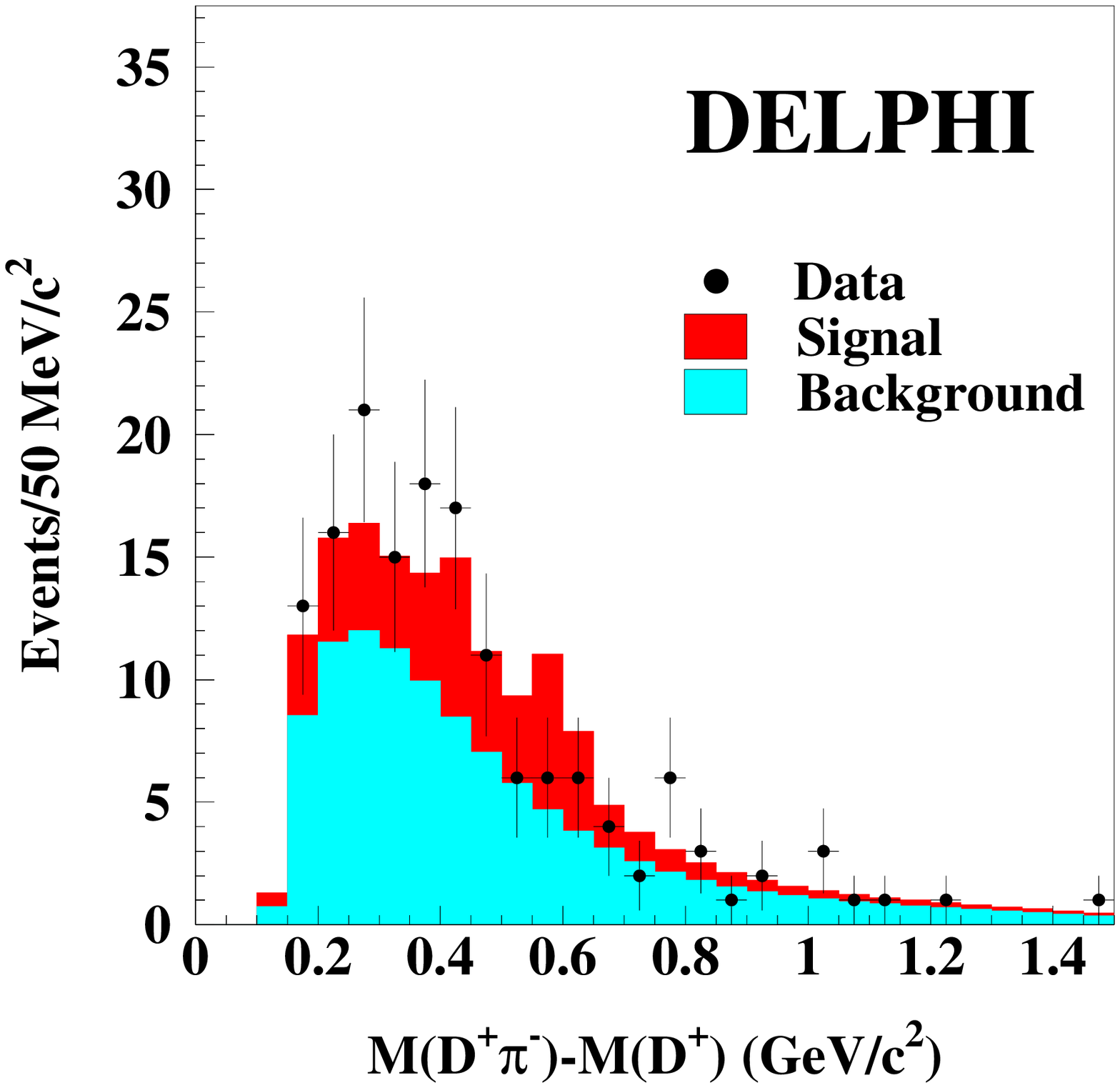 ,width=8.5cm}}\\
    \mbox{\epsfig{file=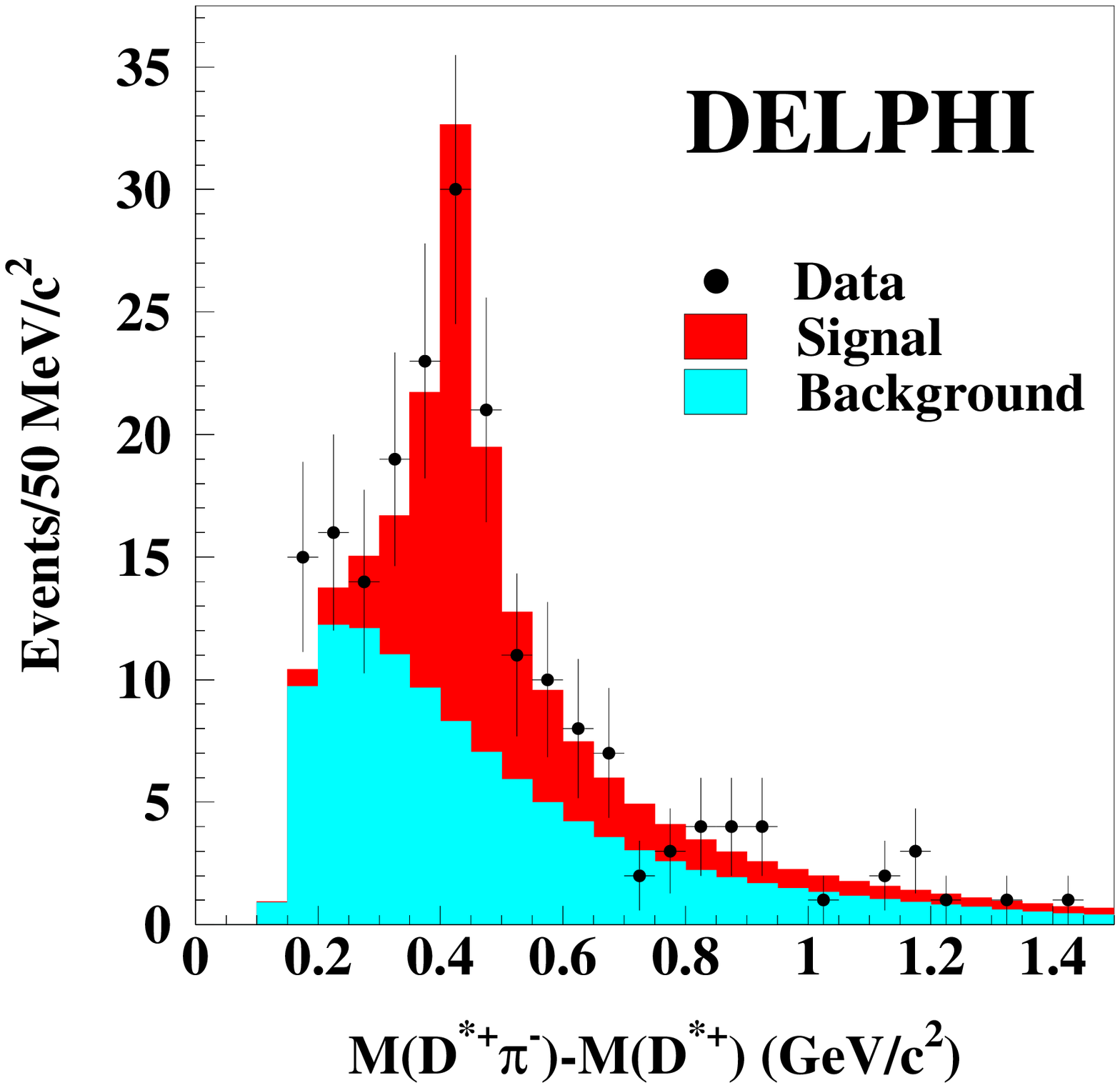 ,width=8.5cm}
    }
\end{center}
  \caption[]{\it {$\Delta_m$ distributions for right-sign events corresponding
to the three ${\rm D}^{(*)}\pi$ combinations.The fitted signal contributions
are superimposed.}
   \label{fig:massdss}}
\end{figure}

\begin{figure}[hbtp!]
  \begin{center}
    \mbox{\epsfig{file=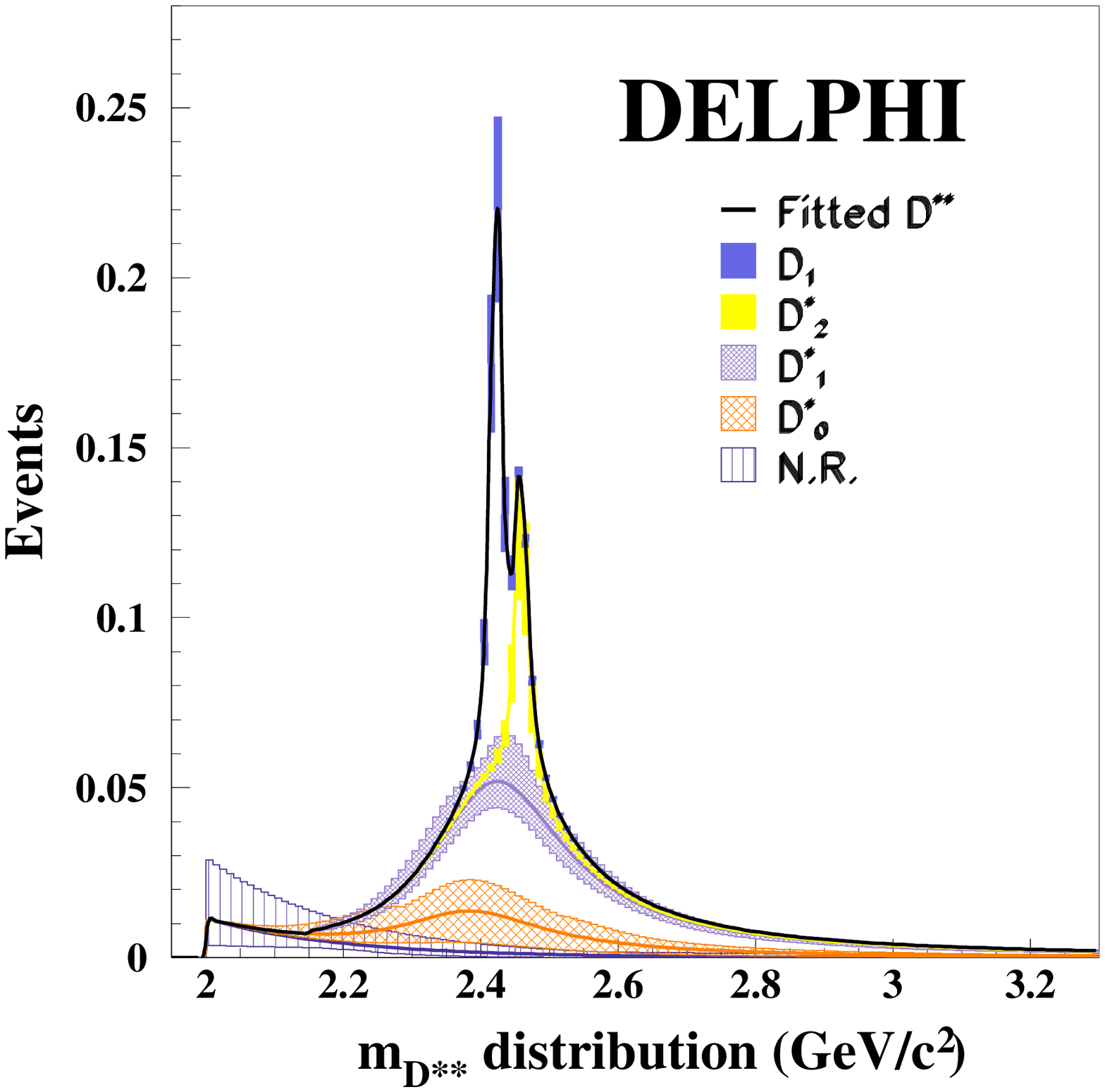,width=9cm}}
  \end{center}
  \caption[]{\it {$\Dsstar$ fitted mass distribution. The different 
components are shown within bands corresponding to the uncertainties 
of the fitted parameters.}
   \label{fig:dsstarmass}}
\end{figure}

Results of the fit are given in Table~\ref{tab:brmes}.
%, 
%the mass $m_{D_0^*}$  of the 
%${\rm D}_0^*$ has been kept fixed at 2.4 $\GeVcd$. 
The corresponding 
$\Delta_m$ mass distributions 
are shown in Figure~\ref{fig:massdss} and the fitted $\Dsstar$
mass distribution, comprising all components, is given in Figure
\ref{fig:dsstarmass}.

Among the fitted parameters, three sets can be considered:
\begin{itemize}
\item quantities corresponding to significant measurements in the present analysis:
the production rate, mass and width of the broad D$_1^*$ state;
\item quantities introduced to parametrize components which are possibly contributing
in the hadronic final state, but for which the present statistics do not 
allow a significant measurement. These components have been introduced as they 
can correspond to different measured ${\rm D}^{(*)} \pi$ mass distributions. 
For instance it is not
equivalent to consider a measured D$\pi$ mass as originating from the
non-resonant component or from a D$\pi\pi$ decay. These parameters are the D$_0^*$
production rate and width and the two quantities describing the possible non-resonant
component;
 
\item quantities constrained by external measurements, in the fit, such as the production fractions
of narrow D$_1$ and D$_2^*$ states.
\end{itemize}

In addition, two parameters ($m_{D_0^*}$ and $b_{\pi\pi}$) have been kept fixed in the fit
and then varied within a specified range to evaluate their contribution to systematic
uncertainties. In the following some remarks on these results are made.

\begin{table}[htb]
\begin{center}
  \begin{tabular}{|c|c|}
    \hline
 parameter & Fitted value    \\
         & $\pm$ stat. err. $\pm$ syst.   \\ 
    \hline
$b_{D_1^*}$ &$(1.24 \pm 0.25\pm 0.27 )\%$  \\
$m_{D_1^*}$ &$2445 \pm 34\pm 10 ~\MeVcd$  \\
$\Gamma_{D_1^*}$ &$234 \pm 74\pm 25~\MeVcd$  \\
\hline
$b_{D_0^*}$ &$(0.42 \pm 0.33\pm 0.22 )\%$  \\
$\Gamma_{D_0^*}$ &$260 \pm 130\pm 130~\MeVcd$  \\
$b_{NR}$ &$(0.23 \pm 0.35\pm 0.44)\%$  \\
$s_{NR}$ &$5 \pm 7 ~(\GeVcd)^{-1}$  \\
%$b_{\pi\pi}$ &$(0.19 \pm 0.13\pm 0.)\%$  \\
\hline
$b_{D_1}$  &$(0.56 \pm 0.10)\%$ (constrained) \\
$b_{D_2^*}$ &$(0.30 \pm 0.08)\%$ (constrained) \\
\hline
  \end{tabular}
  \caption[]{\it { Fitted values for parameters of 
the $\Dsstar$ production characteristics.}
  \label{tab:brmes}}
\end{center}
\end{table}

%In the D$\pi$ final state, data favours a non-resonant component maximal 
%close to threshold with the remaining ${\rm D}_0^*$ component compatible with 
%zero within a 0.2$\%$ uncertainty. The present statistics are not sufficient to unambiguously establish
%the presence of these non-resonant states.

\subsubsection{Broad states ${\rm D}_{0}^{*}$ and ${\rm D}_{1}^{*}$}
Most of the $\Dstar \pi$ component originates from the broad ${\rm D}_{1}^{*}$
state whose mass and width have been measured. These two last values can 
be compared with results obtained for this particle by 
the CLEO \cite{ref:cleobroad} and BELLE \cite{ref:bellebroad} collaborations, 
using ${\rm B}\to {\rm D}^{(*)}\pi\pi$ decays:
\bc
$m_{D_1^*}=2461^{+41}_{-34}\pm10 \pm32\,\MeVcd; \qquad \Gamma_{D_1^*}=290^{+101}_{-~79}\pm26 \pm36\,\MeVcd$ \cite{ref:cleobroad} 
\ec
\bc
$m_{D_1^*}=2427\pm26 \pm20 \pm15\,\MeVcd;\qquad \Gamma_{D_1^*}=384^{+107}_{-75}\pm24 \pm70\,\MeVcd$ \cite{ref:bellebroad} 
\ec
This is the first time that the parameters of this resonance have been measured in semileptonic
B hadron decays.

The present statistics is not sufficient 
to evaluate separately the contributions from the D$_0^*$ 
(or other broad resonances), from
the non-resonant component and from a possible D$\pi\pi$ decay mode of 
$\Dstarstar$ states.

\subsubsection{The non-resonant component}
The fitted rate of the non-resonant component is compatible with zero and this is in agreement
with expectations. The contribution from the $\Dstar$ pole to the non-resonant
D$\pi$ component can be evaluated knowing the total branching fraction for the decay
$\Bdb \rightarrow \Dstarp \ell^- \overline{\nu}_{\ell}$, the minimal value used for 
$\Delta_m$ to select the events and the value of the $\Dstar$ width in the hypothesis
that the mass is distributed according to a Breit-Wigner: 
%{\scriptsize{
%\begin{equation}
%\left . {\rm BR}(\Bdb \rightarrow \Do \pi^+ \ell^- \overline{\nu}_{\ell})
%\right |^{\Dstar pole}_{NR}=\frac{2}{3}\left (1-
%\frac{2}{\pi}\arctan{2 \frac{\left . m(\Do \pi^+)\right |_{min.}-m(\Dstarp)}{\Gamma_{\Dstarp}}}\right)
%\times {\rm BR}(\Bdb \rightarrow \Dstarp \ell^- \overline{\nu}_{\ell}) 
%\simeq 0.12\%  
%\end{equation}
%}}
%% *** A.O. ** 4/5/04 *** modified to avoid small fonts ***

\begin{eqnarray}\label{eq:brnrteo}
\left . {\rm BR}(\Bdb \rightarrow \Do \pi^+ \ell^- \overline{\nu}_{\ell})
\right |^{\Dstar pole}_{NR} \hspace{0.65\textwidth}\nonumber \\ 
\hspace{0.1\textwidth} =~\frac{2}{3}\left (1- \frac{2}{\pi}\arctan{2 \frac{\left . m(\Do \pi^+)\right |_{min.}-m(\Dstarp)}{\Gamma_{\Dstarp}}}\right) 
\times {\rm BR}(\Bdb \rightarrow \Dstarp \ell^- \overline{\nu}_{\ell}) 
\simeq 0.12\%  \nonumber 
\end{eqnarray}
corresponding to a minimum $\Do \pi^+$ mass situated at 10 widths from 
the pole.
For $\left . {\rm BR}(\Bm \rightarrow \Dp \pi^- \ell^- \overline{\nu}_{\ell})
\right |^{\Dstar pole}_{NR}$, there is a natural cutoff corresponding to
the sum of the $\Dp$ and $\pi^-$ masses and the value becomes $0.05\%$.
%\begin{equation}
%{\small
%\left . {\rm BR}(\Bm \rightarrow \Dp \pi^- \ell^- \overline{\nu}_{\ell})
%\right |^{\Dstar pole}_{NR}=\frac{2}{3}
%\frac{2}{\pi}\arctan{2 \frac{\left . m(\Dp \pi^-)\right |_{min.}-m(\Dstaro)}{\Gamma_{\Dstaro}}}
%\times {\rm BR}(\Bm \rightarrow \Dstaro \ell^- \overline{\nu}_{\ell}) 
%\simeq 0.1\% \nonumber }
%\end{equation}

The contribution from the $\Bstar$ pole can be also included in a controlled
way by considering that the slow pion emission, by mesons containing
a heavy quark, is governed by a universal parameter ${\hat g}$
whose value is fixed by the measured $\Dstar$ width \cite{ref:ghat}:
\begin{equation}
\Gamma \left ( \Dstar \rightarrow {\rm D} \pi\right )=
\frac{{\hat g}^2}{4 \pi f_{\pi}^2} p_{\pi}^3
%{\hat g}= 0.59 \pm 0.01 \pm 0.07.
\end{equation} 
where $f_{\pi}$ is the pion decay constant and $p_{\pi}$ is the momentum
in the $\Dstar$ rest frame.
The interference term between the $\Dstar$ and $\Bstar$ pole 
contributions is also fixed and,
using the values given above, this gives:

\begin{equation}
\left . {\rm BR}(\Bdb \rightarrow \Do \pi^+ \ell^- \overline{\nu}_{\ell})
\right |^{\Dstar~+~\Bstar poles}_{NR} 
\simeq 0.07\%. \nonumber
\end{equation}
 
The main uncertainty, in the non-resonant contribution, comes from the 
interference with the ${\rm D}_0^*$. Varying the branching fraction,
${\rm BR}(\Bdb \rightarrow {\rm D}_0^* \ell^- \overline{\nu}_{\ell})$ 
between 0.1 and 0.8$\%$, the non-resonant contribution changes between
$+0.02\%$ and $-0.09\%$ assuming that the interference has a negative
sign below the ${\rm D}_0^*$ pole. The corresponding variation becomes
$+0.1\%$ to $+0.2\%$ for the other sign. 

\subsubsection{Total $\Dsstar$ production rate}
Summing all fitted components, the total rate for $\Dsstar$ production amounts to:
\begin{equation}
{\rm BR}(\Bdb \rightarrow \Dsstar \ell^- \overline{\nu}_{\ell})=
(2.7 \pm 0.7 \pm 0.2)\%.
\label{eq:dsstarrate}
\end{equation}

\subsubsection{Systematic uncertainties}
In addition to those considered in Table \ref{tab:ratesyst}, systematic 
uncertainties have been evaluated for the following effects:

\begin{itemize}
\item the $b_{\pi\pi}$ branching fraction has been varied within the range 
$(20 \pm 15)\%$. The central value corresponds to an estimate obtained
by comparing the phase-space integrals for the $\Dstar \pi$
and D$\rho$ decay modes of a $\Dstarstar$ state. This variation is also in agreement
with the value obtained in data for this parameter when it is left free
to vary in the fit: $(19 \pm 13)\%$; 
\item the central value of the ${\rm D}_0^*$ mass, $m_{D_0^*}$,
which is kept fixed during the fit has been varied between 2.3 and 2.5 
$\GeVcd$;

\item parameters used to define the shape of the combinatorial background distributions
in $\Delta_m$ (see Equation~(\ref{eq:bckg})) have been 
varied. The value of $\alpha$ has been changed between 0.3 and 0.7 and the degree
of the polynomial function was also taken as 3 or 5 ;

\item the contribution from $\Bsb$ semileptonic decays, which have been modelled
similarly to the non-strange $b$-mesons. Properties
of narrow ${\rm D}_s^{**}$ states published in \cite{ref:PDG02}
have been used. For broad states it has been assumed  that their
masses were displaced, relative to the corresponding non-strange
states by the same amount as for narrow states. $\Delta_m$
mass distributions for ${\rm D}^{(*)} \Kp$ final states have also been
evaluated by considering that the $\Kp$ was reconstructed as a $\pi^+$.
For excited charm states produced in semileptonic decays of the
$\Lb$, no simulation has been used, considering the lack of information
on such decays. It has been assumed that the contribution from
these final states was about 50$\%$ of the variation observed when
$\Bsb$ decays are introduced. An uncertainty corresponding to
twice this variation has been used to account for $\Bsb$ and $\Lb$
contributions;

\item considering the recent result from the BELLE collaboration which has measured
a larger width for ${\rm D}_2^*$ mesons~\cite{ref:bellebroad}  than quoted in \cite{ref:PDG02}, 
the effect of changing this value from 20 to 40 $\MeVcd$ has been  evaluated.

\end{itemize}

Individual contributions from these sources of systematics have been listed
when evaluating moments of the hadronic mass distribution. It has also been verified 
that possible additional systematic sources such as:
\begin{itemize}
\item[-] the relative branching fraction of ${\rm D}_2^*$ mesons into $\Dstar \pi$ which
has been varied according to the expected value:  $0.29\pm 0.07$ \cite{ref:argcle};

\item[-]  the expected mass reconstruction accuracy which has been varied by $30\%$
\end{itemize}
have negligible effects on hadronic mass moments.
Uncertainties related to the control of the shape of the discriminant variable
distribution have a small effect ($4\%$) on the measured $\Dstarstar$ production rate
and negligible contributions to hadronic mass moments.

\section{Moments of the hadronic mass distribution in $b$-hadron
semileptonic decays}
\label{sec:mx}

Moments of the $\Dsstar$ mass distribution can be evaluated from the
results of the fit discussed above. Results are given in Table~\ref{tab:dssmom}.
Statistical uncertainties have been obtained by propagating those
on fitted parameters, using their full covariance matrix.
Systematics are dominated by the uncertainty on the possible contribution
from D$\pi\pi$ decays and could be reduced in future when experimental results
on this decay channel become available.

\begin{table}[htb]
\begin{center}
{\small
  \begin{tabular}{|c|r|r|r|r|r|r|}
    \hline
 &$<m_{D^{**}}>$ &$<m_{D^{**}}^2>$  & $<m_{D^{**}}^4>$ & $<m_{D^{**}}^6>$
& $<m_{D^{**}}^8>$ & $<m_{D^{**}}^{10}>$\\ 
 &$(\GeVcd)$ &$(\GeVcd)^2$  & $(\GeVcd)^4$ & $(\GeVcd)^6$
& $(\GeVcd)^8$ & $(\GeVcd)^{10}$\\
    \hline
value &$2.483 \quad $  & $6.22  \quad $  &$40.1  \quad $  &$270.6 \quad $  &$1932  \quad $  &$14732 \quad $ \\
\hline
stat. uncert. &$ \pm0.033 \quad $  & $ \pm0.16 \quad $  & $\pm2.0 \quad $  &$\pm20.9 \quad $ & $\pm206 \quad $  &$\pm2039 \quad $  \\
\hline
$b_{\pi\pi}$&$ \pm 0.030 \quad $  & $\pm 0.14 \quad $  &$\pm 1.6 \quad $  &$\pm 14.4 \quad $ & $\pm 122 \quad $  & $\pm 1032 \quad $  \\
$m_{D_0^*}$ &$ \pm 0.008 \quad $  & $\pm 0.04 \quad $  &$\pm 0.3 \quad $  &$\pm 2.5 \quad $   &$\pm 18 \quad $   & $\pm 36 \quad $  \\
backg. param &$ 0.003 \quad $  & $0.02 \quad $  &$0.2 \quad $  &$2.6 \quad $    &$28 \quad $  & $291 \quad $ \\
$\Bsb,~\Lb$&$ \pm 0.010 \quad $  & $\pm 0.04 \quad $  &$\pm 0.5 \quad $  &$\pm 4.3 \quad $  &$\pm 38 \quad $   &$\pm 334 \quad $  \\
%disc. var.&$ 0.024 \quad $  & $0.11 \quad $  &$1.4 \quad $  &$15. \quad $   &$4 \quad $  &$25 \quad $ \\
casc. rate &$ 0.001 \quad $  & $0.01 \quad $  &$0.1 \quad $  &$0.4 \quad $   & $4 \quad $  & $25 \quad $  \\
$d_{\pm}$ dist. &$ 0.002 \quad $  & $0.01 \quad $  &$0.2 \quad $  &$2.0 \quad $   &$22 \quad $  &$244 \quad $ \\
%$BR(D_2^* \rightarrow \Dstar \pi)$ &$ 0.000 \quad $  & $0.00 \quad $  &$0.0 \quad $  &$0.4 \quad $   & &\\
%mass resol. &$ 0.000 \quad $  & $0.00 \quad $  &$0.0 \quad $  &$0.2 \quad $   & &\\
$\Gamma(D_2^*)=40\MeVcd$ &$  -0.002 \quad $  & $-0.01 \quad $  &$-0.2 \quad $  &$-2.0 \quad $   &$-19 \quad $   &$-200 \quad $  \\
\hline
Tot. syst.&$0.033 \quad $  & $0.15  \quad $  &$1.7  \quad $  &$15.7  \quad $   &$135 \quad $  & $1167 \quad $ \\
\hline
  \end{tabular}
  \caption[]{\it {Measured moments of the $\Dsstar$ mass distribution. When the sign 
of the variation is not given this is because the corresponding quoted systematic error originates
from several sources corresponding to different signs which are given in the text.}
  \label{tab:dssmom}}
}
\end{center}
\end{table}

In determining the moments of the complete hadronic mass distribution
in $b$-hadron semileptonic decays, $b \rightarrow {\rm D}~{\rm and}~
\Dstar \ell^- \overline \nu_{\ell}$ channels have been included.

For the first channel, values of branching fractions 
given in~\cite{ref:PDG02}, have been used.
As $\Bdb$ and $\Bm$ are expected to have the same partial
decay width into the ${\rm D} \ell^- \overline \nu_{\ell}$ channel, we get:
\begin{equation}
{\rm BR}(\Bdb \rightarrow \Dp \ell^- \overline \nu_{\ell})=
(2.06 \pm 0.20)\%.
\label{eq:bdlnu}
\end{equation}

For the second channel, the value given in~\cite{ref:hfag} 
at the time of the Winter 2003 conferences has been used: 
%complemented by the more recent results from BELLE, CLEO and DELPHI;
%obtaining:
\begin{equation}
{\rm BR}(\Bdb \rightarrow \Dstarp \ell^- \overline{\nu}_{\ell})=
(5.27 \pm 0.19)\%.
\label{eq:bdstarlnu}
\end{equation}
%which is larger than the PDG2002 average of ($4.60 \pm 0.21\%)$.

The inclusive semileptonic branching fraction 
(${\rm BR}(\Bdb \rightarrow c \ell^- \overline{\nu}_{\ell})$) has also been 
included as a constraint:
\begin{equation}
{\rm BR}(\Bdb \rightarrow c \ell^- \overline{\nu}_{\ell})=
\frac{\tau(\Bdb)}{\tau(b)}
{\rm BR}(b \rightarrow c \ell^- \overline{\nu}_{\ell})=
(10.25 \pm 0.30)\%.
\label{eq:aver}
\end{equation}
This last value has been obtained using the average values determined at LEP
for the quantities entering into Equation~(\ref{eq:aver}), which are
given in Tables \ref{tab:external} and \ref{tab:meast}.
As the inclusive semileptonic 
branching fraction is the sum of the D, $\Dstar$ and
$\Dsstar$ contributions, the expected rate for $\Dsstar$ production
can be derived from these values:
\begin{equation}
{\rm BR}(\Bdb \rightarrow \Dsstar \ell^- \overline{\nu}_{\ell})=
(2.9 \pm 0.4)\%
\end{equation}
This is in agreement with the rate measured directly as given in 
Equation~(\ref{eq:dsstarrate}).

Moments of the hadronic mass distribution have then been derived as:
\begin{equation}
<m_H^n> = p_D ~m_D^n + p_{D^*} ~m_{D^*}^n + p_{D^{**}} ~<m_{D^{**}}^n>
\end{equation}
where $p_{D_i}=
\frac{{\rm BR}(\Bdb \rightarrow {\rm D}_i \ell^- \overline{\nu}_{\ell})}
{{\rm BR}(\Bdb \rightarrow c \ell^- \overline{\nu}_{\ell})}$.
The value of $p_{D^{**}}$ has been obtained by imposing the constraint
$1=p_D + p_{D^*}  + p_{D^{**}}$ and including the measurement of the $\Dsstar$ 
production rate as given in Equation~(\ref{eq:dsstarrate}). 
%It can be more convenient to express these moments by reference 
%to the spin-average D meson mass \footnote{The spin-average D meson mass
%is the weighted average of the D and $\Dstar$
%masses, the weights being equal to the number of spin components of the two
%states: $m_{spin}=\frac{m_D + 3  m_{D^*}}{4}~=~1.97375~\GeVcd$.}
%($M_n^H=<(m_H^2-m_{spin}^2)^n>$)
%or to the measured average hadronic mass
%($M_n^{\prime H}=<(m_H^2-<m_H^2>)^n>$).

Results, following the notations in Equation~(\ref{eq:defmomh}), are given in 
Tables~\ref{tab:hadmom1} and \ref{tab:hadmom2}.
Systematic uncertainties 
related to measurements of the branching fractions 
in Equations~(\ref{eq:bdlnu}, 
%\ref{eq:bdstarlnu} and
\ref{eq:aver}) are also given in Tables~\ref{tab:hadmom1}
and~\ref{tab:hadmom2}.

Effects induced by the variation of the analysis efficiency versus the
mass of $\Dsstar$ states have been evaluated to correspond to an increase
of 0.7, 1.5 and 2.5$\%$  for $M_1^H$, $M_2^{\prime H}$  and 
$M_3^{\prime H}$ respectively.
Considering the present level of statistical and systematic uncertainty 
of actual measurements,
these corrections have not been included in the quoted central values
for hadronic moments.

\begin{table}[htb]
{\small
\begin{center}
  \begin{tabular}{|c|r|r|r|r|r|}
    \hline
 & $M_1^H\quad $ &$M_2^H \quad$  &$M_3^H \quad$  &$M_4^H \quad$  & $M_5^H \quad $\\
% & {\tiny $<m_H^2-m_{spin}^2>$} &{\tiny $<(m_H^2-m_{spin}^2)^2>$} &
%{\tiny $<(m_H^2-<m_H^2>)^2>$} &{\tiny $<(m_H^2-<m_H^2>)^3>$} \\
 &$(\GeVcd)^2$  & $(\GeVcd)^4$ & $(\GeVcd)^6$
& $(\GeVcd)^8$ & $(\GeVcd)^{10}$\\
    \hline
value & $  0.647  \quad $  & $  1.98 \quad $ & $  7.4 \quad $ & $  35.7  \quad $ & $  205 \quad $\\
\hline
stat. uncert. & $\pm 0.046 \quad $  & $\pm 0.23 \quad $ & $\pm 1.3 \quad $ & $\pm 7.9 \quad $ & $\pm1080 \quad $\\
\hline
Ext. BR  &$  0.079 \quad $ & $  0.22 \quad $ &$  0.8 \quad $ &$  4.1 \quad $ & $  23.4 \quad $ \\
$b_{\pi\pi}$  &$ \pm0.039 \quad $ & $\pm0.15 \quad $ &$\pm0.6 \quad $ &$\pm2.8 \quad $ & $\pm16.4 \quad $ \\
$m_{D_0^*}$ &$ \pm0.015 \quad $ & $\pm0.04 \quad $ &$\mp0.0 \quad $ &$\mp0.7 \quad $ & $\mp7.0 \quad $ \\
backg. param &$  0.007 \quad $ & $  0.04 \quad $ &$  0.2 \quad $ &$  1.2 \quad $ &$  8.0 \quad $ \\
$\Bsb,~\Lb$&$ \pm0.007 \quad $ & $\pm0.03 \quad $ &$\pm0.2 \quad $ &$\pm0.9 \quad $ &$\pm5.2 \quad $ \\
%disc. var.&$  0.033 \quad $ & $  0.147 \quad $ &$$ &$$ & \\
%casc. rate &$  0.006 \quad $ & $  0.035 \quad $ &$$ &$$ & \\
$d_{\pm}$ dist. &$  0.005 \quad $ & $  0.03 \quad $ &$  0.2 \quad $ &$  1.1 \quad $ & $  7.4 \quad $\\
$\Gamma(D_2^*)=40\MeVcd $ &$  -0.004 \quad $ & $  -0.02 \quad $ &$  -0.1 \quad $ &$  -0.9 \quad $ &$  -5.8 \quad $ \\
\hline
Tot. syst.&$  0.090 \quad $ & $  0.27 \quad $ &$  1.1 \quad $ &$  5.4 \quad $ & $  32.3 \quad $\\
\hline
  \end{tabular}
  \caption[]{\it {Measured moments of the hadronic mass distribution,
by reference to the spin averaged D-hadron mass,
in $b$-hadron semileptonic decays.}
  \label{tab:hadmom1}}
\end{center}
}
\end{table}

\begin{table}[htb]
{\small 
\begin{center}
  \begin{tabular}{|c|r|r|r|r|}
    \hline
 & $M_2^{\prime H}\quad$ &$M_3^{\prime H}\quad$ &$M_4^{\prime H}\quad$ &$M_5^{\prime H}\quad$ \\
% & {\tiny $<m_H^2-m_{spin}^2>$} &{\tiny $<(m_H^2-m_{spin}^2)^2>$} &
%{\tiny $<(m_H^2-<m_H^2>)^2>$} &{\tiny $<(m_H^2-<m_H^2>)^3>$} \\
 & $(\GeVcd)^4$ & $(\GeVcd)^6$ & $(\GeVcd)^8$ & $(\GeVcd)^{10}$\\
    \hline
value & $1.56 \quad $  & $4.05 \quad $ & $21.1 \quad $ & $116.0  \quad $ \\
\hline
stat. uncert. & $\pm 0.18 \quad $  & $\pm 0.74 \quad $ & $\pm 4.5 \quad $ & $\pm 27.0 \quad $ \\
\hline
Ext. BR  &$ 0.12 \quad $ & $0.15 \quad $ &$1.1 \quad $ &$5.0 \quad $  \\
$b_{\pi\pi}$  &$ \pm0.10 \quad $ & $\pm0.17 \quad $ &$\pm1.0 \quad $ &$\pm5.2 \quad $  \\
$m_{D_0^*}$ &$ \pm0.02 \quad $ & $\mp0.10 \quad $ &$\mp0.8 \quad $ &$\mp6.0 \quad $  \\
backg. param &$ 0.03 \quad $ & $0.12 \quad $ &$0.7 \quad $ &$4.2 \quad $  \\
$\Bsb,~\Lb$&$\pm 0.02 \quad $ & $\pm0.06 \quad $ &$\pm0.4 \quad $ &$\pm2.3 \quad $  \\
%disc. var.&$ 0.02 \quad $ & $0.11 \quad $ &$ \quad $ &$$  \\
%casc. rate &$ 0.006 \quad $ & $0.035 \quad $ &$$ &$$  \\
$d_{\pm}$ dist. &$ 0.02 \quad $ & $0.11 \quad $ &$0.7 \quad $ &$4.0 \quad $ \\
$\Gamma(D_2^*)=40\MeVcd$&$ -0.02 \quad $ & $-0.10 \quad $ &$-0.5 \quad $ &$-3.1 \quad $  \\
\hline
Tot. syst.&$ 0.16 \quad $ & $0.32 \quad $ &$2.1 \quad $ &$11.7 \quad $  \\
\hline
  \end{tabular}
  \caption[]{\it {Measured moments of the hadronic mass distribution,
by reference to the average mass squared,
in $b$-hadron semileptonic decays.}
  \label{tab:hadmom2}}
\end{center}
}
\end{table}

Error correlation matrices are given in Appendix \ref{append:hadmom}.

\section{Moments of the lepton energy distribution in $b$-hadron
semileptonic decays}
\label{sec:el}

\subsection{Inclusive reconstruction of $b$-hadron semileptonic decays}

Selected events have been divided into 
two hemispheres using the thrust axis. The secondary hadronic system 
accompanying the 
lepton in the semileptonic decay has been reconstructed using an iterative procedure 
applied to 
the particles belonging to the same hemisphere as the tagged lepton.

Charged particles, belonging to the hemisphere of the candidate lepton, with 
$p>0.7$ GeV/$c$, and with at least one associated hit in the Vertex Detector 
and positive impact parameter with respect to the primary vertex have been 
considered.  These have been sorted in decreasing order of their probability 
of being B decay products based on their impact parameter significance and 
considering the particle that crosses the jet direction the furthest away 
from the primary vertex, or, when the crossing is not well-defined, the most 
energetic particle, to be the most likely B decay product.
 
The charged particles have been iteratively tested for forming a secondary vertex.  
The procedure has been iterated while the following conditions have been 
fulfilled:  invariant mass below $2.9$ GeV/$c^2$, distance from the primary 
vertex less than 3 cm but at least 2.8 times the uncertainty and on the 
positive side, and $\Delta \chi^2< 3$ after inclusion of each particle in the 
seed vertex fit.

In those cases where no secondary vertex was found, single particles have 
been accepted when fulfilling one of the following criteria, in decreasing 
order of quality: a charged particle with $p> 3$ GeV/$c$ having a crossing 
point with the jet axis at least 1$\sigma$ downstream from the primary vertex 
and also downstream from the lepton candidate crossing point, but less 
than 15 cm from the 
primary vertex; a charged particle with $p>3$ GeV/$c$ with the largest impact 
parameter significance and positive impact parameter sign; the most energetic 
charged particle within 0.6 rad of the jet axis. 

The remaining charged particles in the hemisphere of the lepton with $p>0.5$ 
GeV/$c$ and at least one vertex detector hit not yet associated with the 
secondary vertex are then considered.  Each of them is tested to belong 
to the vertex, 
and the one with the smallest $\chi^2$ contribution and giving a vertex mass 
closest to the D mass is included in the vertex, and all the remaining 
particles are then tested against this new vertex.  This process is continued 
until the remaining particles have large contributions to the vertex $\chi^2$ 
and increase the mass difference between the vertex and the D mass.

To improve the purity of the vertex, each of the particles associated with it is 
reconsidered, if there are more than two particles in the vertex, the 
$\chi^2$/n.d.f of the vertex exceeds 2 and the mass of the vertex is more 
than the D mass. 
%1.9 GeV/$c^2$
If removing a particle from the vertex 
improves the $\chi^2$/n.d.f and the mass after the removal is closer to the 
D mass, the particle is permanently removed from the vertex. 
The average charged multiplicity of the vertices is $2.9$.
As the last stage, identified $K^0_s$s within 0.8 rad and $\pi^0$s within 1 
rad of the lepton direction have been tested for association based on their 
energy, rapidity and contribution to the vertex mass.  

\begin{figure}[hbtp!]
\begin{center}
\epsfig{file=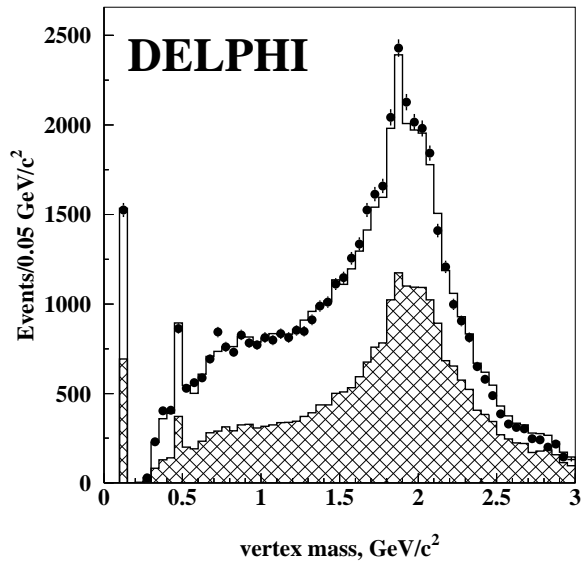, height=8.5cm}
\caption { \it{Mass distribution of the reconstructed charm vertex.  
The peaks at 0.14 and 0.5 GeV/$c^2$ correspond to vertices consisting of 
a single $\pi$ or $K$, respectively.  
%Right: mass distribution of the reconstructed B. 
Points are the data, while the histogram is the simulation with the signal 
clear and the background hatched.
%% ** A.O. ** 4/5/04 ** change sentence ** 
%is MC simulated signal,
% in black the background. 
}
\label{fig:vrm2} }
\end{center}
\end{figure}

On average, $78\%$ of 
the particles associated with the vertex were true D decay products, and 
$74\%$ of the decay products were correctly associated with the vertex.  
The mass distribution of the reconstructed vertex is shown in 
Figure~\ref{fig:vrm2}.

The hadronic energy is obtained from the energy of the reconstructed secondary
 vertex, corrected with a linear function of the reconstructed vertex mass, 
when the reconstructed mass is below the $\Do$ meson mass.  
The direction of the hadronic system is taken from the momentum sum of the 
particles included in the vertex.

For each decay, the energy of the B hadron has been estimated as the energy 
sum of the secondary hadronic system, the identified lepton and the 
neutrino energy.  

The energy obtained in this way was corrected by a function of the reconstructed hadronic 
energy, determined from simulation, with the maximum correction being $\pm$6~GeV. 
The neutrino energy was computed from the missing energy and momentum in both 
hemispheres corrected by a function of the missing mass in the event, 
determined from simulation.
The resolution of the neutrino energy in ${\rm B} \rightarrow {\rm X} \ell \overline{\nu}_{\ell}$ 
decays was estimated to be $\pm2.9$~GeV. 
Neutrino energies larger than 1 GeV were required. 

\begin{center} 
\begin{figure}[hbtp!]
    \mbox{\hspace{-1cm} \epsfig{file=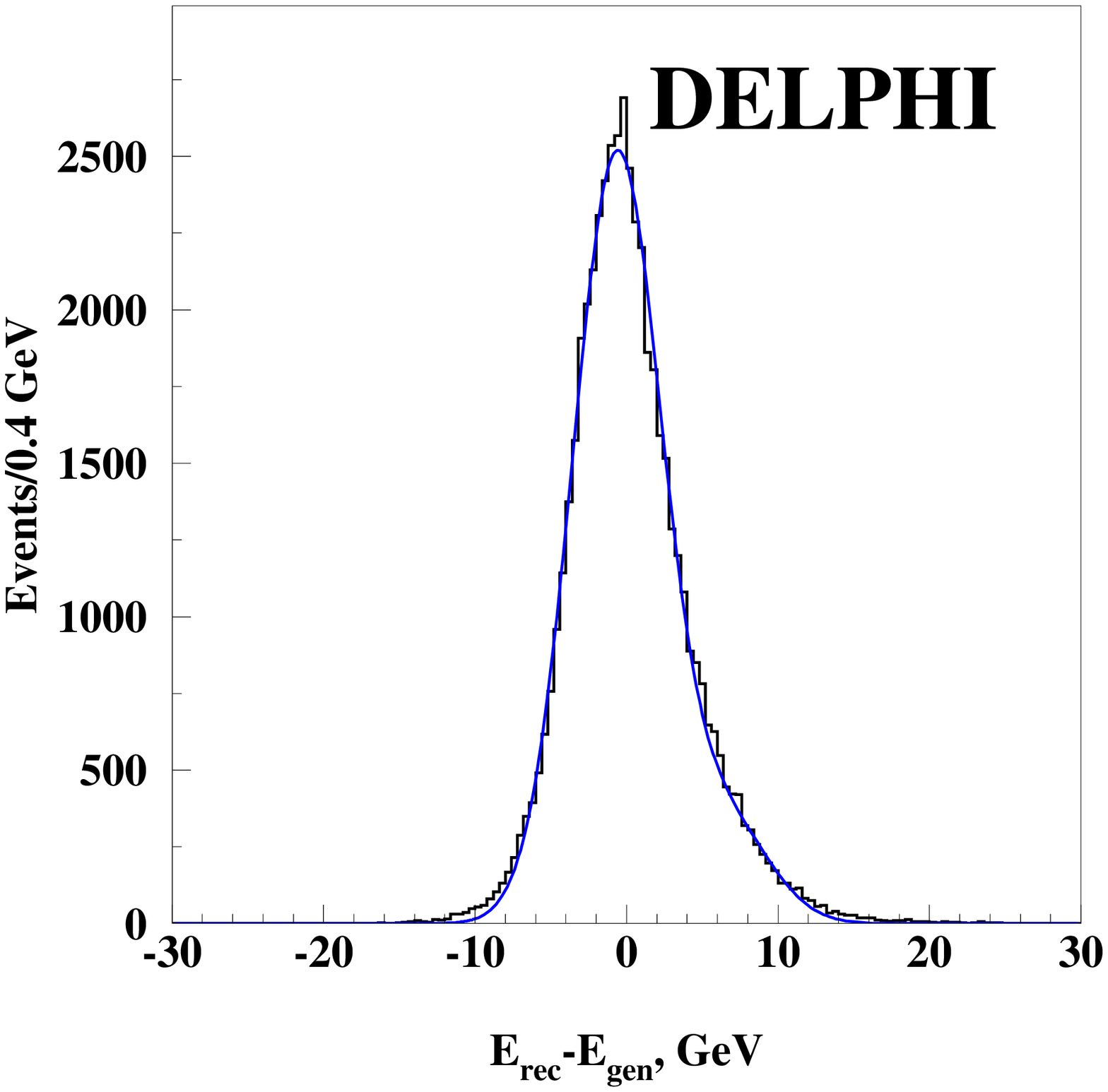,width=6.2cm}
    \hspace{-1cm} \epsfig{file=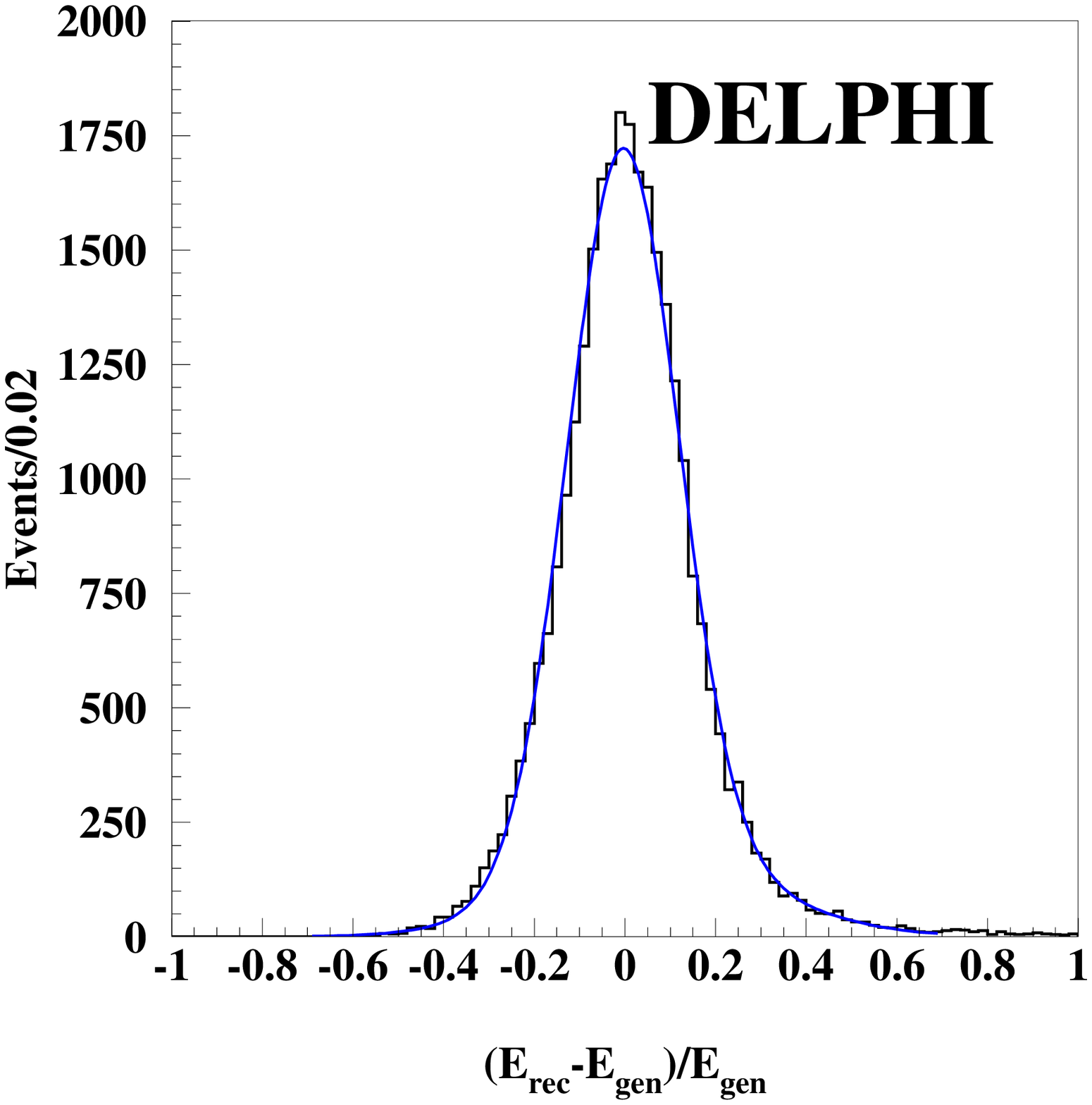,width=6.2cm}
    \hspace{-1cm} \epsfig{file=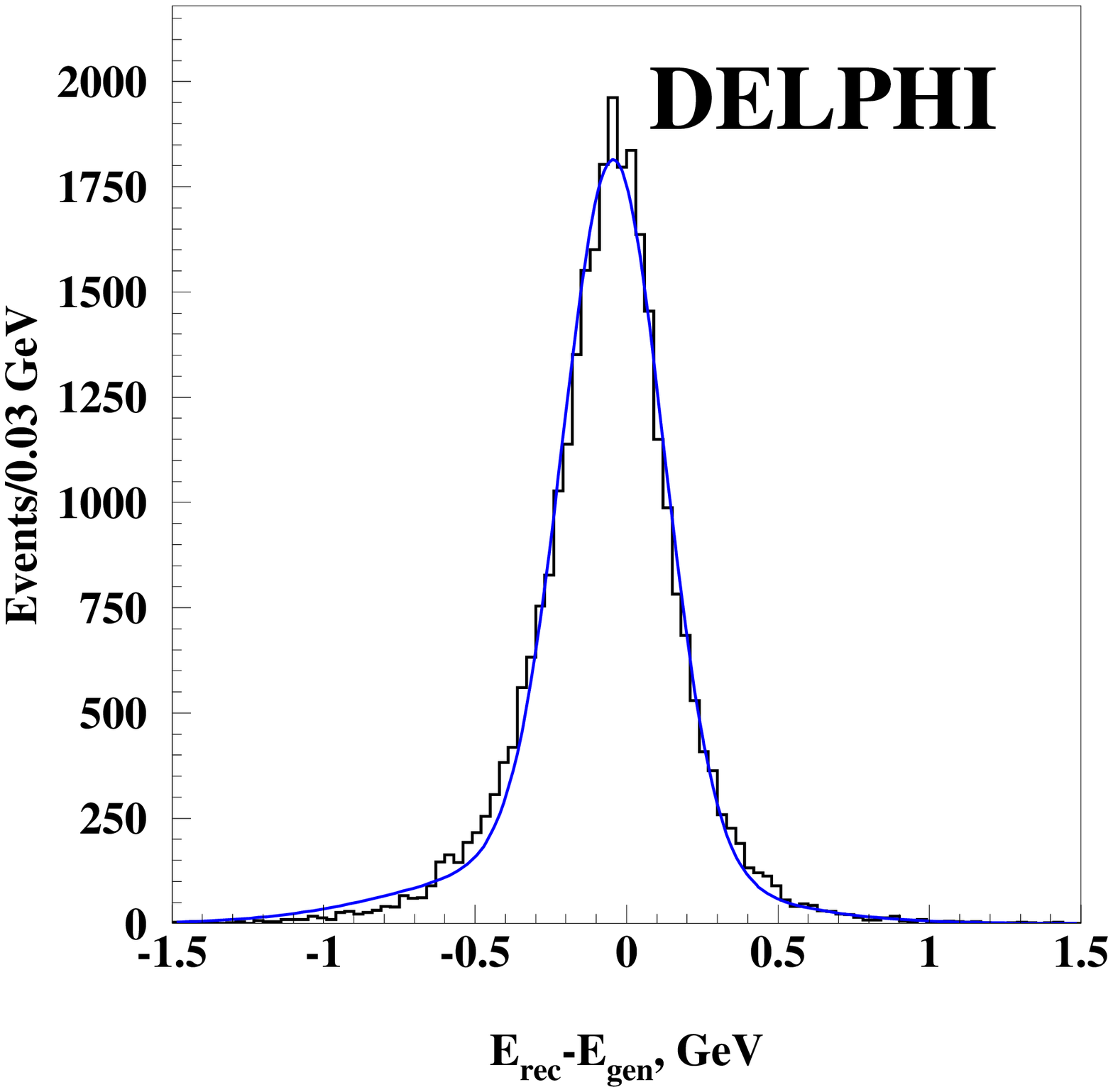,width=6.2cm}}
%\begin{tabular}{c c c}
%\epsfig{file=emisres_aug04.eps,width=5.0cm,height=6.5cm} &
%%%\epsfig{file=beneres_oct03b.eps,width=5.0cm,height=6.5cm} & 
%\epsfig{file=beneres_aug04.eps,width=5.0cm,height=6.5cm} & 
%%\epsfig{file=erecres2_oct03b.eps,width=5.0cm,height=6.5cm} \\
%\epsfig{file=erecres2_aug04.eps,width=5.0cm,height=6.5cm} \\
%\end{tabular}
\caption[] { \it{The resolution on the missing energy (left), the fractional 
resolution on the reconstructed B hadron energy (centre) and 
resolution on the $E^*_{\ell}$ energy (right) estimated 
with simulation for signal ${\rm B} \rightarrow {\rm X}_c \ell \overline{\nu}_{\ell}$ events.}
\label{fig:res1} }
\end{figure}
\end{center}

The resulting resolution of the B energy was found to be $12\%$ for 80\% of all
inclusive semileptonic B decays and $24\%$ for the remaining decays.
%The mass distribution of the reconstructed B is shown in Figure~\ref{fig:Bmass}.
The resolution on the missing energy and on the reconstructed B hadron 
energy estimated with simulation for signal 
${\rm B} \rightarrow {\rm X}_c \ell \overline{\nu}_{\ell}$ events are shown in 
Figure~\ref{fig:res1}.

The direction of the momentum vector of the reconstructed ${\rm X} \ell \overline{\nu}_{\ell}$ 
system was adjusted by up to $\pm 30$ mrad with respect to the lepton direction 
using a function of the reconstructed B mass. Another estimate for the B  
hadron direction was obtained from the B line of flight 
reconstructed from the position of the vertex formed by the lepton with the 
identified charm charged decay products. 
An estimator, which combines these two independent measurements
%The B hadron direction has been estimated using both the momentum vector of 
%the reconstructed ${\rm X} \ell \overline{\nu}_{\ell}$ system and the B decay flight 
%reconstructed by the positions of the vertex formed by the lepton with the 
%identified charm charged decay products. 
%The direction of the B hadron estimated using the momentum vector
% was adjusted up to $\pm 30$ mrad using a function of the lepton direction 
%and the reconstructed B mass.  
%An estimator, which combines these two independent measurements 
according to their expected resolutions as a function of the reconstructed 
energies and the decay distance respectively, was defined. Resolutions of 
$14$(15)~mrad have been achieved in $\phi$ ($\theta$)
for 60\% of all inclusive semileptonic B decays and $40$~mrad
 for the remaining decays.

The identified lepton was then boosted back to the reconstructed B rest frame
and its energy $E^*_{\ell}$ computed in this frame. This resulted in an 
average  resolution of $170$~MeV on $E^*_{\ell}$ for 82\% of all
inclusive semileptonic B decays
and 510~MeV for the remaining decays.
%Figure~\ref{fig:res1}).

\subsection{Signal separation from background sources}

In the reduction of the $b \rightarrow c \rightarrow \ell$ and other 
backgrounds it
is essential to avoid biases of the lepton energy spectrum.  
The separation was 
therefore performed using two discriminating variables, one based
on the topology of the event and the other on charge correlations between 
the lepton and the other particles in the event, which are not sensitive to 
the lepton energy.
%Neither of these variables 
%nor their combination is sensitive to the lepton energy, as shown in 
%Figure~\ref{fig:spectra}.

The topological variable uses information on the lepton impact parameter
with respect to the reconstructed secondary vertex, the topology of the tracks
other than the lepton in the hemisphere, the number of particles not associated 
with the vertex in the hemisphere, the number of particles in the vertex and the 
$\chi^2$ of the vertex.

\begin{figure}[hbtp!]
\begin{center}
\mbox{
\epsfig{file=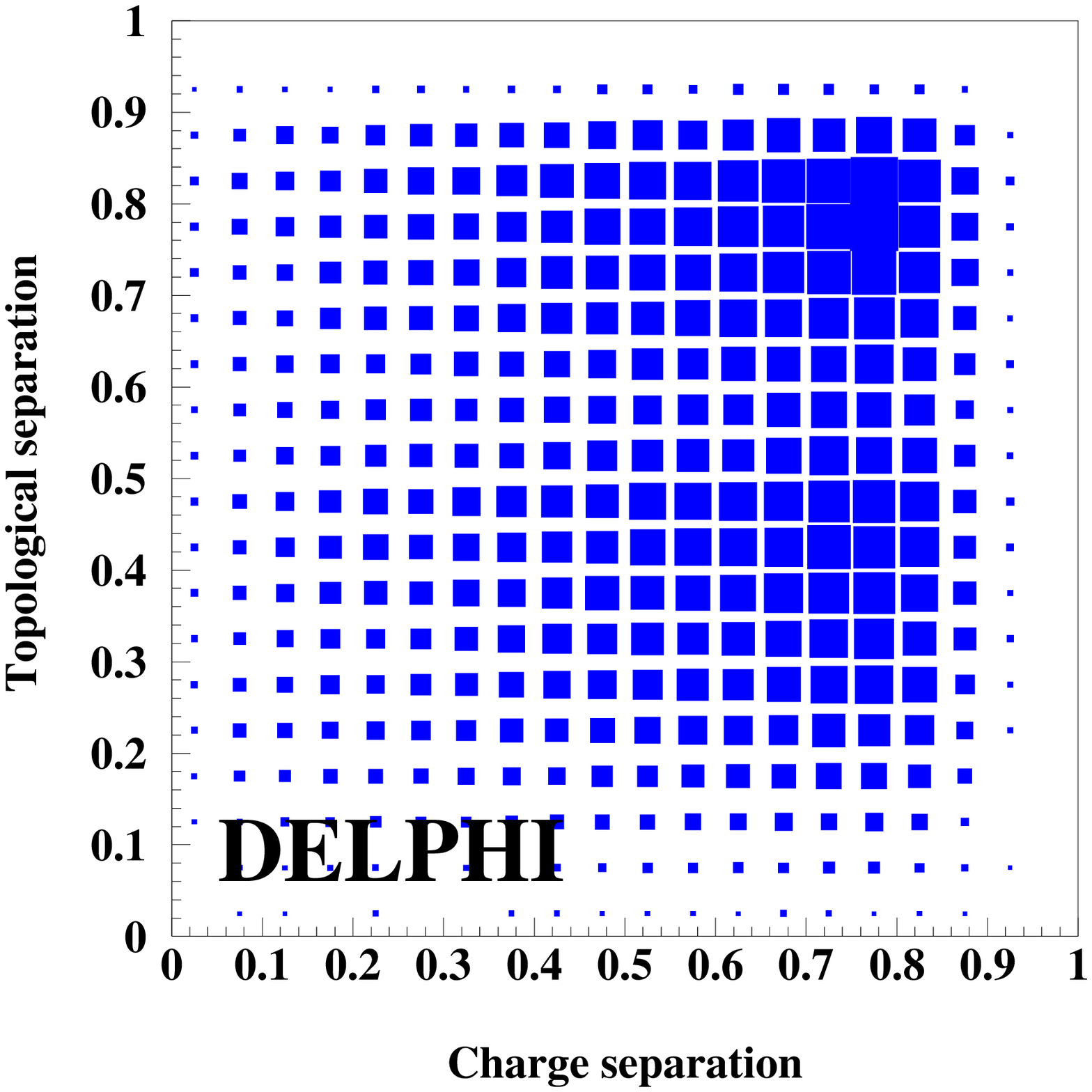,width=8.5cm} 
\epsfig{file=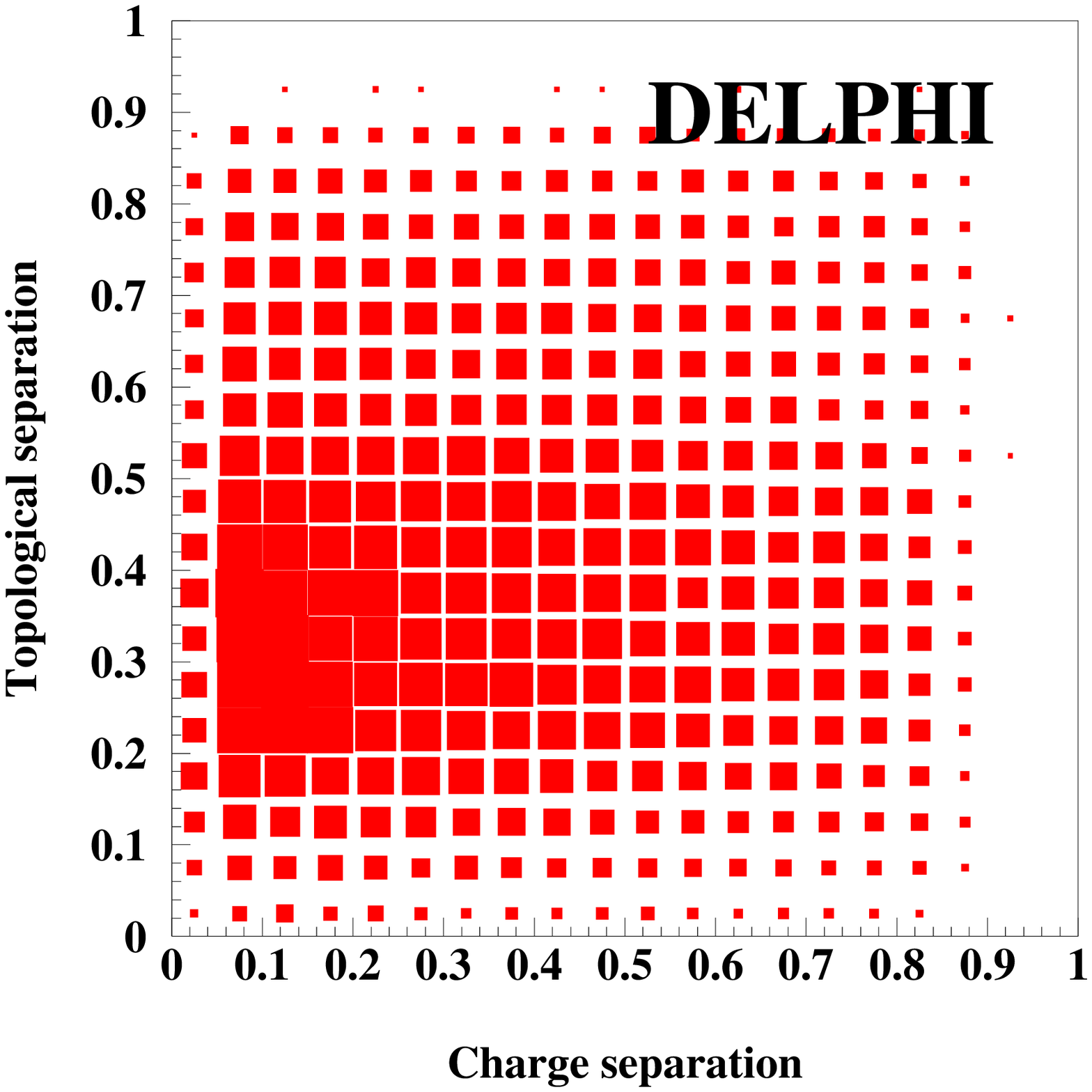,width=8.5cm}} \\
\caption[]{\it{ Two-dimensional
distribution of the topological and charge correlation variables for the signal
${\rm B} \rightarrow {\rm X}_c \ell \overline{\nu}_{\ell}$ (left) and the main background
${\rm B} \rightarrow {\rm X}_c \rightarrow X \ell \nu$ (right).}
\label{fig:vsep} }
\end{center}
\end{figure}

The charge variable consists of a probability built from the correlation of the
charge of the lepton and those of the reconstructed secondary vertex, of other 
vertices in the same and opposite hemispheres, of the jet charge of the 
opposite hemisphere and of the leading kaon candidate.
The kaon candidate was identified based on kaon neural network output
from the MACRIB package \cite{ref:macrib}.
Two-dimensional distributions are shown in Figure~\ref{fig:vsep} for signal
and background, respectively.
The final separation variable ({\sc VSep}) corresponds to the likelihood of an 
event to be a signal event, based on the location in the two-dimensional 
distribution of the topological and charge correlation variables.
%The final separation variable ({\sc VSep}) is then obtained from the 
%two-dimensional distribution of the topological and charge correlation 
%variables. 

\begin{figure}[hbtp!]
\begin{center}
\epsfig{file=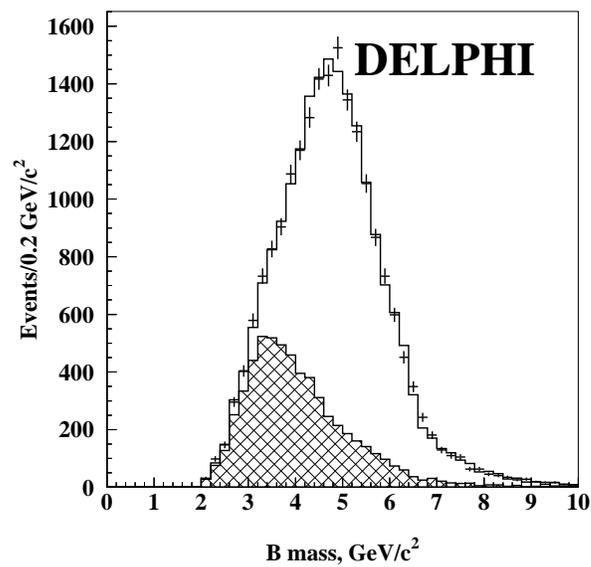,width=8.5cm} 
\caption[]{\it{Reconstructed B mass. Points are the data, histogram is 
the simulation with the signal clear and the background hatched.
} \label{fig:Bmass} }
\end{center}
\end{figure}

A further rejection of background is obtained by requiring a minimum value for the
reconstructed B mass. In Figure~\ref{fig:Bmass} the distribution of the B mass
is shown in data and simulation, for signal and background events.
A B mass larger  than 3.9 GeV/$c^2$ was required.
A final sample of 14364 leptons was selected, with a purity in 
${\rm B} \rightarrow {\rm X}_c \ell \overline{\nu}_{\ell}$ decays of 81\%. 

\begin{figure}[hbtp!]
\begin{center}
\mbox{
\epsfig{file=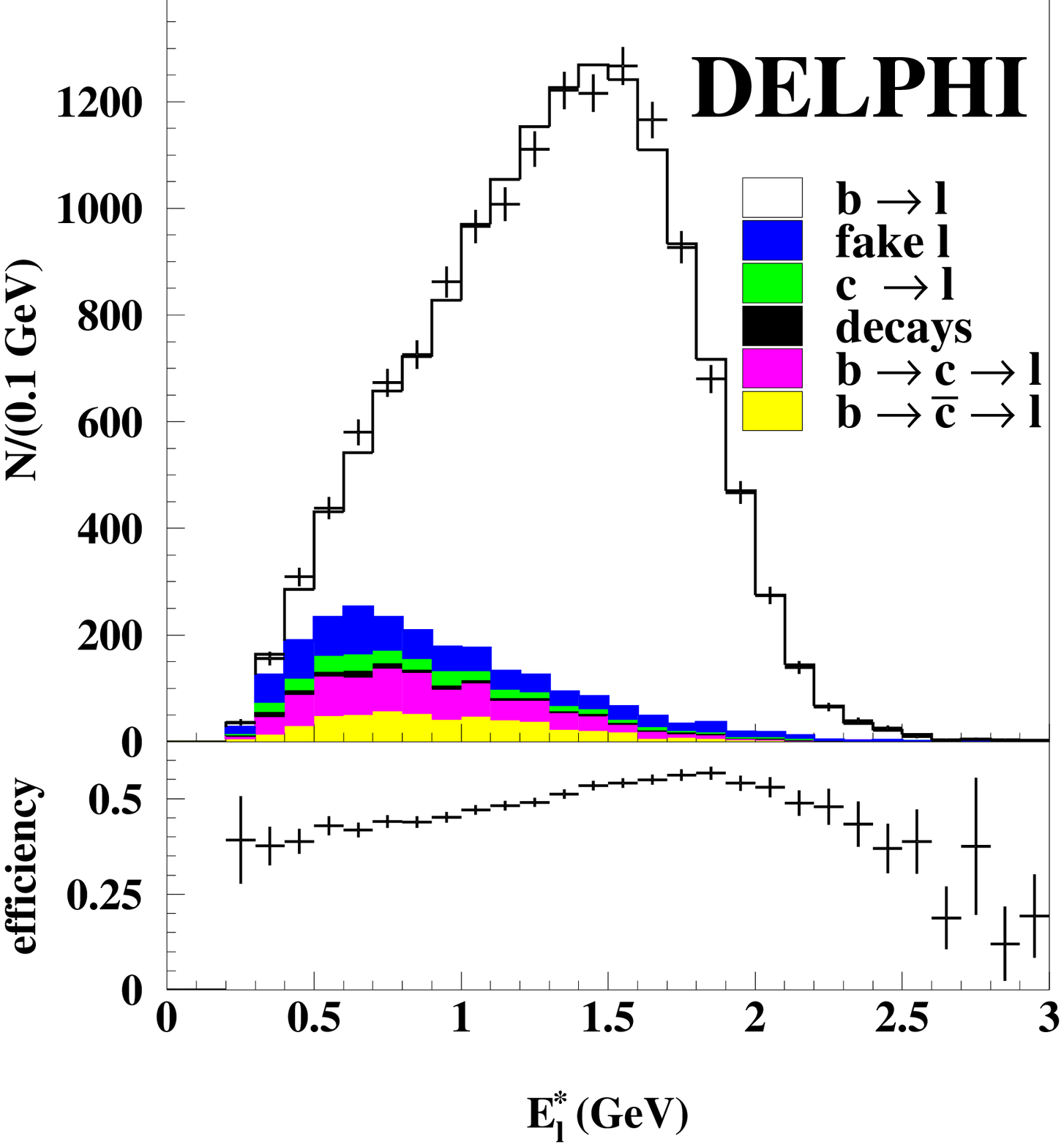,width=8.5cm} 
\epsfig{file=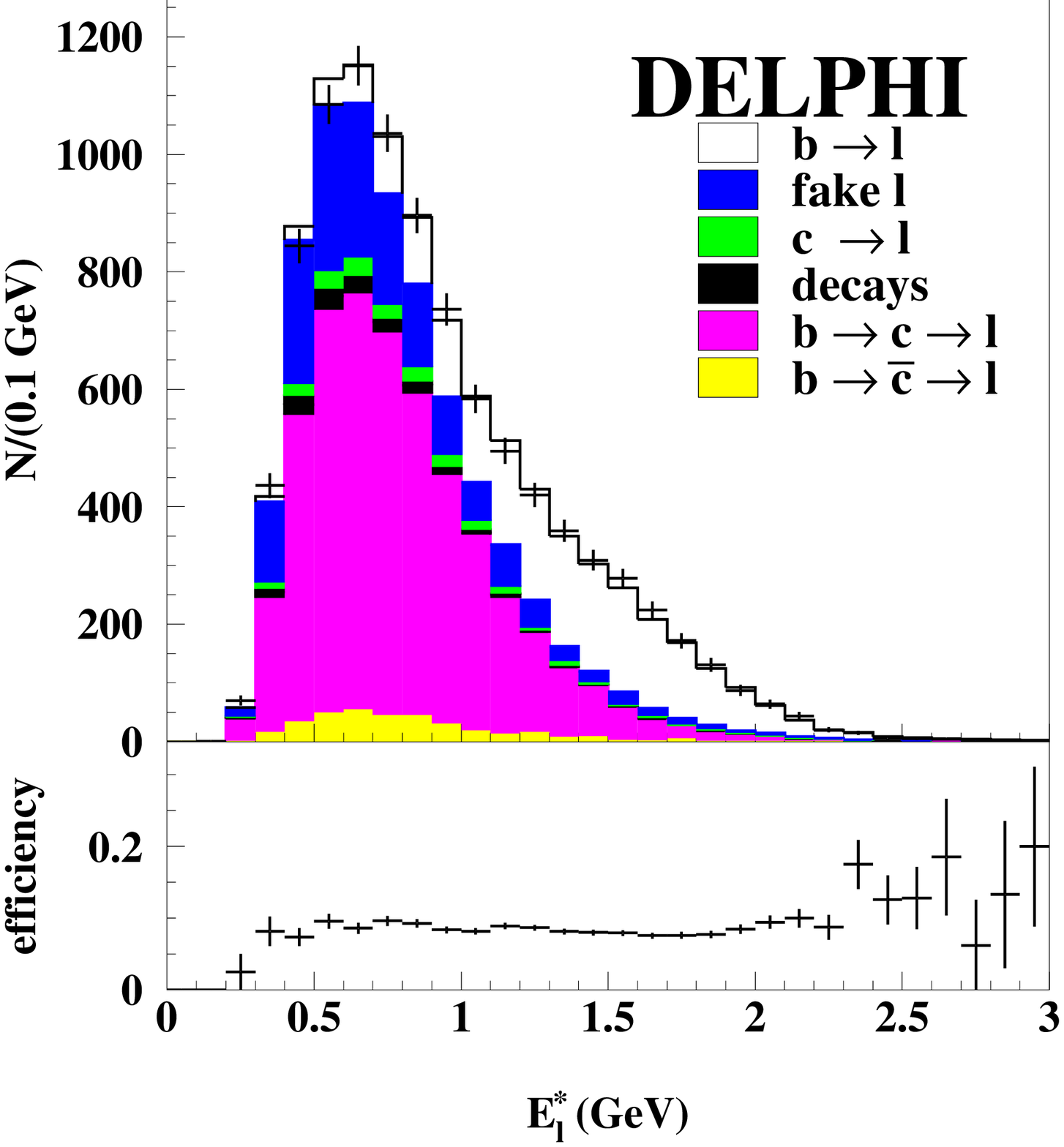,width=8.5cm} }
\caption[]{\it { The resulting $E^*_{\ell}$ spectrum for samples 
enriched (left) and depleted (right) in 
${\rm B} \rightarrow {\rm X}_c \ell \overline{\nu}_{\ell}$ 
decays using the separating variable. 
The data are shown as crosses while the simulation is shown 
as histograms. 
The different sources of background, shown shaded, are from bottom 
to top:
$b \rightarrow \bar{c} \rightarrow \ell$, 
$b \rightarrow c \rightarrow \ell$, 
other lepton sources including decays in flight and converted photons, 
$c \to \ell$, 
misidentified hadrons. 
In the lower plots the efficiencies for the selection are shown, 
as a function of  $E^*_{\ell}$.
} \label{fig:spectra} }
%%
%\caption[]{\it { The resulting $E^*_{\ell}$ spectrum for samples enriched (left)
% and depleted (right) in ${\rm B} \rightarrow {\rm X}_c \ell \overline{\nu}_{\ell}$ decays using the 
%separating variable. The different sources highlighted are 
%${\rm B} \rightarrow c \rightarrow \ell$, 
%$b \rightarrow \bar{c} \rightarrow \ell$, $c \to \ell$, other lepton sources, 
%misidentified hadrons, decays in flight and converted photons from bottom 
%to top. In the lower plots the efficiencies for the selection are shown, 
%as a function of  $E^*_{\ell}$.
%} \label{fig:spectra} }
\end{center}
\end{figure}

Figure~\ref{fig:spectra} shows the lepton spectrum, after {\sc VSep} and 
B mass cuts, and the
corresponding efficiencies as a function of  $E^*_{\ell}$.
A lepton sample, depleted in signal by using an anti-cut on {\sc VSep}
is also shown, as a check of the shape of the simulated backgrounds 
with the data.

\subsection{Study of the lepton energy distribution}

The original lepton spectrum has been extracted from the reconstructed 
distribution by 
a spectrum re-weighting technique. This consisted of determining the 
resolution matrix 
relating the generated to the reconstructed spectrum for simulated signal 
events. Using 
this matrix, the coefficients of a re-weighting function for the generated 
spectrum have 
been fitted to minimize the $\chi^2$ between the resulting spectrum and 
that observed in the data. 
The efficiency correction has been taken into account at this stage.
%A study of the distortions induced on the parton-level lepton spectrum by 
%variations of the $b$-quark mass $m_b$ and kinetic term $\mu_{\pi}^2$ has 
%shown that
%a ratio of polynomials represents a suitable re-weight function. 
The procedure has been carefully tested on lepton spectra generated for 
different values
of the $m_b$ and $\mu_{\pi}^2$ parameters and smeared according to the 
resolution matrix.

In order to increase the statistics in the signal description,
%a sample of 3.1 million simulated $b \bar b$ events has also been used in the 
the sample of simulated $b \bar b$ events (see Table \ref{tab:stat}) has 
also been used in the 
construction of the resolution matrix.

A regularized unfolding method~\cite{blobel} has also been applied as a 
cross-check, but 
the re-weighting method has been preferred for its simplicity.

\subsection{Results on lepton spectra}
\begin{figure}[hbtp!]
\begin{center}
\mbox{
\epsfig{file=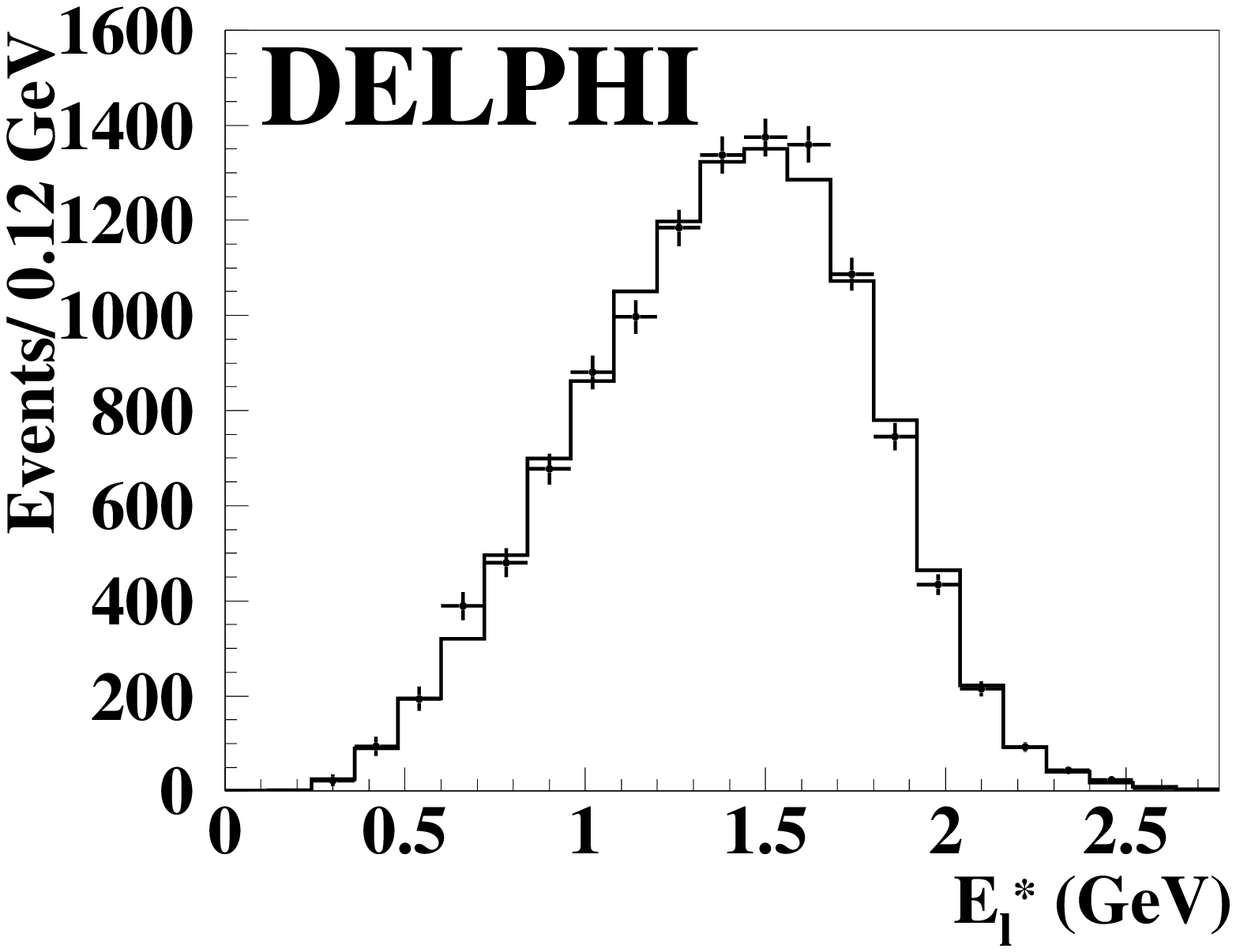,width=8.5 cm} 
\epsfig{file=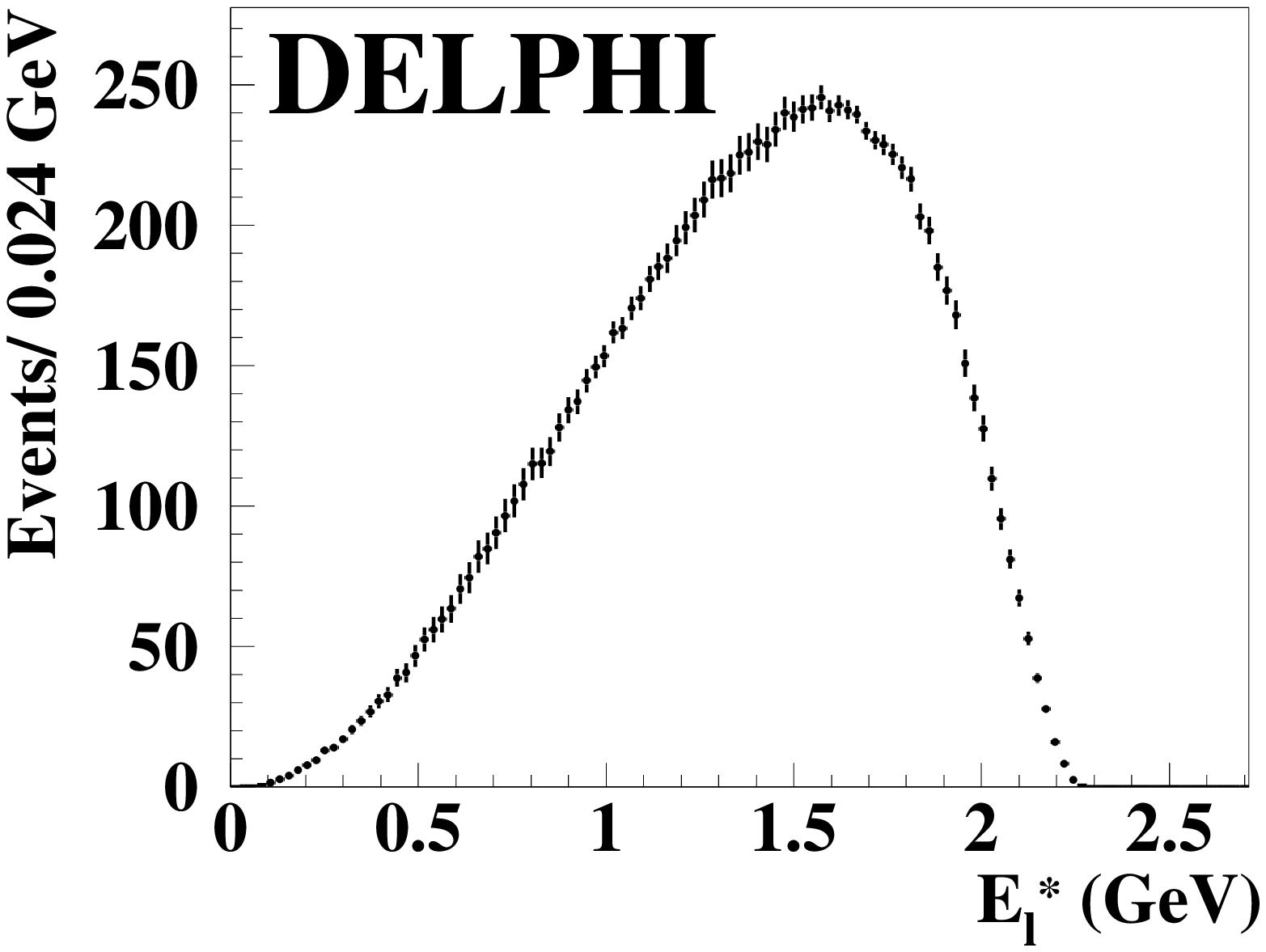,width=8.5 cm}} \\
\caption[]{\it {The background subtracted $E^*_{\ell}$ spectrum (left) for the 
selected signal sample and the unfolded lepton energy spectrum (right).
} \label{fig:fit}}
\end{center}
\end{figure}

The resulting lepton spectrum is shown in Figure~\ref{fig:fit}. 

The first, second and third moments, $M_1^{\ell}$, $M_2^{\prime\ell}$ and 
$M_3^{\prime\ell}$, have been computed. 
In order to reduce the systematic uncertainties, 
the second and third moments have been computed with respect to the average value. 

%%% final values (June 4)
%\begin{table}
%\begin{center}
%\begin{tabular}{|c|l|}
%\hline
%$M_1^{\ell}=<E_{\ell}^*>$ &~1.3782 $\pm$ 0.0073~GeV   \\
%$M_2^{\prime\ell}=<(E_{\ell}^*-<E_{\ell}^*>)^2> $&~0.1838 $\pm$ 0.0058~GeV$^2$\\ 
%$M_3^{\prime\ell}=<(E_{\ell}^*-<E_{\ell}^*>)^3> $ &-0.0301 $\pm$ 0.0015~GeV$^3$\\ 
%\hline
%\end{tabular}
%\caption[]{\it {Corrected lepton moments.}
%\label{table:results} }
%\end{center}
%\end{table}

%%%%%%%
\begin{table}[htb]
{\small 
\begin{center}
  \begin{tabular}{|c|c|c|c|}
    \hline
 & $M_1^{\ell}$ & $M_2^{\prime\ell}$ & $M_3^{\prime\ell}$\\
 & $(\GeV)$ & $(\GeV)^2$ & $(\GeV)^3$ \\
    \hline
value & $1.3782$  & $0.1838 $ & $-0.0301 $  \\
\hline
stat. uncert. & $\pm 0.0073$  & $\pm 0.0058 $ & $\pm 0.0015$ \\
\hline
B species                     & $\pm 0.0027$& $\mp 0.0017$ & $\mp 0.0005$\\ 
${\rm B}\rightarrow {\rm D},~\Dstar,~\Dstarstar \ell \overline{\nu}_{\ell}$ 
                              & $\pm 0.0010$& $\mp 0.0005$ & $\pm 0.0001$\\
B fragmentation               & $\pm 0.0027$& $\mp 0.0020$ & $\mp 0.0007$ \\
${\rm B}\rightarrow {\rm X}_u \ell \overline{\nu}_{\ell}$
                              & $\pm 0.0008$& $\pm 0.0003$ & $\mp 0.0001$\\
e.m. radiation                & $\pm 0.0035$& $\mp 0.0001$ & $\mp 0.0004$\\
\hline
Bkg modelling                  & $\pm 0.0026$ & $\mp 0.0011$  & $\mp 0.0005$\\
B direction reconstruction    & $\pm 0.0027$ & $\mp 0.0018$  & $\mp 0.0006$\\
B energy reconstruction       & $\pm 0.0027$ & $\pm 0.0011$  & $\mp 0.0003$\\
B mass cut                    & $\pm 0.0051$ & $\mp 0.0031$  & $\mp 0.0017$\\
Unfolding                     & $\pm 0.0031$ & $\mp 0.0028$  & $\pm 0.0029$\\
\hline
Tot. syst.                    & $\pm 0.0092$ & $\pm 0.0055$  & $\pm 0.0036$\\
\hline
  \end{tabular}
  \caption[]{\it {Corrected lepton moments and sources of systematic 
uncertainties.}
  \label{table:results}}
\end{center}
}
\end{table}
%%%%%%%

The statistical correlation matrix for these three moments is 
given in Appendix \ref{append:leptmom}.

In order to relate the measured moments to those computed for a B meson, some
corrections need to be applied. 

Firstly the effect of radiation needs to be corrected 
for. This was done using correction factors computed separately for electrons 
and muons following ref.~\cite{Atwood:1989em} and gives a shift of +7.0 MeV,
-0.2 $\times 10^{-3}$ GeV$^2$ and -0.7$\times 10^{-3}$ GeV$^3$ 
on the first, second and third moments, respectively. 
Half of these shifts have been used as an estimation of the systematic 
uncertainty on this correction.
Since $e^+e^- \rightarrow b \bar b$ events at LEP result in the production 
of an admixture of $b$-hadron species, a correction factor accounting 
for the bias due to the semileptonic decays of the heavier $b$-hadrons was applied, 
using the  fraction of $\Bsb$ and of $b$-baryon left in the selected sample 
according to the simulation prediction. The uncertainty in the prediction is
considered in the systematic uncertainty evaluation.
Finally, the presence of $b \rightarrow {\rm X}_u \ell \overline{\nu}_{\ell}$ decays
results in a similar bias of the lepton spectrum toward higher energies, 
due to the larger phase space of this decay compared to 
$b \rightarrow {\rm X}_c \ell \overline{\nu}_{\ell}$. This was also 
corrected for. 
%The sum of these corrections accounts for shifts of 1~MeV - 5~MeV 
%for the three measured moments.
Results, after corrections, are given in Table~\ref{table:results}. 

Several sources of systematic uncertainties have been investigated and 
the results are summarized in Table~\ref{table:results}.
The sources related to the modelling used in the simulation are:
the fractions of the different D species in the decays, the different 
B species and the $b$ fragmentation function. 
For the central value of the moments the same branching fractions for
${\rm B}\rightarrow {\rm D},~\Dstar,~\Dstarstar$ as in Section 4, the {\it b}-hadron fractions from 
reference \cite{ref:PDG02} and the results of reference \cite{Karlsrue}
for the $b$-hadron fragmentation distribution 
have been used, respectively.  
The variations quoted therein have been used for evaluating the
systematic uncertainties reported in Table~\ref{table:results}. 

The uncertainty related to the background modelling has been evaluated
by changing the simulation prediction for the cascade decays 
within the uncertainties of the branching ratios given in  
\cite{ref:PDG02} and changing the misidentification efficiency according to
 Sections \ref{sec:muid} and \ref{sec:elecid}.
It has also been checked by comparing the results obtained by using 
the background  shape as extracted from the anti-tagged data with 
that using the prediction from the simulation.

The uncertainty due to the B reconstruction accounts for variations 
in the B reconstructed energy and direction.

%% Marta d2
The uncertainty related to the unfolding procedure was evaluated
by  varying the reweighting function used in the fit and the binning.
%% end

The stability of the results with respect to changes in the selection 
cut applied on the {\sc VSep} variable and reconstructed B mass have
 been checked.
Changes of the {\sc VSep} cut inducing variations of the accepted statistics 
up to a factor of 1.5 and of the signal purity over a range from 76\% to 89\% 
have been considered and found to give results stable within their 
statistical uncertainty. 
%(see Figure~\ref{fig:stability} ).
The minimum value of the B mass required has been moved between 2 and 5 GeV,
with a corresponding change of purity between 71\% and 86\%.
The maximum variation with respect to the central value
has been used as an estimation of the systematic uncertainty in the
agreement between data and simulation on the B mass reconstruction. 

Results obtained separately on the electron sample and the muon sample have 
been compared.
The difference in the first moments amounts to 20$\pm$13$\pm$3 MeV, where the 
first uncertainty is statistical and the second is the 
uncorrelated systematic uncertainty on the background subtraction.
The difference expected from simulation is 7 MeV. 
The differences on the second and third moments are fully compatible with 
the statistical uncertainty.

%\begin{table}
%\begin{center}
%{\small
%\begin{tabular}{|l|c|c|c|}
%\hline
%Source &  $\delta M_1^{\ell}$ & $\delta M_2^{\prime \ell}$ & 
%$\delta M_3^{\prime \ell}$ \\
%                              & (GeV)      & (GeV$^2$)   & (GeV$^3$)  \\
%\hline \hline
%B species                     & $\pm$0.0027& $\mp$0.0017 & $\mp$0.0005\\ 
%${\rm B}\rightarrow {\rm D},~\Dstar,~\Dstarstar \ell \overline{\nu}_{\ell}$ 
%                              & $\pm$0.0010& $\mp$0.0005 & $\pm$0.0001\\
%B fragmentation               & $\pm$0.0027& $\mp$0.0020 & $\mp$0.0007 \\
%${\rm B}\rightarrow {\rm X}_u \ell \overline{\nu}_{\ell}$
%                              & $\pm$0.0008& $\pm$0.0003 & $\mp$0.0001\\
%e.m. radiation                & $\pm$0.0035& $\mp$0.0001 & $\mp$0.0004\\
%\hline
%Bkg modelling                  & $\pm$0.003 & $\mp$0.001  & $\mp$0.001\\
%B direction reconstruction    & $\pm$0.003 & $\mp$0.002  & $\mp$0.001\\
%B energy reconstruction       & $\pm$0.003 & $\pm$0.001  & $\mp$0.001\\
%B mass cut                    & $\pm$0.005 & $\mp$0.003  & $\mp$0.002\\
%Unfolding                     & $\pm$0.003 & $\mp$0.003  & $\pm$0.003\\
%\hline\hline
%Total                         & $\pm$0.0093 & $\pm$0.0065  & $\pm$0.0039\\
%\hline
%\end{tabular}
%}
%\caption[]{\it { Sources of systematic uncertainties}
%\label{tab:syst}}
%\end{center}
%\end{table}

\section{Interpretation of the results}
\label{sec:fit}

%
%\begin{table}
%\begin{center}
%\begin{tabular}{|c|l|}
%\hline
%$<E_{\ell}^*>$ &(~1.383$\pm$0.012$\pm$0.009)~GeV   \\
%$<(E_{\ell}^*-<E_{\ell}^*>)^2>$ &(~0.192$\pm$0.005$\pm$0.008)~GeV$^2$\\ 
%$<(E_{\ell}^*-<E_{\ell}^*>)^3>$ &(-0.029$\pm$0.005$\pm$0.006)~GeV$^3$\\ \hline
%\end{tabular}
%\caption[]{\it {Moments of the lepton energy spectrum, 
%{\sc Delphi} Preliminary Results.}
%\label{table:results}}
%\end{center}
%\end{table}

A $\chi^2$ fit to the three leptonic 
($M_1^{\ell},~M_2^{\prime \ell}~{\rm and} ~M_3^{\prime \ell}$, see Table \ref{table:results})
and three hadronic mass moments ($M_1^H,~M_2^{\prime H}~{\rm and}~M_3^{\prime H}$, 
see Tables \ref{tab:hadmom1} and \ref{tab:hadmom2}) 
has been performed, using two theoretical frameworks. 
In the fit we also impose additional constraints derived from 
independent determinations. 
We follow the framework presented in \cite{ref:amsterl},
updated with the new results discussed above, and using recent calculations 
given in \cite{ref:paolo_uraltsev}. Parallel fits have also been performed 
by other groups using several frameworks \cite{ref:fits_referee}.

%The present analysis has been described
%with more details in \cite{ref:physlmom}.
 
In the kinetic mass scheme, we fit the full set of parameters: $m_b(1~{\rm{GeV}})$, 
$m_c(1~{\rm{GeV}})$, $\mu_{\pi}^2$ and $\tilde{\rho}_D^3$.
%and $\rho_{LS}^3$. 
Expressions relating moments and these parameters can be 
found in \cite{ref:amsterl}. 
%%%%%
%% PR 2707
%%
We impose  $\mu_G^2$=0.35$\pm$0.07~GeV$^2$
%\cite{Uraltsev:2001ih} 
and 
$\rho_{LS}^3$=-0.15$\pm$0.10~GeV$^3$ \cite{ref:paolo_uraltsev}.
%\cite{ref:bigiold}.
Two mass constraints have also been applied:
%$m_b(1~{\rm{GeV}})$=4.57$\pm$0.10~$\GeVcd$~\cite{mb}, 
$m_b(1~{\rm{GeV}})$=4.61$\pm$0.17~$\GeVcd$ and
%and, to be  conservative,  
%$m_c(1~{\rm{GeV}})$=1.05$\pm$0.30~$\GeVcd$.
$m_c(1~{\rm{GeV}})$=1.14$\pm$0.10~$\GeVcd$ as derived 
from the values quoted in \cite{ref:pdg2004}, which were given using a 
different renormalization scheme.
%The more stringent is that on $m_b(1~{\rm{GeV}})$. It must be noted that this 
%constraint is largely equivalent to that derived from the first moment of the photon 
%energy spectrum in $b \to s \gamma$ in other studies~\cite{cleo_mom2}. 
 Results are obtained for $\alpha_s(m_b) = 0.22 \pm 0.04$ \footnote{For hadron moments a value of $\alpha_s(m_b) = 0.3 \pm 0.1$ has been used to account
for missing terms in theoretical expressions.} and are shown in Table \ref{tab:5}.
In order to study the effect of the bounds on $m_{b,c}$ introduced, 
the fit has been repeated without these constraints. 
Results are shown in Table \ref{tab:5wc}. 
Theoretical uncertainties have been evaluated following the procedure explained
in \cite{ref:paolo_uraltsev}; namely 20$\%$ (30$\%$) errors have been
assumed for terms corresponding to $1/m_b^2$ ($1/m_b^3$) corrections
and adding in quadrature variations corresponding to the uncertainty on 
$\alpha_s(m_b)$.
Corresponding theoretical uncertainties attached to $m_b$ and $m_c$
are due to those on $\alpha_s(m_b)$. 

%The results are consistent, although the accuracy on the masses 
%degrades. In particular, without the constraint on $m_b$, 
%we find $m_b(1~{\rm{GeV}})$=4.58$\pm$0.14~GeV
%\footnote{This value corresponds to 
%$\overline{m_b}(\overline{m_b})^{\overline{MS}}=4.22 \pm 0.13~\GeVcd$.}. 
%It is interesting to observe that the mass 
%constraints applied are of the scale of the fit 
%sensitivity. Also, 
The central values of the heavy quark 
masses are in agreement with independent determinations~\cite{mb,melyel}.
The difference between the values of the two heavy quark masses,
which are highly correlated,  
is $m_b(1~{\rm{GeV}})-m_c(1~{\rm{GeV}})$=
3.422~$\pm$~0.034~$\pm$~0.028~$\GeVcd$ \,(3.382~$\pm$~0.051~$\pm$~0.087~$\GeVcd$ if no constraint on the quark masses is imposed).

In the approach based on pole masses \cite{ref:falketal}, 
the fit extracts $\bar{\Lambda}$, $\lambda_1$, $\lambda_2$, $\rho_1$
and $\rho_2$. We fix ${\cal{T}}_i = 0.0~$GeV$^3$ and impose two constraints on 
$M_{B^*}-M_B$ and $M_{D^*}-M_D$
%$=2\lambda_2/M_B (1+\bar{\Lambda}/M_B)-\rho_2/M_B^2+(T_2+T_4)/M_B^2$, 
which effectively reduces by two the number of free parameters. The
results are given in Table~\ref{tab:6}. 

\begin{figure}[hbtp!]
\begin{center}
\begin{tabular}{c c}
 \epsfig{file=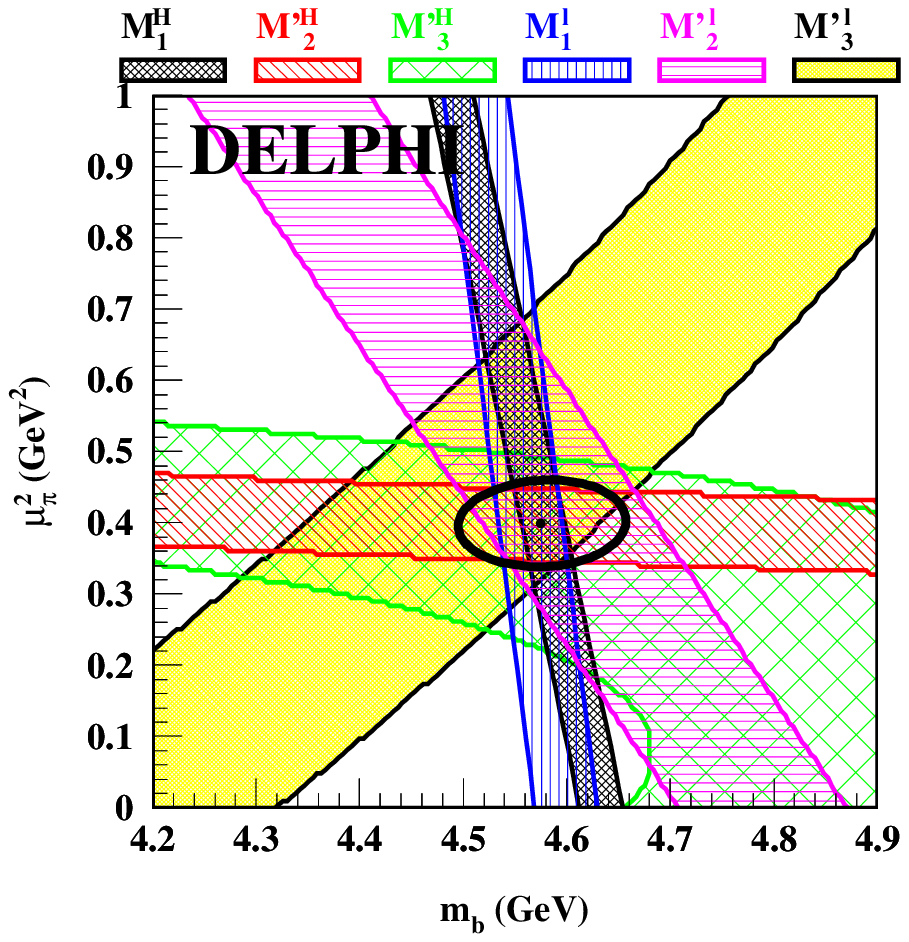,width=8.cm} \hspace{-0.5cm} &
\epsfig{file=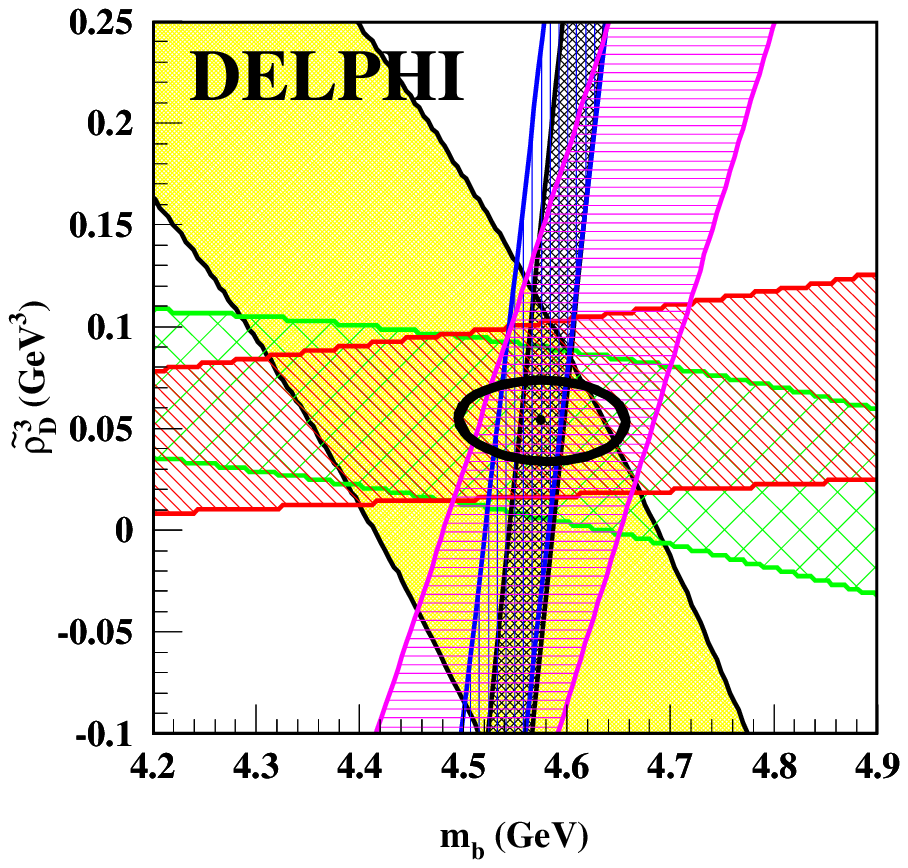,width=8.cm} \\
\end{tabular}
\caption[]{\it {The projection of the constraints of the six measured moments on the 
$m_b(1~{\rm{GeV}})$-$\mu_{\pi}^2(1~{\rm{GeV}})$ (left) and 
$m_b(1~{\rm{GeV}})$-$\tilde{\rho}_D^3$ (right) planes. The bands correspond to the total 
measurement accuracy and are given by keeping all the other 
parameters at their central values. The ellipses represent the 1~$\sigma$ contours and include correlations between the parameters.}
\label{fig:1}}
\end{center}
\end{figure}

\begin{figure}[hbtp!]
\begin{center}
\begin{tabular}{c c}
 \hspace{-0.5cm} \epsfig{file=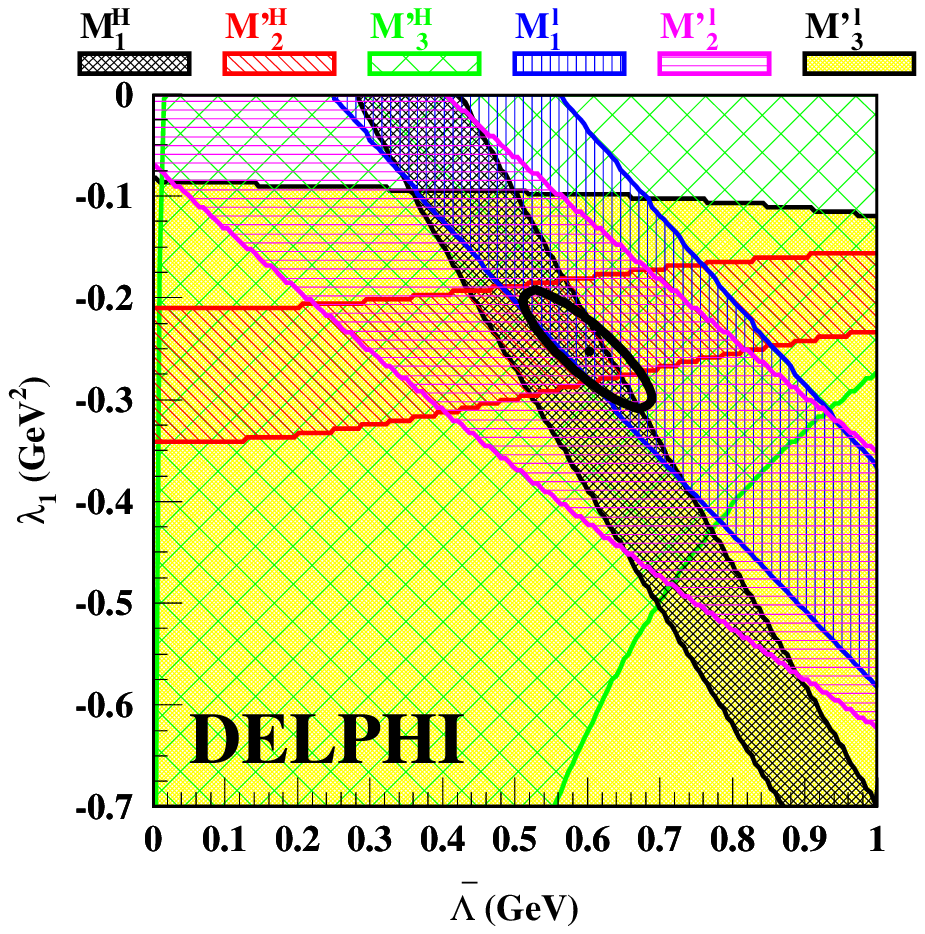,width=8.cm} \hspace{-0.5cm}&
\epsfig{file=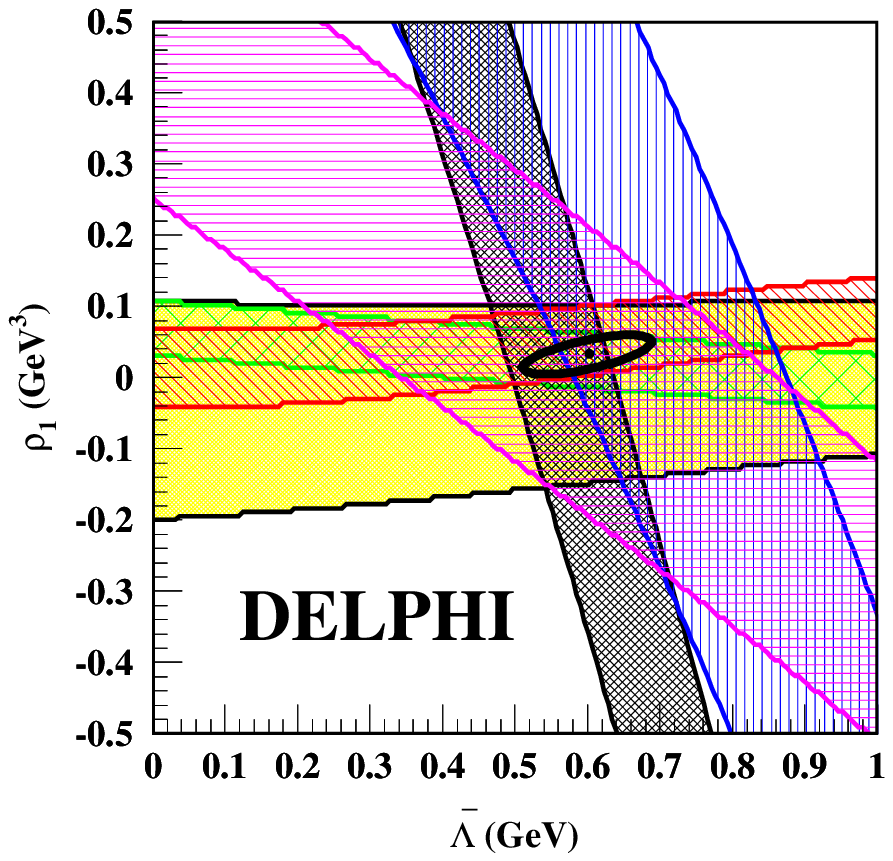,width=8.cm} \\
\end{tabular}
\caption[]{\it {The projection of the constraints of the six measured moments on the 
$\bar{\Lambda}$-$\lambda_1$ (left) and $\bar{\Lambda}$-$\rho_1$ (right) planes. The 
bands correspond to the total measurement accuracy and are given by keeping all the 
other parameters at their central values. The ellipses represent the 1~$\sigma$  contours and include correlations between the parameters.}
\label{fig:2}}
\end{center}
\end{figure}

%% PR d2
Up to  first order corrections in $\alpha_s$, parameters corresponding
to non-perturbative QCD corrections, entering in the two 
approaches, are related:
\begin{equation}
\mu_{\pi}^2 = -\lambda_1 -\frac{{\cal T}_1+3{\cal T}_2}{m_b};~
\mu_G^2 = 3\lambda_2 +\frac{{\cal T}_3+3{\cal T}_4}{m_b};~
\tilde{\rho}_D^3 = \rho_1;
~\rho_{LS}^3 = 3 \rho_2.
\end{equation}
The parameter $\overline{\Lambda}$ enters in the expression relating
heavy quark and heavy meson masses, which, for pseudo-scalar
mesons reads:
\begin{equation}
M_B=m_b + \overline{\Lambda} +\frac{\mu_{\pi}^2-\mu_G^2}{2 m_b}
+ \frac{\tilde{\rho}_D^3+\rho_{LS}^3-\rho_{NL}^3}{4 m_b^2}
+ {\cal O}\left ( \frac{1}{m_b^3}\right ),
\end{equation}
where $\rho_{NL}^3$ corresponds to a linear combination of ${\cal T}_{1-4}$.
%% end PR d2
Projections of the constraints from the six moments in the $m_b$-$\mu_{\pi}^2$ and 
$m_b$-$\tilde{\rho}_D^3$ planes are shown in Figure \ref{fig:1} and those in the 
$\bar{\Lambda}$-$\lambda_1$ and $\bar{\Lambda}$-$\rho_1$ planes in 
Figure \ref{fig:2}. 
The  $\chi^2/{\rm{n.d.f.}}$ of the fits is  0.4 and 0.2 in the two 
formulations. Since the contributions proportional to $\rho_{LS}^3$ in 
the moment expressions are numerically suppressed, the fit is only marginally sensitive to its size and the result is determined by the constraint applied. By removing this, 
the fit would give $\rho_{LS}^3 =-0.4 \pm 0.4 $~GeV$^3$.
In contrast, the value of the leading $1/m_b^3$ correction
(parametrised by $\tilde{\rho}^3_D$) can 
be determined with satisfactory accuracy and its range agrees with
theoretical expectations~\cite{Uraltsev:2001ih}. These Figures illustrate the 
importance of the second hadronic moment to 
determine $\mu_{\pi}^2(1~{\rm{GeV}})$ and of the second and third hadronic moments to extract $\tilde{\rho}_D^3$.

\begin{table}
\begin{center}
\begin{tabular}{|l|c c c c c|}
\hline
Fit & Fit & Fit & Syst. & Syst. & \\
Parameter & Values & Uncertainty & moments & theory & \\ \hline
$m_b(1~{\rm{GeV}})$& 4.591 & $\pm$ 0.062 & $\pm$ 0.039 &$\pm$ 0.005 &$\GeVcd$\\
$m_c(1~{\rm{GeV}})$& 1.170 & $\pm$ 0.093 & $\pm$ 0.055 &$\pm$ 0.005 &$\GeVcd$\\
$\mu_{\pi}^2(1~{\rm{GeV}})$ 
                   & 0.399 & $\pm$ 0.048 & $\pm$ 0.034 &$\pm$ 0.087& GeV$^2$\\
$\tilde{\rho}_D^3$ & 0.053 & $\pm$ 0.017 & $\pm$ 0.011 &$\pm$ 0.026 & GeV$^3$ \\ \hline
%$\rho_{LS}^3$     &-0.17 & $\pm$ 0.05 & $\pm$ 0.13 &$\pm$ 0.005 & GeV$^3$ \\  \hline
\end{tabular}
\caption{\it {Results of the fit in the $m_b(\mu)$, $m_c(\mu)$ and 
$\mu_{\pi}^2(\mu)$ formalism.}
\label{tab:5}}
\end{center}

\end{table}

\begin{table}
\begin{center}
\begin{tabular}{|l|c c c c c|}
\hline
Fit & Fit & Fit & Syst. & Syst. & \\
Parameter & Values & Uncertainty & moments & theory & \\ \hline
$m_b(1~{\rm{GeV}})$& 4.67 & $\pm$ 0.11 & $\pm$ 0.19 &$\pm$ 0.03 &$\GeVcd$\\
$m_c(1~{\rm{GeV}})$& 1.29 & $\pm$ 0.17 & $\pm$ 0.27 &$\pm$ 0.04 &$\GeVcd$\\
$\mu_{\pi}^2(1~{\rm{GeV}})$ 
                   & 0.41 & $\pm$ 0.05 & $\pm$ 0.04 &$\pm$ 0.08& GeV$^2$\\
$\tilde{\rho}_D^3$ & 0.05 & $\pm$ 0.02 & $\pm$ 0.01 &$\pm$ 0.03 & GeV$^3$ \\ \hline
%$\rho_{LS}^3$     &-0.19 & $\pm$ 0.11 & $\pm$ 0.089 &$\pm$ 0.005 & GeV$^3$ \\  \hline
\end{tabular}
\caption{\it {Results of the fit in the $m_b(\mu)$, $m_c(\mu)$ and 
$\mu_{\pi}^2(\mu)$ formalism, without contraints on $m_b(1~{\rm{GeV}})$ and $m_c(1~{\rm{GeV}})$. Values of the fitted masses correspond to 
$\overline{m_b}(\overline{m_b})^{\overline{MS}}= 4.31~\pm~0.20~\GeVcd $ and
$\overline{m_c}(\overline{m_c})^{\overline{MS}}= 1.37~\pm~0.24~\GeVcd $. 
}
\label{tab:5wc}}
\end{center}
\end{table}

\begin{table}
\begin{center}
\begin{tabular}{|l|c c c c c|}
\hline
Fit & Fit & Fit & Syst. & Syst. & \\
Parameter & Values & Uncertainty & moments & theory &\\ \hline
$ \bar\Lambda$&~0.601 & $\pm$ 0.065 & $\pm$ 0.061 & $\pm$ 0.05& GeV~\\
$\lambda_1$   & -0.252 & $\pm$ 0.054 & $\pm$ 0.018 & $\pm$ 0.07& GeV$^2$\\
$\lambda_2$   & ~0.117 & $\pm$ 0.003 & $\pm$ 0.002 & $\pm$ 0.00& GeV$^2$\\
$\rho_1$      & ~0.032 & $\pm$ 0.021 & $\pm$ 0.010 & $\pm$ 0.04& GeV$^3$ \\ \hline
$\rho_2$      & ~0.085 & $\pm$ 0.011 & $\pm$ 0.017 & $\pm$ 0.21& GeV$^3$ \\  \hline
\end{tabular}
\caption{\it {Results of the fit in the $\bar{\Lambda}$-$\lambda_1$ formalism.}
\label{tab:6}}
\end{center}

\end{table}

In Tables \ref{tab:5} and \ref{tab:6}, the first column of systematic uncertainties
corresponds to systematics on moments; correlated errors between 
the different moments have been included in the fit. 
The second column is due to systematics from theory.
% which corresponds to
%systematic uncertainties due to ranges of residual parameters which have been fixed 
%and missing terms in the expansions which have been estimated. 
For the kinetic mass formalism we propagate the uncertainty on $\alpha_s$ 
and follow the suggestions of \cite{ref:paolo_uraltsev} to account for missing 
corrections.
%evaluate the effect of removing the BLM corrections \cite{ref:blm}
%from the lepton moments. 

%In this scheme this is a small effect and higher order
%perturbative corrections are expected to be under control. 
%Dimensional estimates suggest that 
%$1/m_b^4$ effects do not exceed the present experimental
%resolution.  

For the 
$\bar{\Lambda}$-$\lambda_1$ formalism we include in the theory 
systematics the effect of 
${\cal{T}}_i = (0.0 \pm 0.50)^3$,
%=\pm 0.125$~GeV$^3$, 
$\alpha_s$ = 0.22$\pm$0.04  
and we also estimate the effect of the missing corrections to third moments as
%$M_B^6(0.001\pm0.0005)\,\beta_0\,(\alpha_s/\pi)^2$
%and $M_B^6(0.003\pm0.003)\left (\bar\Lambda/\overline M_B\right )
%\left ( \alpha_s/\pi \right )$.
explained in \cite{ref:amsterl}.

%There are several facets of these results to be looked at. One interesting 
%piece of information comes from the correlation between $m_c$ and 
%$m_b$ extracted from the fit. It corresponds 
%to $m_c(1~{\rm{GeV}}) = 1.63 \times (m_b(1~{\rm{GeV}})-3.91)$. 
%Therefore a competitive value of the charm mass can be extracted from a 
%precise determination of $m_b$. Using, for instance, 
%$m_b(1~{\rm{GeV}}) = (4.60 \pm 0.05)$~GeV would give 
%$m_c(1~{\rm{GeV}}) = (1.13 \pm 0.09)$~GeV. This can be 
%compared to the present typical lattice uncertainties which range 
%between 50 and 120~MeV~\cite{mc2}. 

\subsection{Implications for $|V_{cb}|$}
The value of $\Vcb$ obtained from the total semileptonic decay width depends on the 
OPE parameters extracted above. We discuss now the implications of our results for 
$\Vcb$, using the input parameters given in Table~\ref{tab:meast},
which correspond to measurements obtained at LEP. 
%The uncertainties 
%on the ${\rm BR}(b \rightarrow X \ell^- \nu)$ have been increased compared to 
%ref.~\cite{pdg} for not using the heavy quark forward-backward asymmetries in the 
%LEP global electroweak fit and to account for the $\pm15\%$ uncertainty on the 
%equality of s.l.\ partial width of $b$ baryons and mesons.

\begin{table}[ht!]
\begin{center}
\begin{tabular}{|c|c|c|}
\hline
Measurement & Value & Reference\\
\hline
$b$-hadron lifetime & $1.573 \pm 0.007~{\rm ps}$ & HFAG Winter 2003~\cite{ref:hfag}\\
${\rm BR}(b \rightarrow X \ell^- \nu)$ & $(10.65 \pm 0.23)~\%$ & LEPEWWG/2003-01 \cite{ref:LEPEWWG}\\
${\rm BR}(b \rightarrow {\rm X}_u \ell^- \nu)$ & $(0.17 \pm 0.05)~\%$&PDG 2002~\cite{ref:PDG02} \\
\hline
\end{tabular}
\caption[]{\it {Input values, obtained at LEP, used for the determination of $|V_{cb}|$.}
\label{tab:meast}}
\end{center}
\end{table}

The determination of $\Vcb$ and the contributions of the various parameters in 
the kinetic mass scheme are described in~\cite{urabig}. This approach has been 
preferred to the pole mass scheme as it does not rely on an expansion
in $1/m_c$ and also because corrections contributing at order $1/m_b^3$ have 
been fixed by experiment. 
%An approximate 
%formula which displays the dependence on the different parameters is:
%{\small
%\begin{eqnarray}
%\Vcb = \Vcb_0 \!\!& &\left [ 1 -0.65 \left ( m_b(1) -4.6~\GeVcd \right )
%+0.40 \left ( m_c(1) -1.15~\GeVcd \right ) \right. \nonumber \\
%& & +0.01 \left ( \mu_{\pi}^2 -0.4~\GeVd \right )
%+0.10 \left ( \rho_D^3 -0.12~\GeVt \right ) \nonumber \\
%& & \left. +0.06 \left ( \mu_G^2-0.35~\GeVd \right ) 
%-0.01 \left ( \rho_{LS}^3 +0.17~\GeVt \right ) \right ].
%\label{eq:vcbnum}
%\end{eqnarray}
%}
%In summary, we obtain:
Using the expression of $\Vcb$ quoted in \cite{ref:bigi2}
\footnote{We have used the Equation~(\ref{eq:bw}) given in \cite{ref:bigi2}.}, 
it gives:
\begin{equation}
\Vcb = 0.0421 \times \left ( 1 \pm 0.014_{\,meas.}
\pm 0.014_{\,fit} 
\pm 0.015_{\,th.} \right ),
\label{eq:vcb}
\end{equation}
where the first uncertainty reflects the accuracy on the semileptonic width 
determination. This experimental uncertainty, corresponding to LEP-alone 
results, can be reduced to $\pm 1\%$ by including recent measurements obtained
at B-factories which gave~\cite{ref:hfag}: \\
${\rm BR}(b \rightarrow X \ell^- \overline \nu_\ell)=(10.73 \pm 0.28)\%$, 
in agreement with
the LEP measurement and having a  similar accuracy. The second uncertainty,
given in Equation~(\ref{eq:vcb}), corresponds to uncertainties from the fit
of the parameters, obtained in Table \ref{tab:5} and not including the theoretical uncertainties given in the last column. The third uncertainty has been 
taken from the result given in \cite{ref:bigi2}.

\section{Conclusions}

Production characteristics of $\Dsstar$ mesons in $b$-hadron semileptonic
decays have been studied using exclusively reconstructed decay channels.

The total production fraction has been measured to be:
\begin{equation}
{\rm BR}(\Bdb \rightarrow \Dsstar \ell^- \overline{\nu}_{\ell})=
(2.7 \pm 0.7 \pm 0.2)\%. \nonumber
\end{equation}

Decay final states are dominated by the ${\rm D}^{(*)}\pi $ channel
and no-evidence for a signal in channels with two pions has been obtained: 
\begin{equation}
{\rm BR}(b \rightarrow \Do \pi^+ \pi^- \ell^- \overline{\nu}_{\ell})~=~
{\rm BR}(b \rightarrow \Dp \pi^+ \pi^-\ell^- \overline{\nu}_{\ell})~
<~0.18~\%~{\rm at}~ 90~\%~{\rm C.L.} \nonumber
\end{equation}
\begin{equation}
{\rm BR}(b \rightarrow \Dstarp \pi^+ \pi^-\ell^- \overline{\nu}_{\ell})
~<~0.13~\%~{\rm at}~ 90~\%~{\rm C.L.}\nonumber
\end{equation}

The dominant contributing channel is the broad ${\rm D}_1^*$ whose mass and
total width have been measured to be:
\begin{eqnarray}
m_{D_1^*} & = & 2445 \pm 34 \pm 10~\MeVcd \nonumber\\
\Gamma_{D_1^*} & = & 234 \pm 74 \pm 25~\MeVcd. \nonumber
\end{eqnarray}
%% ** A.O. ** 4/5/04 ** establish firmly -> firmly establish **
Broad ${\rm D}\pi$ final states favour a production which is maximum close to 
threshold, as is expected from non-resonant production, 
but the present statistics do not allow this feature to be firmly established.

Moments of the hadronic mass distribution corresponding to $\Dsstar$ states 
in $b$-hadron semileptonic decays
%and of the total hadronic mass distribution 
have been measured:
\begin{eqnarray}
<m_{D^{**}}> &=&2.483 \pm 0.033 \pm 0.033~\GeVcd \nonumber\\
<m_{D^{**}}^2> &=&6.22 \pm 0.16\pm 0.15~(\GeVcd)^2 \nonumber\\
<m_{D^{**}}^4> & =&40.1 \pm 2.0\pm 1.7~(\GeVcd)^4 \nonumber \\
<m_{D^{**}}^6> & =&271 \pm 21 \pm 16~(\GeVcd)^6 \nonumber \\
<m_{D^{**}}^8> & =& (19.3 \pm 2.1 \pm 1.4)\,10^{2}~(\GeVcd)^8 \nonumber \\
<m_{D^{**}}^{10}> & =&(14.7 \pm 2.0\pm 1.2)\,10^{3}~(\GeVcd)^{10} \nonumber
\end{eqnarray}

Using these results and world averaged measurements for the exclusive 
$b$-hadron semileptonic decay fractions into a D or a $\Dstar$ meson, various
moments of the full hadronic mass distribution have been obtained
in Section \ref{sec:mx}. 

%
%and
%\begin{eqnarray}
%<m_H^2-m_{spin}^2> &= & (0.647 \pm 0.046 \pm  0.093 )~(\GeVcd)^2  \nonumber\\
%<(m_H^2-m_{spin}^2)^2> &= & (1.98 \pm 0.23 \pm 0.29 )~(\GeVcd)^4  \nonumber \\
%<(m_H^2-<m_H^2>)^2> &= & (1.56 \pm 0.18 \pm 0.17 )~(\GeVcd)^4  \nonumber \\
%<(m_H^2-<m_H^2>)^3> &= & (4.05 \pm 0.74 \pm 0.31 )~(\GeVcd)^6 \nonumber 
%\end{eqnarray}

Moments of the lepton energy spectrum in semileptonic B decays have also been 
measured as:

\begin{center}
$<E_{\ell}^*>$ =                  ~1.3782 $\pm$ 0.0073 $\pm$ 0.0092~GeV   \\
$<(E_{\ell}^*-<E_{\ell}^*>)^2>$ = ~0.1838 $\pm$ 0.0058 $\pm$ 0.0055~GeV$^2$\\ 
$<(E_{\ell}^*-<E_{\ell}^*>)^3>$ = -0.0301 $\pm$ 0.0015 $\pm$ 0.0036~GeV$^3$.\\
\end{center}

These results are interpreted in terms of constraints on the values of
heavy quark 
masses, of the $b$-quark kinetic energy and of the parameters
contributing at order $1/m_b^3$ in theoretical expressions for the
$b$-hadron semileptonic partial decay width. The values obtained are:
\begin{eqnarray}
m_b(1~{\rm{GeV}}) & =& 4.591 \pm 0.062  \pm 0.039 \pm 0.005~\GeVcd \nonumber \\
m_c(1~{\rm{GeV}})& =& 1.170  \pm 0.093 \pm 0.055 \pm 0.005~\GeVcd \nonumber \\
\mu_{\pi}^2(1~{\rm{GeV}})& =& 0.399  \pm 0.048 \pm 0.034 \pm 0.087~{\rm GeV}^2 \nonumber \\
\tilde{\rho}_D^3 & = &0.053 \pm 0.017 \pm 0.011 \pm 0.026~{\rm GeV}^3, \nonumber 
\end{eqnarray}
and include corrections at order $1/m_b^3$.

Using these results, and inclusive measurements of the $b$-hadron lifetime and semileptonic 
branching 
fraction obtained at LEP, an accurate determination of the value of the $\Vcb$
element has been obtained:
\begin{equation}
\Vcb = 0.0421 \times \left ( 1 \pm 0.014_{\,meas.}
\pm 0.014_{\,fit} 
\pm 0.015_{\,th.} \right ). \nonumber
\end{equation}
The first uncertainty becomes $\pm 1\%$ if measurements 
of $\Bdb$ and $\Bm$ lifetime and semileptonic decay rates,
obtained at the $\Upsilon$(4S), are included.

\section{Acknowledgements}
%%%For the conf note:
%We had useful discussions on the properties of the non-resonant 
%D$^{(*)}\pi$ component with A. Le Yaouanc.
%I. Bigi gave us guidance through the theoretical formalism.
%They are warmly thanked. 
%

We would like to thank P. Gambino and N. Uraltsev for helpful discussions
during this work and for providing us with the necessary theoretical 
expressions. We had also useful discussions on the properties of the 
non-resonant D$^{(*)}\pi$ component with A. Le Yaouanc. I. Bigi is warmly 
thanked for his guidance through the theoretical formalism.

 We are greatly indebted to our technical 
collaborators, to the members of the CERN-SL Division for the excellent 
performance of the LEP collider, and to the funding agencies for their
support in building and operating the DELPHI detector.
We acknowledge in particular the support of \\
Austrian Federal Ministry of Education, Science and Culture,
GZ 616.364/2-III/2a/98, \\
FNRS--FWO, Flanders Institute to encourage scientific and technological 
research in the industry (IWT), Belgium,  \\
FINEP, CNPq, CAPES, FUJB and FAPERJ, Brazil, \\
Czech Ministry of Industry and Trade, GA CR 202/99/1362,\\
Commission of the European Communities (DG XII), \\
Direction des Sciences de la Mati$\grave{\mbox{\rm e}}$re, CEA, France, \\
Bundesministerium f$\ddot{\mbox{\rm u}}$r Bildung, Wissenschaft, Forschung 
und Technologie, Germany,\\
General Secretariat for Research and Technology, Greece, \\
National Science Foundation (NWO) and Foundation for Research on Matter (FOM),
The Netherlands, \\
Norwegian Research Council,  \\
State Committee for Scientific Research, Poland, SPUB-M/CERN/PO3/DZ296/2000,
SPUB-M/CERN/PO3/DZ297/2000, 2P03B 104 19 and 2P03B 69 23(2002-2004)\\
JNICT--Junta Nacional de Investiga\c{c}\~{a}o Cient\'{\i}fica 
e Tecnol$\acute{\mbox{\rm o}}$gica, Portugal, \\
Vedecka grantova agentura MS SR, Slovakia, Nr. 95/5195/134, \\
Ministry of Science and Technology of the Republic of Slovenia, \\
CICYT, Spain, AEN99-0950 and AEN99-0761,  \\
The Swedish Natural Science Research Council,      \\
Particle Physics and Astronomy Research Council, UK, \\
Department of Energy, USA, DE-FG02-01ER41155, \\
EEC RTN contract HPRN-CT-00292-2002.\\

\newpage

\appendix

\section{Error matrices for hadronic mass moments}
\label{append:hadmom}

In these matrices, which refer to moments
$ M_1^H$, $M_2^H$, $M_3^H$, $M_4^H$, $M_5^H$, $M_2^{\prime H}$, $M_3^{\prime H}$, $M_4^{\prime H}$ and $M_5^{\prime H}$ given in this order,
the diagonal elements are the errors and non-diagonal elements correspond to 
correlation coefficients.

\noindent Statistical error matrix:
\begin{center}
$\left(
\begin{array}{c c c c c c c c c}
0.0455 & 0.947 & 0.865 & 0.812 &0.921 &0.907 &0.725 &0.724 & 0.698\\
  & 0.232     & 0.978 & 0.951 &0.996 &0.994 &0.904 &0.899 & 0.878 \\
  &       &   1. 29   & 0.994 &0.992 &0.994 &0.973 &0.969 & 0.956\\
  &       &       &   7.92    &0.974 &0.975 &0.990 &0.990 & 0.983\\
  &       &       &           &1080. &0.998 &0.936 &0.933 & 0.916 \\ 
  &       &       &           &      &0.176 &0.943 &0.936 & 0.918\\ 
  &       &       &           &      &      &0.736 &0.998 & 0.993\\ 
  &       &       &           &      &      &      &4.53  & 0.998\\ 
  &       &       &           &      &      &      &      & 27.0\\ 
\end{array}
\right )$
\end{center}
 Error matrix for systematics:
\begin{center}
$\left(
\begin{array}{c c c c c c c c c}
0.0896 & 0.981 & 0.949 & 0.905 &0.863 &0.946 &0.655 &0.664 & 0.540\\
  & 0.273     & 0.981 & 0.945 &0.908 &0.990 &0.749 &0.747 & 0.630 \\
  &       &   1.09   & 0.987 &0.967 &0.980 &0.848 &0.853 & 0.758\\
  &       &       &   5.42    &0.995 &0.952 &0.910 &0.918 & 0.844\\
  &       &       &           &32.3 &0.920 &0.937 &0.950 & 0.891 \\ 
  &       &       &           &      &0.161 &0.801 &0.789 & 0.680\\ 
  &       &       &           &      &      &0.318 &0.992 & 0.974\\ 
  &       &       &           &      &      &      &2.06  & 0.986\\ 
  &       &       &           &      &      &      &      & 11.7\\ 
\end{array}
\right )$
\end{center}

 Total error matrix:
\begin{center}
$\left(
\begin{array}{c c c c c c c c c}
0.100 & 0.945 & 0.844 & 0.759 &0.440 &0.872 &0.533 &0.544 & 0.481\\
  & 0.358     & 0.966 & 0.914 &0.664 &0.984 &0.763 &0.765 & 0.712 \\
  &       &   1.69   & 0.987 &0.778 &0.987 &0.900 &0.903 & 0.866\\
  &       &       &   9.59    &0.820 &0.957 &0.954 &0.959 & 0.934\\
  &       &       &           &1080. &0.755 &0.870 &0.860 & 0.851 \\ 
  &       &       &           &      &0.239 &0.853 &0.849 & 0.804\\ 
  &       &       &           &      &      &0.802 &0.997 & 0.990\\ 
  &       &       &           &      &      &      &4.98  & 0.996\\ 
  &       &       &           &      &      &      &      & 29.5\\ 
\end{array}
\right )$
\end{center}

\newpage

\section{Error matrices for lepton energy moments}
\label{append:leptmom}
In these matrices, which refer to moments
$ M_1^l$, $M_2^{\prime \ell}$ and $M_3^{\prime \ell}$
given in this order,
the diagonal elements are the errors and non-diagonal elements correspond to 
correlation coefficients.

\noindent Statistical error matrix:
\begin{center}
$\left(
\begin{array}{r r r }
0.0073 & -0.6041 & -0.3435  \\
       &  0.0058 & -0.5381 \\
       &         &  0.0015 \\  
\end{array}
\right )$
\end{center}
 Error matrix for systematics:
\begin{center}
$\left(
\begin{array}{r r r }
0.0092 & -0.7823   & -0.2427  \\
       &  0.0055    & 0.0342 \\
       &           &  0.0036 \\  
\end{array}
\right )$
\end{center}
 Total error matrix:
\begin{center}
$\left(
\begin{array}{r r r }
0.0118 & -0.6942  &  -0.2578  \\
       & 0.0080  &  -0.1286 \\
       &         &   0.0039 \\  
\end{array}
\right )$
\end{center}


\newpage

\begin{thebibliography}{ref99}

\bibitem{ref:bigiold}
I. Bigi, M. Shifman and N.G. Uraltsev, Ann. Rev. Nucl. Part. Sci. {\bf 47} (1997) 591 and references therein.


\bibitem{ref:bigi2}
D. Benson, I.I. Bigi, T. Mannel and N. Uraltsev, Nucl. Phys. {\bf B665} (2003) 367, [hep-ph/0302262]. 

\bibitem{ref:sirlin}
A. Sirlin,  Nucl. Phys. {\bf B71} (1964) 29, and Rev. Mod. Phys. {\bf 50} 
(1978) 573. Erratum-ibid.  {\bf 50} (1978) 905.

\bibitem{ref:chay}
J. Chay, H. Georgi and B. Grinstein, Phys. Lett.  {\bf B247} (1990) 399.

\bibitem{Bigi:1992su}
I.~I.~Y.~Bigi, N.~G.~Uraltsev and A.~I.~Vainshtein,
%``Nonperturbative corrections to inclusive beauty and charm decays:
% QCD versus phenomenological models,''
Phys.\ Lett.\ {\bf B293} (1992) 430.
Erratum-ibid.\ {\bf B297} (1993) 477, [hep-ph/9207214].
%%CITATION = HEP-PH 9207214;%%

\bibitem{Uraltsev:2001ih}
N.~Uraltsev,
Phys.\ Lett.\ {\bf B545} (2002) 337
[hep-ph/0111166]. 

\bibitem{ref:amsterl}
M. Battaglia {\it et al.}  Phys. Lett. {\bf B556} (2003) 41, [hep-ph/0210319].

\bibitem{urabig}
N. Uraltsev, Mod. Phys. Lett. {\bf A17} (2002) 2317,
[hep-ph/0210413].

\bibitem{cleo_mom1}
%J.~Bartelt {\it et al.} (CLEO Collaboration), CLEO-CONF 98-21.
N.E. Adam  {\it et al.} (CLEO Collaboration), Phys. Rev. {\bf D67} (2003) 032001.

\bibitem{cleo_mom2}
D.~Cronin-Hennessy {\it et al.} (CLEO Collaboration), Phys. Rev. Lett.
{\bf 87} (2001) 251808.

\bibitem{cleo_mom3}
S.E. Csorna {\it et al.} (CLEO Collaboration), Phys. Rev. {\bf D70} (2004) 032002
[hep-ex/0403052]. 

\bibitem{babar_mom0}
B. Aubert {\it et al.} (BaBar Collaboration), BABAR-CONF-02/029, 
SLAC-PUB-9314, [hep-ex/0207084].

\bibitem{babar_mom1}
B. Aubert {\it et al.} (BaBar Collaboration), Phys. Rev. {\bf D69} (2004) 111103, 
[hep-ex/0403031].

\bibitem{babar_mom2}
B. Aubert {\it et al.} (BaBar Collaboration), Phys. Rev. {\bf D69} (2004) 111104, 
[hep-ex/0403030].

\bibitem{belle_mom1}
K. Abe {\it et al.} (Belle Collaboration), BELLE-CONF-0474, [hep-ex/0409015].

\bibitem{belle_mom2}
K. Abe {\it et al.} (Belle Collaboration), BELLE-CONF-0426, [hep-ex/0408139].

\bibitem{cdf_mom}
D. Acosta  {\it et al.} (CDF Collaboration), Phys. Rev. {\bf D71} (2005) 051103,
[hep-ex/0502003].

\bibitem{ref:dsstarargus}
H. Albrecht { \it et al.}, (ARGUS Collaboration), 
Z. Phys. {\bf C57} (1993) 533.

\bibitem{ref:dsstaraleph}
D. Buskulic { \it et al.}, (ALEPH Collaboration), 
Z. Phys. {\bf C73} (1997) 601.

\bibitem{ref:dsstarrosner}
J.L. Rosner, Comments Nucl. Part. Phys. {\bf 16} (1986) 109.

\bibitem{ref:PDG02}
K. Hagiwara {\it et al.}, (Particle Data Group), Phys. Rev. {\bf D66} (2002) 010001.

\bibitem{ref:narrowcleo}
A. Anastassov { \it et al.}, (CLEO Collaboration), 
Phys. Rev. Lett. {\bf 80} (1998) 4127.

\bibitem{ref:quarkmod}
S. Godfrey and N. Isgur, Phys. Rev. {\bf D32} (1985) 189.

\bibitem{ref:goity}
J.L. Goity and W. Roberts, Phys. Rev. {\bf D51} (1995) 3459.


\bibitem{ref:dsstardelphi}
P. Abreu { \it et al.}, (DELPHI Collaboration), 
Phys. Lett. {\bf B475} (2000) 407.

%\bibitem{ref:cleomom}
%D. Cronin-Hennessy { \it et al.}, (CLEO Collaboration), Phys. Rev. Lett. {\bf 87} (2001) 251808.
%\bibitem{Falk} 
%A.~F.~Falk and M.~Luke, Phys. Rev. {\bf D57} (1998) 424.

\bibitem{ref:luc}
T. Sj\"ostrand, Comp. Phys. Comm. {\bf 82} (1994) 74.

\bibitem{ref:delsim}
P. Abreu{ \it et al.}, (DELPHI Collaboration), Nucl. Instr. and Meth. {\bf A378} (1996) 57,
Erratum-ibid. {\bf A396} (1997) 281.

\bibitem{ref:btag}
J. Abdallah { \it et al.}, (DELPHI Collaboration), Eur. Phys. J. {\bf C32} (2004) 185.\\



%\bibitem{muonid}
%G.R. Wilkinson, { \em Improvements to the Muon Identification in the 
%94C2 Short DST Production } DELPHI 97-37 PHYS 690. 

%\bibitem{eleid}
%C. Kreuter, {\em Electron Identification using a Neural Network}
%DELPHI 96-169 PHYS 658. 

\bibitem{ref:companion}
J. Abdallah { \it et al.}, (DELPHI Collaboration), Eur. Phys. J. {\bf C33} 
(2004) 213, [hep-ex/0401023]
%DELPHI Collaboration, DELPHI 2002-074 CONF 608, contribution to ICHEP2002.

\bibitem{ref:hmcmll}
R. Barlow and C. Beeston, Comp. Phys. Comm. {\bf 77} (1993) 219.

\bibitem{twocharma}
R. Barate {\it et al.}, (ALEPH Collaboration), Eur. Phys. J. {\bf C4} (1998) 387.

\bibitem{twocharmb}
B. Aubert {\it et al.},  (BaBar Collaboration), Phys. Rev. {\bf D68} (2003) 
092001, [hep-ex/0305003]. 
%P. Robbe, BaBar Collaboration, Ph. D Thesis (2002).

%\bibitem{ref:PDG00}
%D.E. Groom {\it et al.}, Eur. Phys. J. {\bf C15} (2000) 1.

%\bibitem{ref:dsstaraleph}
%D. Buskulic {\it et al.}, ALEPH Collaboration, Z. Phys.{\bf C73} (1997) 601.

\bibitem{ref:lephf}
Combined results on $b$-hadron production rates, lifetimes, oscillations
and semileptonic decays (ALEPH, CDF, DELPHI, L3, OPAL, SLD),
CERN-EP/2001-050, [hep-ex/0112028].

\bibitem{ref:argcle}
H. Albrecht {\it et al.}, (ARGUS Collaboration), Phys. Lett. {\bf B232} (1989) 398;\\
P. Avery  {\it et al.}, (CLEO Collaboration), Phys. Rev. {\bf D41} (1990) 774, 
Phys. Lett. {\bf B331} (1994) 236 and Erratum-ibid. {\bf B342} (1995) 453.

\bibitem{ref:cleobroad}
S. Anderson {\it et al.} (CLEO Collaboration), Nucl.\ Phys.\ {\bf A663} (2000) 647. 

\bibitem{ref:bellebroad}
K. Abe {\it et al.} (BELLE Collaboration), Phys. Rev. {\bf D69} (2004) 112002, [hep-ex/0307021]. 

\bibitem{ref:ghat}
A. Anastassov {\it et al.} (CLEO Collaboration), Phys. Rev. {\bf D65} (2002) 032003.

\bibitem{ref:hfag}
Heavy Flavour Averaging Group (http://www.slac.stanford.edu/xorg/hfag/).

\bibitem{ref:macrib}
Z. Albrecht, M. Feindt, M. Moch, 'MACRIB- High 
Efficiency, High Purity Hadron Identification for DELPHI', DELPHI internal 
note, 99-150 RICH 95 (1999),  [hep-ex/0111081]. 

%\bibitem{emiss}
%P. Abreu {\it et al.} (DELPHI Collaboration), Zeit.\ Phys.\ {\bf C 71} 
%(1996), 539.

\bibitem{blobel}
V.~Blobel, 
OPAL technical Note TN361 (1996) 

[http://www-zeus.desy.de/$\sim$desler/blobel.html].

\bibitem{Atwood:1989em}
D.~Atwood and W.~J.~Marciano,
%``Radiative Corrections And Semileptonic B Decays,''
Phys.\ Rev.\ {\bf D41} (1990) 1736.
%%CITATION = PHRVA,D41,1736;%%

\bibitem{Karlsrue}
A Study of the $b$-Quark Fragmentation Function with the DELPHI Detector 
at LEP I, G. Barker {\it et al.} (DELPHI Collaboration), ICHEP2002,
DELPHI note 2002-069-CONF-603.
%******************************************************************

%\bibitem{ref:bigietal}
%I.I. Bigi {\it et al.}, Phys. Lett. {\bf B293} (1992) 430 
%[(E) {\it ibid.} {\bf B297} (1993) 477]; 
%I.I. Bigi {\it et al.}, Phys. Rev. Lett. {\bf 71} (1993) 496;
%A.V. Manohar and M.B. Wise, Phys. Rev. {\bf D49} (1994) 1310;
%B. Blok {\it et al.}, Phys. Rev. {\bf D49} (1994) 3356
%[(E) {\it ibid.} {\bf D50} (1994) 3572];
%T. Mannel, Nucl. Phys. {\bf B413} (1994) 396.  

%\bibitem{ref:mammoth}
%M. Feindt, W. Oberschulte, C. Weiser, DELPHI note, 96-52 PROG 216. 

%\bibitem{ref:leptons}
%K.D. Brand, I. Roncagliolo, F. Simonetto, 'Electron Identification for Electro-
%Weak b,c Physics',  DELPHI note 96-23 PHYS 598.\\
%P. Abreu {\it et al.}, DELPHI Collaboration. Eur. Phys. J {\bf C20} (2001) 455


%\bibitem{ref:firstvcb}
%P. Abreu {\it et al.}, DELPHI Collaboration, Z. Phys. {\bf C71} (1996) 539.

%******************************************************************

%\bibitem{ref:physlmom}
%M. Battaglia, M. Calvi, P. Gambino, A. Oyanguren, P. Roudeau, L. Salmi, 
%J. Salt, A. Stocchi and N. Uraltsev,
%Phys. Lett. {\bf B556} (2003) 41, [hep-ph/0210319].


\bibitem{ref:paolo_uraltsev}
P.~Gambino, N.~Uraltsev, Eur. Phys. J. {\bf C34} (2004) 181. 

\bibitem{ref:fits_referee}
C.W.~Bauer {\it et al.} Phys.\,Rev.\ {\bf D70} (2004) 094017, [hep-ph/0408002];

B. Aubert {\it et al.} (BaBar Collaboration), Phys. Rev. Lett. {\bf 93} (2004) 011803, [hep-ex/0404017]; O. Buchm\"{u}ller and H. Fl\"{a}cher, [hep-ph/0507253]. 

\bibitem{ref:pdg2004}
S. Eidelman {\it et al.}, (Particle Data Group), Phys. Lett. {\bf B592} (2004) 1.

\bibitem{mb}
Heavy Quarkonium Dynamics, A.~Hoang, [hep-ph/0204299].

\bibitem{melyel}
K.\,Melnikov and A.\,Yelkhovsky, Phys.\,Rev.\ {\bf D59} (1999)
114009.


\bibitem{ref:falketal}
A.F.~Falk and M.E.~Luke, Phys.\,Rev.\ {\bf D57} (1998) 424, [hep-ph/9708327]. 

%\bibitem{mc2}
%D.~Becirevic, V.~Lubicz, G.~Martinelli,
%Phys.\ Lett.\ B {\bf 524} (2002) 115, 
%[arXiv:hep-ph/0107124]; 
%J.~Rolf and S.~Sint
%Nucl.\ Phys.\ (Proc.\ Suppl.) 106 (2002) 239,
%[arXiv:hep-ph/0110139].

%\bibitem{ref:ston1}
%S. Stone, ``Experimental results in Heavy Flavor Physics'', [hep-ph/0310153],
%plenary talk at International Europhysics Conference on High Energy Physics 
%EPS (July 17th-23rd 2003) in Aachen, Germany.

%\bibitem{ref:artus1}
%See for instance the section on ``Determination of $\Vcb$'' 
%by M. Artuso and E. Barberio in PDG 2002 (K. Hagiwara {\it et al.} Phys. Rev. 
%{\bf D66} (2003) 010001 or M. Artuso in ``B meson semileptonic decays''
%(FPCP Conference, Paris 2003) [hep-ph/0309104].

%\bibitem{ref:ural1}
%I.I. Bigi, M. Shifman, N. Uraltsev and A. Vainshtein, Phys. Rev. {\bf D56}
%(1997) 4017.

%\bibitem{ref:buchalla}
%G. Buchalla, ``Heavy Quark Theory'', lectures given at the 55$^{th}$
%Scottish Universities Summer School in Physics, University of St. Andrews, Scotland
%(7-23 August 2001), [hep-ph/0202092].

%\bibitem{ref:bigidual}
%I.I. Bigi and N. Uraltsev, Int. J. Mod. Phys. {\bf  A16} (2001) 5201, 
%[hep-ph/0106346].

%\bibitem{ref:isgurdual}
%N. Isgur, Phys. Lett. {\bf  B448} (1999) 111, [hep-ph/9811377].

%\bibitem{ref:leyaou}
%A. Le Yaouanc {\it et al.}, Phys. Lett. {\bf  B480} (2000) 119, 
%[hep-ph/0003087];
%  Phys. Rev. {\bf  D62} (2000) 074007, [hep-ph/0004246];
% Phys. Lett. {\bf  B517} (2001) 135, [hep-ph/0103339];
% Phys. Lett. {\bf  B488} (2000) 153, [hep-ph/0005039].

%\bibitem{ref:blm}
%S.J.~Brodsky, G.P.~Lepage and P.B.~Mackenzie, Phys. Rev. {\bf D28} (1983) 228.

\bibitem{ref:LEPEWWG}
A combination of preliminary electroweak measurements and constraints on the 
Standard Model, CERN-EP/2003-091, [hep-ex/0312023].

%\bibitem{urabig}
%N.~Uraltsev, {\it QCD corrections in $\Gamma_{\rm sl}(B)$}, Bicocca-FT-02-22,
%[hep-ph/0210413].  

%\bibitem{ref:pdgvcb}
%Mini-review on the determination of $\Vcb$,
%S. Eidelman {\it et al.}, Phys. Lett. {\bf B592} (2004) 1.

%\bibitem{ref:marinat}
%M. Artuso, Plenary talk at EPS HEP 99, Tampere, Finland, July 1999,
%[hep-ph/9911347].

%\bibitem{ref:ural1}
%I.I. Bigi, M. Shifman, N. Uraltsev and A. Vainshtein, Phys. Rev. {\bf D56}
%(1997) 4017.

%\bibitem{ref:ckmw}
%Workshop on the CKM unitarity Triangle, CERN Geneva 2002, CERN-EP/2003-002-rev,
%[hep-ph/0304132].

%\bibitem{ref:ston1}
%S. Stone, ``Experimental results in Heavy Flavor Physics'', [hep-ph/0310153],
%plenary talk at International Europhysics Conference on High Energy Physics 
%EPS (July 17th-23rd 2003) in Aachen, Germany.

%\bibitem{ref:bigidual}
%I.I. Bigi and N. Uraltsev, Int. J. Mod. Phys. {\bf  A16} (2001) 5201, 
%[hep-ph/0106346].

%\bibitem{ref:artus1}
%See for instance the section on ``Determination of $\Vcb$'' 
%by M. Artuso and E. Barberio in PDG 2002 (K. Hagiwara {\it et al.} Phys. Rev. 
%{\bf D66} (2003) 010001 or M. Artuso in ``B meson semileptonic decays''
%(FPCP Conference, Paris 2003) [hep-ph/0309104].

%\bibitem{ref:buchalla}
%G. Buchalla, ``Heavy Quark Theory'', lectures given at the 55$^{th}$
%Scottish Universities Summer School in Physics, University of St. Andrews, Scotland
%(7-23 August 2001), [hep-ph/0202092].

%\bibitem{ref:isgurdual}
%N. Isgur, Phys. Lett. {\bf  B448} (1999) 111, [hep-ph/9811377].

%\bibitem{ref:leyaou}
%A. Le Yaouanc {\it et al.}, Phys. Lett. {\bf  B480} (2000) 119, 
%[hep-ph/0003087];
%Phys. Rev. {\bf  D62} (2000) 074007, [hep-ph/0004246];
%Phys. Lett. {\bf  B517} (2001) 135, [hep-ph/0103339];
%Phys. Lett. {\bf  B488} (2000) 153, [hep-ph/0005039].


\end{thebibliography}
\end{document}